\pdfoutput=1

\documentclass[a4paper,11pt]{article}
\usepackage{jheppub}

\usepackage{bm}

\usepackage{color}
\usepackage[usenames,dvipsnames,svgnames,table]{xcolor}

\usepackage{amsmath}
\usepackage{amssymb}
\usepackage{graphicx}
\usepackage{slashed}
\usepackage{soul}
\usepackage{lscape}
\usepackage{graphicx}
\usepackage{xcolor}
\usepackage{multirow}
\usepackage{placeins}

\newcommand{\vspCW}{\ensuremath{v_{S, {\rm CW}}'}}

\begin{document}

\title{Nucleation is More than Critical \\
\large A Case Study of the Electroweak Phase Transition in the NMSSM
}

\author[a]{Sebastian~Baum,}
\author[b,c]{Marcela~Carena,}
\author[d]{Nausheen~R.~Shah,} 
\author[c,e]{Carlos~E.~M.~Wagner,}
\author[b,c]{Yikun~Wang}

\affiliation[a]{Stanford Institute for Theoretical Physics, Stanford University, Stanford, CA 94305, USA}
\affiliation[b]{Fermi National Accelerator Laboratory, P.~O.~Box 500, Batavia, IL 60510, USA}
\affiliation[c]{Enrico Fermi Institute and Kavli Institute for Cosmological Physics, University of Chicago, Chicago, IL 60637, USA}
\affiliation[d]{Department of Physics \& Astronomy, Wayne State University, Detroit, MI 48201, USA}
\affiliation[e]{HEP Division, Argonne National Laboratory, 9700 Cass Ave., Argonne, IL 60439, USA}

\emailAdd{sbaum@stanford.edu}
\emailAdd{carena@fnal.gov}
\emailAdd{nausheen.shah@wayne.edu}
\emailAdd{cwagner@uchicago.edu}
\emailAdd{yikwang@uchicago.edu}

\preprint{FERMILAB-PUB-20-490-T
\\\phantom{0} \hfill EFI-20-18
\\\phantom{0} \hfill WSU-HEP-2004}

\abstract{
Electroweak baryogenesis is an attractive mechanism to generate the baryon asymmetry of the Universe via a strong first order electroweak phase transition. We compare the phase transition patterns suggested by the vacuum structure at the {\it critical temperatures}, at which local minima are degenerate, with those obtained from computing the probability for {\it nucleation} via tunneling through the barrier separating local minima. Heuristically, nucleation becomes difficult if the barrier between the local minima is too high, or if the distance (in field space) between the minima is too large. As an example of a model exhibiting such behavior, we study the Next-to-Minimal Supersymmetric Standard Model, whose scalar sector contains two $SU(2)$ doublets and one gauge singlet. We find that the calculation of the nucleation probabilities prefers different regions of parameter space for a strong first order electroweak phase transition than the calculation based solely on the critical temperatures. Our results demonstrate that analyzing only the vacuum structure via the critical temperatures can provide a misleading picture of the phase transition patterns, and, in turn, of the parameter space suitable for electroweak baryogenesis.
}


\maketitle
\flushbottom

\section{Introduction}

Cosmological observations suggest that our Universe has a large {\it dark energy} component, and that its matter component is dominated by an unknown form of {\it dark matter}~\cite{Aghanim:2018eyx}. Only 5\,\% of the energy budget of the Universe consist of the particles of the Standard Model (SM), mainly its baryons. Extensive tests at particle accelerators and other laboratory experiments have found no (unambiguous) deviations from the SM predictions. However, whereas the SM accurately describes the behavior of the particles making up the ordinary matter, it fails to give an explanation of how they came to be. 

Under the assumption that particles and anti-particles are produced in equal number in the early Universe, the SM predicts that they would have long annihilated each other without leaving any remnant matter today. As first enunciated by Sakharov~\cite{Sakharov:1967dj}, producing a {\it baryon asymmetry}, i.e. more matter than anti-matter, requires baryon number violation, C and CP violation, and out-of-equilibrium processes to all occur at the same time. The SM does provide for C, CP, and baryon number violation through the electroweak interactions and sphalerons, respectively. The Electroweak Phase Transition (EWPT), however, is a smooth crossover in the SM and, thus, is not giving rise to sufficient deviations from thermal equilibrium~\cite{Morrissey:2012db}. In addition, the amount of C and CP violation in the SM is too small to generate the observed baryon asymmetry~\cite{Gavela:1994dt} even if the EWPT were to provide out of equilibrium conditions.

In order to generate the observed baryon asymmetry, sources of CP violation and out-of-equilibrium processes beyond those found in the SM must be realized in nature. One interesting possibility to achieve the latter is via a Strong First Order Electroweak Phase Transition (SFOEWPT), yielding promising conditions for {\it electroweak baryogenesis}. Accommodating a SFOEWPT demands modifications of the Higgs potential. Such modifications may be induced predominantly by thermal effects, as it happens e.g. in the Minimal Supersymmetric extension of the Standard Model (MSSM)~\cite{Carena:1996wj,Delepine:1996vn,Laine:1998qk,Cline:1998hy,Balazs:2004ae,Lee:2004we,Carena:2008vj}, or by zero-temperature effects that have a lasting consequence after thermal effects are taken into account. The latter situation naturally occurs in models of new physics containing additional light scalar particles with sizable couplings to the Higgs. 

To study the phase transition patterns of models with extended Higgs sectors, most previous works solely rely on analyses of the temperature-dependent vacuum structure via the computation of the {\it critical temperature}, $T_c$, at which two (distinct) local minima of the effective potential become degenerate. While the critical temperature is indicative of the thermal history since it is the temperature at which the role of the global minimum passes from one vacuum phase to another, this calculation does not account for the probability of the associated phase transition actually taking place. First order phase transitions proceed via bubble nucleation, and the probability of the system transitioning from the false vacuum to the (new) true vacuum is computed via the {\it bounce action}, the Euclidean space-time integral over the effective Lagrangian, see, e.g., ref.~\cite{Mazumdar:2018dfl} for a review.

Heuristically, bubble nucleation becomes difficult if the barrier separating two local minima becomes too high, or if the distance (in field space) separating the minima is too large. These conditions occur most readily if multiple scalar fields participate in the phase transition. For the EWPT, the possibility of a SM gauge singlet field participating in the phase transition is particularly interesting. While electroweak precision data tightly constrains the couplings and vacuum expectation values (vevs) of any fields charged under the electroweak symmetry, such constraints do not apply to gauge singlets. Since its couplings are free parameters, a gauge singlet field can radically alter the shape of the effective potential, enabling a SFOEWPT. On the other hand, a gauge singlet may  induce large barriers separating local minima and acquire a large vev during the EWPT, increasing the distance between the local minima and reducing the nucleation probability. Therefore, a careful analysis of these effects is necessary in order to determine the region of parameter space leading to a successful SFOEWPT. 

The Next-to-Minimal Supersymmetric extension of the Standard Model (NMSSM)~\cite{Maniatis:2009re,Ellwanger:2009dp} is a well-motivated example of physics beyond the SM that may solve the hierarchy problem of the electroweak scale~\cite{Witten:1981nf,Dimopoulos:1981zb,Witten:1981kv,Kaul:1981hi,Sakai:1981gr} and provide a dark matter candidate~\cite{Cheung:2014lqa,Cao:2015loa,Badziak:2015exr,Ellwanger:2016sur,Cao:2016nix,Cao:2016cnv,Beskidt:2017xsd,Badziak:2017uto,Baum:2017enm,Ellwanger:2018zxt,Abdallah:2019znp}. Its scalar sector contains a (complex) gauge singlet and two $SU(2)$ doublets, thus, it is well-suited for a case study of the comparison of the phase transition patterns suggested by the critical temperature calculation and those obtained from calculating the nucleation probabilities. Moreover, the NMSSM provides a range of possibilities for C and CP violation beyond what is found in the SM. For example, CP violation can occur in the Higgs sector, or between the superpartners of the SM particles. Assuming the latter, CP violation in the Higgs sector is induced only via (small) quantum corrections, and one can study the EWPT in the CP-conserving limit of the scalar potential. 

The EWPT in the NMSSM has been studied previously in the literature. To the best of our knowledge, Pietroni~\cite{Pietroni:1992in} was the first to consider electroweak baryogenesis in the NMSSM, noting that the dimensionful coupling of the singlet to the Higgs doublets, $A_\lambda$, allows for shapes of the scalar potential suitable for a SFOEWPT at tree level. This is to be contrasted with the situation in the MSSM, where a barrier between the trivial and the physical minimum necessary for a SFOEWPT arises only from thermal effects. Subsequent work on the EWPT in the NMSSM includes refs.~\cite{Davies:1996qn, Huber:2000mg, Menon:2004wv, Funakubo:2005pu, Carena:2011jy, Cheung:2012pg, Balazs:2013cia, Huang:2014ifa, Kozaczuk:2014kva, Bi:2015qva, Huber:2015znp, Bian:2017wfv, Akula:2017yfr, Athron:2019teq}, and work on closely related models can be found in refs.~\cite{Kang:2004pp, Kumar:2011np}. Many of these papers focused on numerical scans of the NMSSM parameter space, aiming at identifying regions of parameter space suitable for realizing a SFOEWPT. Analytic studies have been carried out in refs.~\cite{Pietroni:1992in,Menon:2004wv, Funakubo:2005pu, Carena:2011jy, Balazs:2013cia, Huang:2014ifa}. A common idea in these works was to use parameters shaping the potential in the singlet-only direction to characterize the EWPT.

In the NMSSM, in general, there are ten degrees of freedom in the Higgs sector. In practice it suffices to consider the three-dimensional subspace spanned by the CP-even neutral scalar degrees of freedom. Nevertheless, computing the bounce action in this three-dimensional field space is still numerically expensive, and, until now, results for the phase transition based on the nucleation calculation have only been presented for a few benchmark points in parameter space, see, e.g., refs.~\cite{Huber:2000mg, Carena:2011jy, Kozaczuk:2014kva, Bian:2017wfv, Athron:2019teq}. These studies mainly reported small-to-moderate {\it supercooling}, i.e. nucleation temperatures not much smaller than the corresponding critical temperatures for their benchmark points. More importantly, the thermal histories indicated by the critical temperatures agree with the ones obtained by the nucleation calculation. The notable exception is the recent work of Athron {\it et al.}~\cite{Athron:2019teq}, where results for the nucleation temperatures of four benchmark points were presented: For two of those four points, the authors reported small-to-moderate supercooling, while for the two remaining points the authors found that the nucleation condition could not be satisfied and, hence, the transition pattern indicated by the calculation of the critical temperatures was not a good indicator of the thermal history.

In this work, we present results for the EWPT in the NMSSM based on the nucleation calculation for a broad scan of the parameter space. We use \texttt{CosmoTransitions}~\cite{Wainwright:2011kj} for the calculation of the bounce action, and support our results with analytic studies.\footnote{In this work, we use the nucleation temperature, defined as the temperature at which the tunneling probability from the false to the true vacuum is one per Hubble volume and Hubble time, as a proxy for successful nucleation of a phase transition. We do not compute the {\it percolation temperature}, defined as the temperature at which a given fraction (often taken to be $1/e \sim 37\,\%$) of the Universe's volume has transitioned to the true vacuum. Sizable differences between the nucleation and percolation temperature can appear in the case of very large supercooling, however, a calculation of the percolation temperatures is beyond the scope of this work.} We focus on the region of parameter space where {\it alignment-without-decoupling} is realized in the Higgs sector, and on small-to-moderate values of $\tan\beta$, the ratio of the vevs of the scalar $SU(2)$ doublets. This is motivated by the phenomenology of the 125\,GeV Higgs boson observed at the Large Hadron Collider (LHC). In the NMSSM, a mass of 125\,GeV of the SM-like Higgs boson can be achieved in the low-to-moderate $\tan\beta \lesssim 5$ regime without the need for large radiative corrections. The couplings of this state to SM particles are SM-like if it is (approximately) {\it aligned} with the interaction eigenstate that couples like the SM Higgs boson to other SM particles. In the NMSSM, there are two ways to achieve such alignment: i) the {\it decoupling} limit, that requires the non-SM-like interaction eigenstates to have masses much larger than the SM-like interaction state, and ii) the {\it alignment-without-decoupling} limit, where the parameters of the theory conspire to suppress the mixing of the SM-like interaction state with the non-SM-like interaction states~\cite{Carena:2015moc}. The latter is of particular interest for realizing a SFOEWPT in the NMSSM: in the alignment-without-decoupling limit the non-SM-like states can have masses comparable to that of the SM-like Higgs boson, and hence, they can easily alter the shape of the scalar potential in ways relevant for the EWPT.

The null-results from searches for superpartners at the LHC suggests that the squarks and gluinos are heavy and decoupled from the EWPT. We use an effective field theory approach, integrating out all superpartners except for the neutralinos and charginos. This leaves the full SM particle content, an augmented scalar sector consisting of two $SU(2)$ doublets and a complex singlet, and the electroweakinos (composed of the superpartners of the photon, the $Z$- and $W$-bosons, the two Higgs doublets, and the scalar singlet) as dynamical degrees of freedom; similar approaches have been taken in refs.~\cite{Kozaczuk:2014kva, Athron:2019teq, Kumar:2011np}. In order to maintain the location of the physical minimum in field space, the mass of the SM-like Higgs boson, and the alignment of the singlet-like and SM-like interaction eigenstates after including the radiative corrections to the effective potential from these remaining dynamical degrees of freedom, we add a set of (finite) counterterms, see refs.~\cite{Huber:2015znp,Bi:2015qva, Bian:2017wfv} for similar schemes. 

The outline of our work is as follows: We begin by discussing the scalar sector of the NMSSM in section~\ref{sec:NMSSM}. In section~\ref{sec:RadCorr} we discuss the radiative corrections to the scalar sector of the NMSSM, and in section~\ref{sec:TCorr}, the thermal corrections. After analyzing the zero-temperature vacuum structure of the NMSSM in section~\ref{sec:VacStr}, we discuss the phase transition behavior of the NMSSM in section~\ref{sec:thermal_ana}, in particular, we identify the relevant characteristics of the transition patterns for a SFOEWPT, and develop some analytical intuition for the regions of parameter space where phase transitions can successfully nucleate. In section~\ref{sec:Numerical}, we present our numerical results. In section~\ref{sec:Num_BC} we study the region of parameters in which the proper physical minimum is obtained. We compare the results for the phase transitions obtained from the nucleation calculation with the transition patterns suggested by the critical temperature analysis in section~\ref{sec:Num_Tcrit}. In section~\ref{sec:Num_pheno} we comment on the collider and dark matter phenomenology in the region of parameter space where we find SFOEWPTs. We summarize and present our conclusions in section~\ref{sec:conc}. We present five benchmark points in appendix~\ref{app:BP_points}. Explicit formulae for the field-dependent masses, the finite temperature corrections to the masses, and the equations we use to fix the counterterms are listed in appendices~\ref{app:field_masses},~\ref{app:ct_coeff}, and~\ref{app:Daisy}, respectively. 

Let us here already highlight our main result: We find that the phase transition patterns of given parameter points vary substantially between the critical temperature analysis and the nucleation calculation. Thus, calculating only critical temperatures is not enough to identify the regions of parameter space favorable for electroweak baryogenesis. 

The code used to perform our calculations is available at \url{https://github.com/sbaum90/NMSSM_CosmoTrans.git}.

\section{The Next-to-Minimal Supersymmetric Standard Model} \label{sec:NMSSM}

The Next-to-Minimal Supersymmetric Standard Model augments the particle content of the MSSM by a SM gauge-singlet chiral superfield $\widehat{S}$, see refs.~\cite{Maniatis:2009re,Ellwanger:2009dp} for reviews. The best-studied version of the NMSSM is the $\mathbb{Z}_3$-NMSSM. In this model, an additional discrete symmetry is imposed, under which all left-handed chiral superfields transform as $\widehat{\Phi} \to e^{2\pi i/3} \widehat{\Phi}$ and all gauge superfields transform trivially. An interesting consequence of the $\mathbb{Z}_3$ symmetry is that it renders the superpotential of the NMSSM scale invariant; in particular the Higgsino mass parameter $\mu$ arises from the vacuum expectation value (vev) of the scalar component of the singlet superfield, $S$. Thus, the NMSSM alleviates the MSSM's $\mu$-problem. 

Of greater phenomenological interest is that the NMSSM can accommodate a $125\,$GeV SM-like Higgs boson without the need for large radiative corrections to its mass. Furthermore, the presence of the scalar gauge singlet makes a SFOEWPT easily achievable in the NMSSM~\cite{Pietroni:1992in,Davies:1996qn,Huber:2000mg,Menon:2004wv,Funakubo:2005pu,Huber:2006ma,Kumar:2011np,Carena:2011jy,Cheung:2012pg,Balazs:2013cia,Huang:2014ifa,Kozaczuk:2014kva,Bi:2015qva,Huber:2015znp,Bian:2017wfv,Akula:2017yfr,Athron:2019teq}. This should be contrasted with the situation in the MSSM, where, in the presence of a 125\,GeV SM-like Higgs, the scalar potential is constrained such that a SFOEWPT is only possible if the stops are very light~\cite{Carena:1996wj,Delepine:1996vn,Laine:1998qk,Cline:1998hy,Balazs:2004ae,Lee:2004we,Carena:2008vj}. Such stops have been virtually ruled out by the LHC, not only via direct searches but also by the fact that such light stops would lead to a variation of the Higgs production cross section and decay branching ratios that are in conflict with current Higgs precision measurement data~\cite{Curtin:2012aa,Cohen:2012zza,Carena:2012np,Katz:2014bha,Katz:2015uja,Kobakhidze:2015scd,Liebler:2015ddv}. This places severe pressure on the possibility of electroweak baryogenesis in the MSSM. In the NMSSM, the presence of the singlet $S$, the bosonic component of $\widehat{S}$, allows for radically different shapes of the scalar potential, which make a SFOEWPT possible in the NMSSM without the need for light stops.

The superpotential of the $\mathbb{Z}_3$-NMSSM is given by
\begin{equation} \label{eq:W}
	W = \lambda \widehat{S} \widehat{H}_u \cdot \widehat{H}_d + \frac{\kappa}{3} \widehat{S}^3 + W_{\rm Yuk}\;,
\end{equation}
where $\lambda$ and $\kappa$ are dimensionless parameters that can be chosen manifestly real in the CP-conserving case. The superfields $\widehat{H}_d = \left( \widehat{H}_d^0, \widehat{H}_d^- \right)^T$ and $\widehat{H}_u = \left( \widehat{H}_u^+, \widehat{H}_u^0 \right)^T$ are the usual $SU(2)$-doublet Higgs superfields, we use a dot-notation for $SU(2)$ products
\begin{equation}
	\widehat{H}_u \cdot \widehat{H}_d = \widehat{H}_u^+ \widehat{H}_d^- - \widehat{H}_u^0 \widehat{H}_d^0\;,
\end{equation}
and $W_{\rm Yuk}$ indicates the Yukawa terms which are identical to those in the MSSM~\cite{Martin:1997ns}.

Including $F$-, $D$- and soft SUSY-breaking terms, the scalar potential reads
\begin{equation} \begin{split} \label{eq:V0}
	V_0 &= m_{H_d}^2 \left| H_d \right|^2 + m_{H_u}^2 \left|H_u\right|^2 + m_S^2 \left|S\right|^2 + \lambda^2 \left| S \right|^2 \left( \left| H_d \right|^2 + \left| H_u \right|^2 \right) + \left| \lambda H_u \cdot H_d + \kappa S^2 \right|^2 \\
	&\quad + \left( \lambda A_\lambda S H_u \cdot H_d + \frac{\kappa}{3} A_\kappa S^3 + {\rm h.c.} \right) + \frac{g_1^2 + g_2^2}{8} \left( \left| H_d \right|^2 - \left| H_u \right|^2 \right)^2 + \frac{g_2^2}{2} \left| H_d^\dagger H_u \right|^2 \;,
\end{split} \end{equation}
where $m_i^2$ and $A_i$ are soft SUSY-breaking parameters of dimension mass-squared and mass, respectively, and $g_1$ and $g_2$ are the $U(1)_Y$ and $SU(2)_L$ gauge couplings.

The Higgs fields have large couplings amongst themselves, to the electroweak gauge bosons, and to third generation (s)fermions. These couplings lead to sizable radiative corrections to $V_0$, to which we return in section~\ref{sec:RadCorr}. However, many of the properties of the scalar potential can already be seen from the tree level potential, eq.~\eqref{eq:V0}. 

In order to be compatible with phenomenology, the NMSSM must preserve charge. While in the MSSM the scalar potential is sufficiently constrained to make charge-breaking minima very rare (see, e.g., ref.~\cite{Aitchison:2005cf}), the additional freedom of the NMSSM's scalar potential makes such minima a much larger problem. However, ref.~\cite{Krauss:2017nlh} demonstrated numerically that, while charge-breaking minima may be present in the NMSSM, they are virtually always accompanied by additional charge-conserving minima, and the tunneling rate from the metastable physical minimum to these charge-conserving minima is larger than to the charge-breaking minima. Hence, we can neglect such charge-breaking minima; in the following we will assume that for all phenomenologically relevant vacua the vevs can be rotated to have the form
\begin{equation}
	\left\langle H_d \right\rangle = \begin{pmatrix} v_d \\ 0 \end{pmatrix}\;, \qquad \left\langle H_u \right\rangle = \begin{pmatrix} 0 \\ v_u \end{pmatrix}\:, \qquad \left\langle S \right\rangle = v_S \;,
\end{equation} 
breaking $SU(2)_L \times U(1)_Y \to U(1)_{\rm em}$. Without loss of generality, one can furthermore take all vevs to be real-valued: While the $\mathbb{Z}_3$-NMSSM does allow for stationary points in the scalar potential which spontaneously break CP, at tree level such points are either saddle points or local maxima~\cite{Romao:1986jy}. In summary, it suffices to allow the neutral real components of $H_d$, $H_u$, and $S$ to take non-trivial vevs\footnote{Observe that in general the sfermions can get non-trivial vevs as well, potentially giving rise to charge and/or color breaking vacua. We will not entertain this possibility further in this work.} when studying the vacuum structure of the NMSSM. This reduction from a ten-dimensional to a three-dimensional field space makes the task considerably more tractable.

In order to ensure that the scalar potential has a stationary point at the physical minimum, we use the minimization conditions
\begin{equation} \label{eq:mini}
	\left.\frac{\partial V}{\partial H_d}\right|_{\substack{H_d = v_d\\H_u=v_u\\S=v_S}} = \left.\frac{\partial V}{\partial H_u}\right|_{\substack{H_d = v_d\\H_u=v_u\\S=v_S}} = \left.\frac{\partial V}{\partial S}\right|_{\substack{H_d = v_d\\H_u=v_u\\S=v_S}} = 0 \;,
\end{equation}
replacing the squared mass parameters $m_{H_d}^2$, $m_{H_u}^2$, and $m_S^2$ with the vevs $v_d$, $v_u$, and $v_S$ in eq.~\eqref{eq:V0}. In practice, it is convenient to re-parameterize the vevs,
\begin{equation}
	v \equiv \sqrt{v_d^2 + v_u^2}\;, \qquad \tan\beta \equiv v_u/v_d\;, \qquad \mu \equiv \lambda v_S \;.
\end{equation}
The observed mass of the electroweak gauge bosons is reproduced by fixing $v = 174\,$GeV, removing one of the NMSSM's free parameters.

In order to account for the constraints on the NMSSM imposed by the SM-like couplings of the observed 125\,GeV Higgs boson, it is useful to write the Higgs fields in the {\it extended Higgs basis}~\cite{Georgi:1978ri,Donoghue:1978cj,gunion2008higgs,Lavoura:1994fv,Botella:1994cs,Branco99,Gunion:2002zf,Carena:2015moc}\footnote{Note, that there are different conventions in the literature for the Higgs basis differing by an overall sign of $H^{\rm NSM}$ and $A^{\rm NSM}$.}
\begin{align}
   H_d &= \begin{pmatrix} \frac{1}{\sqrt{2}} \left( c_\beta H^{\rm SM} - s_{\beta} H^{\rm NSM} \right) + \frac{i}{\sqrt{2}} \left( - c_\beta G^0 + s_\beta A^{\rm NSM} \right) \\ -c_\beta G^- + s_\beta H^- \end{pmatrix} \;, \\
   H_u &= \begin{pmatrix} s_\beta G^+ + c_\beta H^+ \\ \frac{1}{\sqrt{2}} \left( s_\beta H^{\rm SM} + c_{\beta} H^{\rm NSM} \right) + \frac{i}{\sqrt{2}} \left( s_\beta G^0 + c_\beta A^{\rm NSM} \right) \end{pmatrix} \;, \\
   S &= \frac{1}{\sqrt{2}} \left( H^{\rm S} + i A^{\rm S} \right) \;.
\end{align}
$H^{\rm SM}$, $H^{\rm NSM}$, and $H^{\rm S}$ are the three neutral CP-even interaction states of the Higgs basis, $A^{\rm NSM}$ and $A^{\rm S}$ are the CP-odd states, and $H^\pm$ is the charged Higgs. The neutral and charged Goldstone modes are denoted by $G^0$ and $G^\pm$, respectively, and we used a shorthand notation
\begin{equation}
	s_\beta \equiv \sin\beta \;, \qquad c_\beta \equiv \cos\beta \;.
\end{equation}

In this basis, the couplings to pairs of SM particles take a particularly simple form. Focusing on the CP-even states, the couplings to pairs of down-type and up-type fermions and pairs of vector bosons (VV) are
\begin{align}
	H^{\rm SM}({\rm down,\,up\,,VV}) &= \left( g_{\rm SM}, ~g_{\rm SM}, ~g_{\rm SM} \right) \;, \\ 
	H^{\rm NSM}({\rm down,\,up\,,VV}) &= \left( - g_{\rm SM} \tan\beta, ~g_{\rm SM}/\tan\beta, ~0 \right) \;, \\
	H^{\rm S}({\rm down,\,up\,,VV}) &= \left( 0, ~0, ~0 \right) \;,
\end{align}
where $g_{\rm SM}$ is the corresponding coupling of the SM Higgs boson to pairs of such particles. Thus, $H^{\rm SM}$ has the same couplings to pairs of SM particles as the SM Higgs boson. Furthermore, $H^{\rm SM}$ is the only Higgs boson which couples to pairs of vector bosons. $H^{\rm NSM}$ has $\tan\beta$ enhanced (suppressed) couplings to pairs of down-type (up-type) SM fermions, and $H^{\rm S}$ does not couple to pairs of SM particles. Note that at the physical minimum, only $\langle H^{\rm SM} \rangle = \sqrt{2} v$ and $\langle H^{\rm S} \rangle = \sqrt{2} v_S$ take non-trivial vevs, while $\langle H^{\rm NSM} \rangle = 0$.

The interaction states mix into mass eigenstates. We denote the CP-even mass eigenstates as $\left\{ h_{125}, H, h_S \right\}$, where $h_{125}$ is identified with the 125\,GeV state observed at the LHC, $H$ is the non-SM-like state with the largest $H^{\rm NSM}$ component, and $h_S$ the state with the largest $H^{\rm S}$ component. Similarly, the CP-odd interaction states $A^{\rm NSM}$ and $A^{\rm S}$ mix into two mass eigenstates, which we denote as $A$ and $a_S$. 

In order to ensure compatibility with the observed Higgs boson phenomenology, the $h_{125}$ state must be dominantly composed of $H^{\rm SM}$. Denoting the squared mass matrix for the CP even states as $\mathcal{M}_S^2$ in the basis $\left\{ H^{\rm SM}, H^{\rm NSM}, H^{\rm S} \right\}$, the tree-level mass of the SM-like state is given by
\begin{equation} \label{eq:mh125_tree}
	m_{h_{125}}^2 \simeq \mathcal{M}_{S,11}^2 = m_Z^2 \cos^2(2\beta) + \lambda^2 v^2 \sin^2(2\beta) \;,
\end{equation}
where $m_Z^2 = v^2 \left( g_1^2 + g_2^2 \right) / 2 $ is the $Z$-boson mass. While $m_{h_{125}}$ receives sizable radiative corrections via the stops, see section~\ref{sec:RadCorr}, it is interesting to note that the term proportional to $\lambda^2 v^2$ allows one to obtain $m_{h_{125}} = 125\,$GeV already at tree level for small values of $\tan\beta \lesssim 3$ if $\lambda$ takes values $0.7 \lesssim \lambda \lesssim 1$. Thus, there is no need for large radiative corrections to the mass of the SM-like Higgs, i.e. no need for heavy stops, in the NMSSM. Including moderate corrections from the stops, the required value for the mass of the SM-like Higgs boson is obtained for $0.6 \lesssim \lambda \lesssim 0.8$ in the small-to-moderate $\tan\beta \lesssim 5$ regime.

In order to ensure that the mass eigenstate $h_{125}$ is dominantly composed of $H^{\rm SM}$, the mixing angles of $H^{\rm NSM}$ and $H^{\rm S}$ with $H^{\rm SM}$ must be suppressed. The mixing of $H^{\rm SM}$ with $H^{\rm NSM}$ is suppressed if
\begin{equation} \label{eq:align1}
	\left| \mathcal{M}_{S,12}^2 \right| \ll \left| \mathcal{M}_{S,22}^2 - \mathcal{M}_{S,11}^2 \right| \;,
\end{equation}
and similarly, the mixing of $H^{\rm SM}$ with $H^{\rm S}$ is suppressed if
\begin{equation} \label{eq:align2}
	\left| \mathcal{M}_{S,13}^2 \right| \ll \left| \mathcal{M}_{S,33}^2 - \mathcal{M}_{S,11}^2 \right| \;.
\end{equation}
Here, the $\mathcal{M}_{S,ij}^2$ again are the entries of the squared mass matrix for the CP-even states in the basis $\left\{H^{\rm SM}, H^{\rm NSM}, H^{\rm S}\right\}$. There are two possibilities to achieve such (approximate) {\it alignment} of $h_{125}$ with $H^{\rm SM}$: either, the entries of the squared mass matrix corresponding to such mixing are small, or, the right hand sides of eqs.~\eqref{eq:align1} and~\eqref{eq:align2} become large. The latter option is the so-called {\it decoupling limit}. Realizing alignment in this way implies $\left\{m_H, m_{h_S}\right\} \gg m_{h_{125}}$. As we will see below, a relatively light singlet-like state gives the scalar potential a favorable shape for SFOEWPT. Thus, the former option, the so-called {\it alignment without decoupling} limit, is more interesting for electroweak baryogenesis. 

At tree-level, alignment between the two states originating from the Higgs doublets, eq.~\eqref{eq:align1}, is achieved for
\begin{equation}
	\mathcal{M}^2_{S,12} = - \left( m_Z^2 - \lambda^ 2 v^2 \right) \sin(2\beta) \cos(2\beta) \to 0 \;.
\end{equation}
It is convenient to instead rewrite this condition as
\begin{equation}
	\mathcal{M}^2_{S,12} = \frac{1}{\tan\beta} \left[ \mathcal{M}^2_{S,11} - m_Z^2 \cos(2\beta) - 2 \lambda^2 v^2 \sin^2 \beta \right] \to 0\;,
\end{equation}
because this form is robust against radiative corrections~\cite{Carena:2015moc}. Identifying $\mathcal{M}_{S,11}^2 = m_{h_{125}}^ 2$, one obtains the alignment condition
\begin{equation} \label{eq:Align1}
	\lambda^2 = \frac{m_{h_{125}}^2 - m_Z^2 \cos(2\beta)}{2 v^2 \sin^2\beta} \;.
\end{equation}
For small to moderate values of $\tan\beta$, this condition yields $0.6 \lesssim \lambda \lesssim 0.7$. It is interesting to note that, for moderate values of $\tan\beta \lesssim 5$, this range of $\lambda$ coincides with the range for which one obtains $m_{h_{125}} = 125\,$GeV without the need for large radiative corrections. 

Suppressing the mixing of $H^{\rm SM}$ with $H^{\rm S}$, eq.~\eqref{eq:align2}, yields a second alignment condition from demanding $\mathcal{M}_{S,13}^2 \to 0$, namely
\begin{equation} \label{eq:Align2}
		M_A^2 = \frac{4 \mu^2}{\sin^ 2(2\beta)} \left( 1 - \frac{\kappa}{2\lambda} \sin 2\beta \right)\;, 
\end{equation}
where we introduced the parameter
\begin{equation} \label{eq:MAdef}
	M_A^2 = \frac{2 \mu}{\sin 2\beta} \left( A_\lambda + \frac{\kappa\mu}{\lambda} \right) \;.
\end{equation}
$M_A^2$ is the (squared) mass parameter of $A^{\rm NSM}$ and controls the mass scale of the mostly doublet-like CP-even and CP-odd mass eigenstates as well as the mass scale of the charged Higgs boson. The alignment condition eq.~\eqref{eq:Align2} gives rise to a mass spectrum where, provided $\kappa < \lambda$, the doublet-like mass eigenstates have approximate masses $m_H, m_A, m_{H^\pm} \sim 2 \mu/\sin2\beta$~\cite{Baum:2017gbj,Baum:2019uzg}. 

In the remainder of this work, we will consider the NMSSM in the alignment limit, choosing parameters to satisfy eqs.~\eqref{eq:Align1} and~\eqref{eq:Align2}. While current data~\cite{Sirunyan:2018koj, Aad:2019mbh} allow for some deviation from perfect alignment, the phenomenological impact of such departures on the EWPT in the NMSSM is small. Note also that in refs.~\cite{Baum:2017gbj,Baum:2019uzg} it was demonstrated that, in random parameter scans where the alignment conditions are not {\it a priori} enforced, requiring compatibility with the phenomenology of the observed 125\,GeV Higgs boson selects the region of parameter space where eqs.~\eqref{eq:Align1} and~\eqref{eq:Align2} are (approximately) satisfied.

The NMSSM parameter space is constrained by a number of additional arguments. Let us briefly discuss two of them here, while we derive constraints arising from the stability of the electroweak vacuum in section~\ref{sec:VacStr}. It is well known, that large values of the dimensionless parameters $\lambda$ and $\kappa$ lead to Landau poles. Avoiding the appearance of Landau poles below the GUT scale [$Q_{\rm GUT} \sim \mathcal{O}(10^{16})\,$GeV] entails constraining the values of the NMSSM's couplings, at the electroweak scale, to~\cite{Ellwanger:2009dp}
\begin{equation}
   \sqrt{\lambda^2 + \kappa^2} \lesssim 0.7\;.
\end{equation}
As discussed above, both the SM-like nature of the observed Higgs boson and its mass value lead to a preference of sizable values of $0.6 \lesssim \lambda \lesssim 0.7$ in the NMSSM. Hence, avoiding Landau poles below $Q_{\rm GUT}$ limits the value of $\left|\kappa\right| \lesssim 0.3$ in the alignment limit. Note that the NMSSM with larger couplings (and Landau poles between the TeV and the GUT scale) is known as $\lambda$-SUSY, see, for example, refs.~\cite{Barbieri:2006bg,Cao:2008un,Farina:2013fsa}.

The parameter space is also constrained by avoiding tachyonic masses. The most relevant constraint arises from the singlet-like CP-odd mass eigenstate $a_S$. Taking into account first-order mixing effects, its mass is approximately~\cite{Carena:2015moc}
\begin{equation}
	m_{a_S}^2 \simeq 3 \kappa v^2 \left[ \frac{3\lambda}{2}\sin(2\beta) - \left( \frac{\mu A_\kappa}{\lambda v^2} + \frac{3 \kappa \mu^2}{M_A^2} \right) \right] \;.
\end{equation}
Recalling that alignment requires $M_A^2 \simeq 4\mu^2 / \sin^2(2\beta)$, we can deduce the condition the NMSSM parameters must satisfy to keep $a_S$ from becoming tachyonic:
\begin{equation} \label{eq:a_tachyonic_condition}
	\frac{ \kappa \mu A_\kappa}{v^2} \lesssim \frac{3 \kappa \lambda^2 \sin(2\beta)}{2} \left[ 1 - \frac{\kappa \sin(2\beta)}{2 \lambda} \right] \;.
\end{equation}
For small-to-moderate values of $\tan\beta$ and in the alignment limit, where $0.6 \lesssim \lambda \lesssim 0.7$, the right-hand side of eq.~\eqref{eq:a_tachyonic_condition} is approximately $\kappa \times \mathcal{O}(1)$. Hence, equation~\eqref{eq:a_tachyonic_condition} implies $\mu A_\kappa \lesssim v^2$ for $\kappa > 0$, while for $\kappa < 0$ the condition becomes $\mu A_\kappa \gtrsim v^2$; in particular, disfavoring ${\rm sgn}(\mu A_\kappa)=-1$ for $\kappa<0$. 

\subsection{Radiative Corrections} \label{sec:RadCorr}

The scalar potential receives sizable radiative corrections from the large couplings between the Higgs bosons themselves as well as from their large couplings to the electroweak gauge bosons and the (s)fermions, in particular the (s)tops, see, for example, refs.~\cite{Ellwanger:2009dp,Derendinger:1983bz,King:1995vk,Masip:1998jc}. Since the precise interplay between the higher-order corrections to the Higgs mass and the mass values of the SM particles and their superpartners does not play a relevant role in our study of the EWPT, we shall take only the dominant one loop corrections into account in this work. The null-results from SUSY searches at the LHC suggest that all squarks as well as the gluinos have masses $\gtrsim 1\,$TeV. LHC constraints on new states neutral under QCD are less stringent. Furthermore, to yield a scalar potential sufficiently different from that of the SM to accommodate a SFOEWPT, the Higgs bosons' masses should not be much larger than the electroweak scale. These considerations motivate studying a scenario in which all sfermions\footnote{For simplicity we also take the sleptons to be heavy here. Because the couplings of sleptons to the scalar sector are much smaller than the gauge couplings and the top Yukawa coupling, lighter sleptons would not lead to large radiative corrections to the scalar sector.} and the gluinos are heavy and can be integrated out, yielding an effective theory where the remaining dynamical degrees of freedom are the SM particles, the new Higgs bosons $\left\{H, h_S, A, a_S, H^\pm\right\}$, the five neutralinos $\widetilde{\chi}_i^0$, and the two charginos $\widetilde{\chi}_i^\pm$; see refs.~\cite{Kozaczuk:2014kva,Athron:2019teq, Kumar:2011np} for similar approaches. The parameters of this effective theory are obtained by matching onto the full theory (containing all the NMSSM's degrees of freedom) at an intermediate scale. The leading operator one obtains from this procedure is 
\begin{equation} \label{eq:Lstopcorr}
   \Delta\mathcal{L} = - \frac{\Delta\lambda_2}{2} \left|H_u\right|^4 \;,
\end{equation}
arising from stop loops. At one loop, the coefficient $\Delta\lambda_2$ is related to the parameters of the stop sector via~\cite{Ellis:1991zd,Haber:1993an, Casas:1994us,Carena:1995bx} 
\begin{equation} \label{eq:Dlam2}
   \Delta \lambda_2 = \frac{3}{8\pi^2} h_t^4 \left[ \log\left(\frac{M_S^2}{m_t^2}\right) + \frac{A_t^2}{M_S^2} \left( 1 - \frac{A_t^2}{12 M_S^2}\right) \right] \;,
\end{equation}
where $h_t$ is the top Yukawa coupling determined from the (running) top quark mass $m_t = h_t v \sin\beta$, $M_S$ is the geometric mean of the stop masses, and $A_t $ is the soft trilinear stop-Higgs coupling. We note that for small to moderate values of $\tan\beta$, the top quark superfield has a sizable coupling only to $\widehat{H}_u$ in the superpotential. After the singlet acquires a non-trivial vev, an effective $\mu$-term is generated and additional effective quartic couplings, which involve not only $H_u$ but also $H_d$, arise via stop loops. However, these contributions are suppressed by powers of $\mu/M_S$. We shall work in a region of parameter space where $\left|\mu\right| \ll M_S$ and, hence, the dominant contribution induced by integrating out the stop sector is given by eq.~\eqref{eq:Lstopcorr}. At higher loop orders, the exact relation between $\Delta \lambda_2$ and the parameters in the stop sector is modified, but, for small values of $\left|\mu\right|$, the stop radiative corrections can still be effectively parametrized by $\Delta \lambda_2$ (see, for example, refs.~\cite{Haber:1993an, Carena:1995bx,Lee:2015uza}).

The scalar potential of this effective theory is then given by
\begin{equation} \label{eq:V0eff}
   V_0^{\rm eff} = V_0 + \frac{\Delta\lambda_2}{2} \left|H_u\right|^4\;.
\end{equation}
This new contribution gives sizable corrections to the Higgs mass matrix. In particular, the mass of the SM-like Higgs state is given by
\begin{equation} \label{eq:mh125_stop}
   m_{h_{125}}^2 \simeq \mathcal{M}_{S,11}^2 = m_Z^2 \cos^2(2\beta) + \lambda^2 v^2 \sin^2(2\beta) + 2 \Delta\lambda_2 v^2 \sin^4\beta \;.
\end{equation}
Note that the alignment conditions in eqs.~\eqref{eq:Align1} and ~\eqref{eq:Align2} are not modified by $\Delta\lambda_2$. While the value of $\Delta\lambda_2$ is in principle controlled by the soft parameters in the stop sector, see eq.~\eqref{eq:Dlam2}, in the remainder of this work we use eq.~\eqref{eq:mh125_stop} to set $\Delta\lambda_2$ to reproduce the observed mass of the SM-like Higgs boson, $m_{h_{125}} = 125\,$GeV. 

The radiative corrections to the effective potential from the remaining dynamical degrees of freedom are given by the {\it Coleman-Weinberg} potential~\cite{Coleman:1973jx}
\begin{equation} \label{eq:CW}
   V_{\rm 1-loop}^{\rm CW} = \frac{1}{64\pi^2} \sum_{i=B,F} (-1)^{F_i} n_i \widehat{m}_i^4 \left[ \log\left(\frac{\widehat{m}_i^2}{m_t^2}\right) - C_i \right] \;,
\end{equation}
where $F_i=0$ for bosons and $F_i = 1$ for fermions. The constant $C_i$ takes values $C_i = 3/2$ for scalars, longitudinally polarized vector bosons, and fermions, while for transversal vector bosons $C_i = 1/2$. We denote the field-dependent masses computed from $V_0^{\rm eff}$ by 
\begin{equation}
   \widehat{m}_i^2 = \widehat{m}_i^2(H^{\rm SM}, H^{\rm NSM}, H^{\rm S})\;,
\end{equation}
and work in the Landau gauge; explicit expressions for the $\widehat{m}_i^2$ are collected in appendix~\ref{app:field_masses}. The bosonic fields entering eq.~\eqref{eq:CW} are $B = \left\{h_i, a_i, H^\pm, G^0, G^\pm, Z, W^\pm \right\}$ with $n_B = \left\{1,1,2,1,2,3,6\right\}$ degrees of freedom, respectively. Here, $h_i$ and $a_i$ denote the three neutral CP-even and two CP-odd Higgs bosons, $H^\pm$ the charged Higgs, $G^0$ and $G^\pm$ the neutral and charged Goldstone modes, and $Z$ and $W^\pm$ the electroweak gauge bosons. The fermionic fields entering the Coleman-Weinberg potential are\footnote{We neglect the (small) radiative corrections from the SM fermions other than the top quark.} $F = \left\{\widetilde{\chi}_i^0, \widetilde{\chi}_i^\pm, t\right\}$ with $n_F = \left\{2,4,12\right\}$, where $\widetilde{\chi}_i^0$ and $\widetilde{\chi}_i^\pm$ denote the five neutralinos and two charginos, respectively, and $t$ is the top quark. We have chosen $m_t$ as the renormalization scale, implying that the parameters are defined at such scale. In order to guarantee the one-loop renormalization scale independence and preserve the supersymmetric relations, the parameters at the scale $m_t$ must be related with those at higher energies, up to the supersymmetry breaking scale, by including all particles in the effective theory in the running to higher energies. 

Note that since the Goldstone modes' masses vanish at the physical minimum, their contributions to the Coleman-Weinberg potential lead to divergent contributions to physical masses and coupling coefficients computed from derivatives of the loop-corrected effective potential. This divergence is an artefact of the perturbative calculation~\cite{Martin:2014bca,Elias-Miro:2014pca} and can be dealt with by shifting the masses of the Goldstone modes by an infrared regulator, $\widehat{m}_G^2 \to \widehat{m}_G^2 + \mu_{\rm IR}^2$. In our numerical calculations, we use a value of $\mu_{\rm IR}^2 = 1\,{\rm GeV}^2$; note, however, that in numerical calculations numerical errors on $\widehat{m}_G^2$ typically suffice to ``regulate'' the logarithmically divergent contribution from $\widehat{m}_G^2 \to 0$, even before including an explicit infrared regulator.

Including the Coleman-Weinberg contributions, the (effective) scalar potential at zero temperature is given by
\begin{equation}
   V_1(T=0) = V_0^{\rm eff} + V_{\rm 1-loop}^{\rm CW} \;.
\end{equation}
The Coleman-Weinberg corrections alter the location of the minima as well as the physical masses. We include a set of counterterms
\begin{equation} \label{eq:ct_lag}
   \delta\mathcal{L} = - \delta_{m_{H_d}^2} \left| H_d \right|^2 - \delta_{m_{H_u}^2} \left|H_u\right|^2 - \delta_{m_S^2} \left|S\right|^2 - \delta_{\lambda A_\lambda} \left( S H_u \cdot H_d + {\rm h.c.} \right) - \frac{\delta_{\lambda_2}}{2} \left|H_u\right|^4 \;, 
\end{equation}
to keep the location of the physical minimum at $\left\{H^{\rm SM}, H^{\rm NSM}, H^{\rm S}\right\} = \sqrt{2} \left\{v,0,\mu/\lambda\right\}$, ensure $\mathcal{M}_{S,13}^2 \to 0$, preserving alignment, and maintain $m_{h_{125}} = 125\,$GeV. Note that these counterterms correspond to a redefinition of the soft SUSY-breaking terms\footnote{The counterterm $\delta_{\lambda_2}$ corresponds to a soft SUSY-breaking term in the sense that it can be understood as a counterterm shifting the soft parameters in the stop sector and, in turn, the threshold correction $\Delta\lambda_2$ that we obtain from integrating out the stops.}, see refs.~\cite{Huber:2015znp,Bi:2015qva, Bian:2017wfv} for similar approaches. We list equations for the fixing of the counterterms in appendix~\ref{app:ct_coeff}.

The input parameters for our model are thus
\begin{equation}
   \tan\beta\;, \quad \mu\;,\quad \kappa\;, \quad A_\kappa\;.
\end{equation}
All other parameters are fixed by the various conditions we impose on the model, namely, $\lambda$ and $M_A^2$ are determined by alignment, $\Delta\lambda_2$ by setting $m_{h_{125}} = 125\,$GeV, and the counterterms are fixed by the conditions discussed in the previous paragraph.

\subsection{Thermal Corrections} \label{sec:TCorr}

So far, we have discussed the scalar potential at zero temperature. At finite temperatures, thermal corrections to the potential have to be taken into account. The one-loop finite temperature potential is given by
\begin{equation} \label{eq:V1Tonly}
   V_{\rm 1-loop}^{T\neq0} = \frac{T^4}{2\pi^2} \sum_{i=B,F} (-1)^{F_i} n_i J_{B/F} \left( \frac{\widetilde{m}_i^2}{T^2} \right) \:,
\end{equation}
where analogously to our definition of the Coleman-Weinberg potential, eq.~\eqref{eq:CW}, the sum runs over bosonic and fermionic degrees of freedom, $n_i$ counts the degrees of freedom of species $i$, and $F_i=0$ ($F_i=1$) for bosons (fermions). We denote thermal (field-dependent) masses with a tilde, $\widetilde{m}_i^2$. Compared to the field-dependent masses, which we denote with a hat, $\widehat{m}_i^2$, the thermal masses include the so-called Daisy corrections re-summing hard thermal loops,
\begin{equation}
   \widetilde{m}_i^2 \equiv \widetilde{m}_i^2(H^{\rm SM}, H^{\rm NSM}, H^{\rm S}; T) = \widehat{m}_i^2(H^{\rm SM}, H^{\rm NSM}, H^{\rm S}) + c_i T^2 \;.
\end{equation}
The Daisy coefficients $c_i$ are only non-zero for bosonic fields. Furthermore, only the longitudinal polarization states of vector bosons receive non-zero Daisy corrections, gauge symmetry protects the transversal degrees of freedom. We list the Daisy coefficients for the relevant fields in appendix~\ref{app:Daisy}.

The thermal functions are defined as
\begin{equation}
   J_{B/F}(x^2) = \int_0^\infty dy\,y^2 \log\left( 1 \mp e^{-\sqrt{y^2+x^2}} \right) \;.
\end{equation}

Following ref.~\cite{Curtin:2016urg}, we improve the calculation of the thermal corrections by replacing the field-dependent masses with the thermal masses in the Coleman-Weinberg potential,
\begin{equation}
   V_{\rm 1-loop}^{\rm CW}(\widehat{m}_i^2) \to V_{\rm 1-loop}^{\rm CW}(\widetilde{m}_i^2) = \frac{1}{64\pi^2} \sum_{i=B,F} (-1)^{F_i} n_i \widetilde{m}_i^4 \left[ \log\left(\frac{\widetilde{m}_i^2}{m_t^2}\right) - C_i \right] \;.
\end{equation}
Including the Coleman-Weinberg and the thermal corrections, the temperature-dependent effective potential at one-loop order is given by
\begin{equation} \label{eq:V1T}
   V_1(T) = V_0^{\rm eff} + V_{\rm 1-loop}^{\rm CW}(\widetilde{m}_i^2) + V_{\rm 1-loop}^{T\neq0}(\widetilde{m}_i^2) \;.
\end{equation}

\subsection{Zero-Temperature Vacuum Structure} \label{sec:VacStr}

While the NMSSM's scalar potential is subject to radiative as well as thermal corrections as discussed in sections~\ref{sec:RadCorr} and~\ref{sec:TCorr}, one can already learn much about the possibility of a SFOEWPT from considering the effective potential, $V_0^{\rm eff}$, obtained after integrating out all sfermions and the gluinos and prior to including the Coleman-Weinberg and thermal corrections. In this section, we derive the most interesting regions of NMSSM parameter space for realizing a SFOEWPT from $V_0^ {\rm eff}$. As we shall show later on, these regions of parameter space are only mildly affected by radiative corrections. Recall that in order to study the vacuum structure of the NMSSM, it suffices to consider the three-dimensional field space spanned by the neutral CP-even fields $\left\{H^{\rm SM}, H^{\rm NSM}, H^{\rm S}\right\}$,
\begin{equation}
   V_0^{\rm eff, 3}(H^{\rm SM}, H^{\rm NSM}, H^{\rm S}) \equiv \left.V_0^{\rm eff}\right|_{\substack{A^{\rm NSM} = 0 \\ A^{\rm S} = 0 \\ H^\pm = 0}}\;,
\end{equation}
where $V_0^{\rm eff}$ is the potential given in eq.~\eqref{eq:V0eff}.

As discussed above, the singlet plays a special role for realizing a SFOEWPT. Its coupling to the Higgs doublets, $\lambda$, and its self-coupling, $\kappa$, are free parameters, while the quartic couplings between the Higgs doublets are governed by the gauge couplings (and $\Delta \lambda_2$). Furthermore, as a consequence of $U(1)_Y$ symmetry, $V_0^{\rm eff, 3}$ is invariant under the transformation $H^{\rm SM} \to -H^{\rm SM}$, $H^{\rm NSM} \to -H^{\rm NSM}$, $H^{\rm S} \to H^{\rm S}$. This residual $\mathbb{Z}_2$ symmetry ensures that any extrema in the singlet-only direction, i.e. where $H^{\rm SM} = H^{\rm NSM} = 0$, are also extrema (or saddle points) of $V_0^{{\rm eff},3}$. In the alignment limit (or, more specifically, as long as the second alignment condition, eq.~\eqref{eq:align2}, is satisfied) the scalar potential in the singlet-only direction is given by 
\begin{equation} \label{eq:V3eff_align}
   V_0^{{\rm eff},3}(0,0,H^{\rm S}) \to - \kappa^2 \frac{\mu}{\lambda} \left( \frac{\mu}{\lambda} + \frac{A_\kappa}{2\kappa} \right) (H^{\rm S})^2 + \frac{\kappa A_\kappa}{3\sqrt{2}} (H^{\rm S})^3 + \frac{\kappa^2}{4} (H^{\rm S})^4 \;.
\end{equation}
This potential has extrema at
\begin{equation}
	H^{\rm S} =\left\{0 \;,\quad \frac{\sqrt{2}\mu}{\lambda} \;,\quad -\sqrt{2}\left(\frac{\mu}{\lambda} + \frac{A_\kappa}{2\kappa} \right) \right\} \;.
\end{equation}
The first of these field values corresponds to the trivial minimum of the scalar potential $H^{\rm SM} = H^{\rm NSM} = H^{\rm S} = 0$, and the second value coincides with the vev of $H^{\rm S}$ at the physical minimum $v_S = \mu / \lambda$. The third field value marks a new special location in $H^{\rm S}$ space, which, in the following, we refer to as
\begin{equation} \label{eq:vsprime}
	v_S' \equiv -\left(\frac{\mu}{\lambda} + \frac{A_\kappa}{2\kappa} \right) \;.
\end{equation}

Recall that since we used the minimization conditions, eq.~\eqref{eq:mini}, to replace the $m_i^2$ parameters in the scalar potential with $v$, $\tan\beta$, and $\mu$, the physical minimum $\left\{ H^{\rm SM}, H^{\rm NSM}, H^{\rm S} \right\} = \sqrt{2} \left\{v, 0, \mu/\lambda\right\}$ is also guaranteed to be a stationary point of the scalar potential. Hence, in the alignment limit, all first-order derivatives of $V_0^{{\rm eff},3}$ vanish at
\begin{equation} \label{eq:min_loc}
	\left\{ H^{\rm SM}, H^{\rm NSM}, H^{\rm S} \right\} = \left\{0, 0, 0 \right\} \lor \left\{ 0, 0, \sqrt{2} v_S' \right\} \lor \left\{0, 0, \frac{\sqrt{2}\mu}{\lambda} \right\} \lor \left\{ \sqrt{2} v, 0, \frac{\sqrt{2}\mu}{\lambda} \right\} \;.
\end{equation}
The potential $V_0^{{\rm eff},3}$ may have additional stationary points; we will return to the possibility of such minima below.

In order to constrain the allowed parameter space, we consider the value of the potential at the field values given in eq.~\eqref{eq:min_loc} and demand the physical minimum to be the global minimum. As we will see, the $\left|\mu\right|$ vs. $v_S'/v_S$ plane is a useful projection of the parameter space. In the alignment limit, the potential at the physical minimum takes the value
\begin{align}
   V_0^{{\rm eff},3}(\sqrt{2}v,0,\frac{\sqrt{2}\mu}{\lambda}) &= - \frac{m_Z^2 c_{2\beta}^2 + \lambda^2 v^2 s_{2\beta}^2 + 2 \Delta\lambda_2 v^2 s_\beta^4}{4}v^2 - \frac{\kappa^2 \mu^3}{\lambda^3} \left( \frac{\mu}{\lambda} + \frac{A_\kappa}{3 \kappa} \right) \\
   &= -\frac{1}{4} m_{h_{125}}^2 v^2 - \frac{1}{3} \frac{\kappa^2 \mu^4}{\lambda^4} \left( 1 - 2 \frac{v_S'}{v_S} \right) \;, \label{eq:V3phys}
\end{align}
where we used eqs.~\eqref{eq:mh125_stop} and~\eqref{eq:vsprime} for the second equality.

We can derive a first constraint on the parameter space by demanding the physical minimum to be deeper than the trivial minimum. The scalar potential vanishes at the trivial minimum, $V_0^{{\rm eff},3}(0,0,0) = 0$. Thus, in the alignment limit, demanding $V_0^{{\rm eff},3}(\sqrt{2}v,0,\sqrt{2}\mu\lambda) < V_0^{{\rm eff},3}(0,0,0)$ yields the condition
\begin{equation} \label{eq:cond_triv}
   \frac{v_S'}{v_S} < \frac{1}{2} \left( 1 + \frac{3}{4} \frac{\lambda^4}{\kappa^2} \frac{m_{h_{125}}^2 v^2}{\mu^4} \right) \;.
\end{equation} 

At $\left\{H^{\rm SM}, H^{\rm NSM}, H^{\rm S} \right\} = \left\{ 0, 0, \sqrt{2}\mu/\lambda \right\}$ the potential takes the value
\begin{equation}
	V_0^{{\rm eff},3}(0,0,\frac{\sqrt{2}\mu}{\lambda}) = - \frac{\kappa^2 \mu^3}{\lambda^3} \left( \frac{\mu}{\lambda} + \frac{A_\kappa}{3\kappa} \right) = - \frac{1}{3} \frac{\kappa^2 \mu^4}{\lambda^4} \left( 1 - 2 \frac{v_S'}{v_S} \right) \;.
\end{equation} 
Comparing with eq.~\eqref{eq:V3phys}, we see that this stationary point of the potential is never deeper than the physical minimum; $\left\{H^{\rm SM}, H^{\rm NSM}, H^{\rm S}\right\} = \left\{0,0,\sqrt{2}\mu/\lambda\right\}$ is a saddle point of the scalar potential in the alignment limit. 

On the other hand, at $\left\{H^{\rm SM}, H^{\rm NSM}, H^{\rm S} \right\} = \left\{0, 0, \sqrt{2} v_S' \right\}$, the scalar potential (in the alignment limit) takes the value
\begin{equation} \label{eq:V3vsp}
	V_0^{{\rm eff},3}(0,0,\sqrt{2}v_S') = - \frac{\kappa^2}{3} \left( \frac{\mu}{\lambda} + \frac{A_\kappa}{2\kappa} \right)^3 \left( \frac{3\mu}{\lambda} + \frac{A_\kappa}{2\kappa} \right) = \frac{1}{3} \frac{\kappa^2 \mu^4}{\lambda^4} \left( \frac{v_S'}{v_S} \right)^3 \left( 2 - \frac{v_S'}{v_S} \right) \;.
\end{equation} 
Demanding this minimum to be shallower than the physical minimum, $V_0^{{\rm eff},3}(0,0,\sqrt{2}v_S') > V_0^{{\rm eff},3}(\sqrt{2}v,0,\frac{\sqrt{2}\mu}{\lambda})$, yields the condition
\begin{equation} \label{eq:cond_vS'}
	\left( \frac{v_S'}{v_S}-1 \right)^3 \left( \frac{v_S'}{v_S}+1 \right) < \frac{3}{4} \frac{\lambda^4}{\kappa^2} \frac{m_{h_{125}}^2 v^2}{\mu^4} \;,
\end{equation}
defining a range of $v_S'/v_S$ for which the physical minimum is deeper than the minimum at $\left\{H^{\rm SM}, H^{\rm NSM}, H^{\rm S}\right\} = \left\{0,0,\sqrt{2}v_S'\right\}$. 

\begin{figure}
   \includegraphics[width=0.49\linewidth]{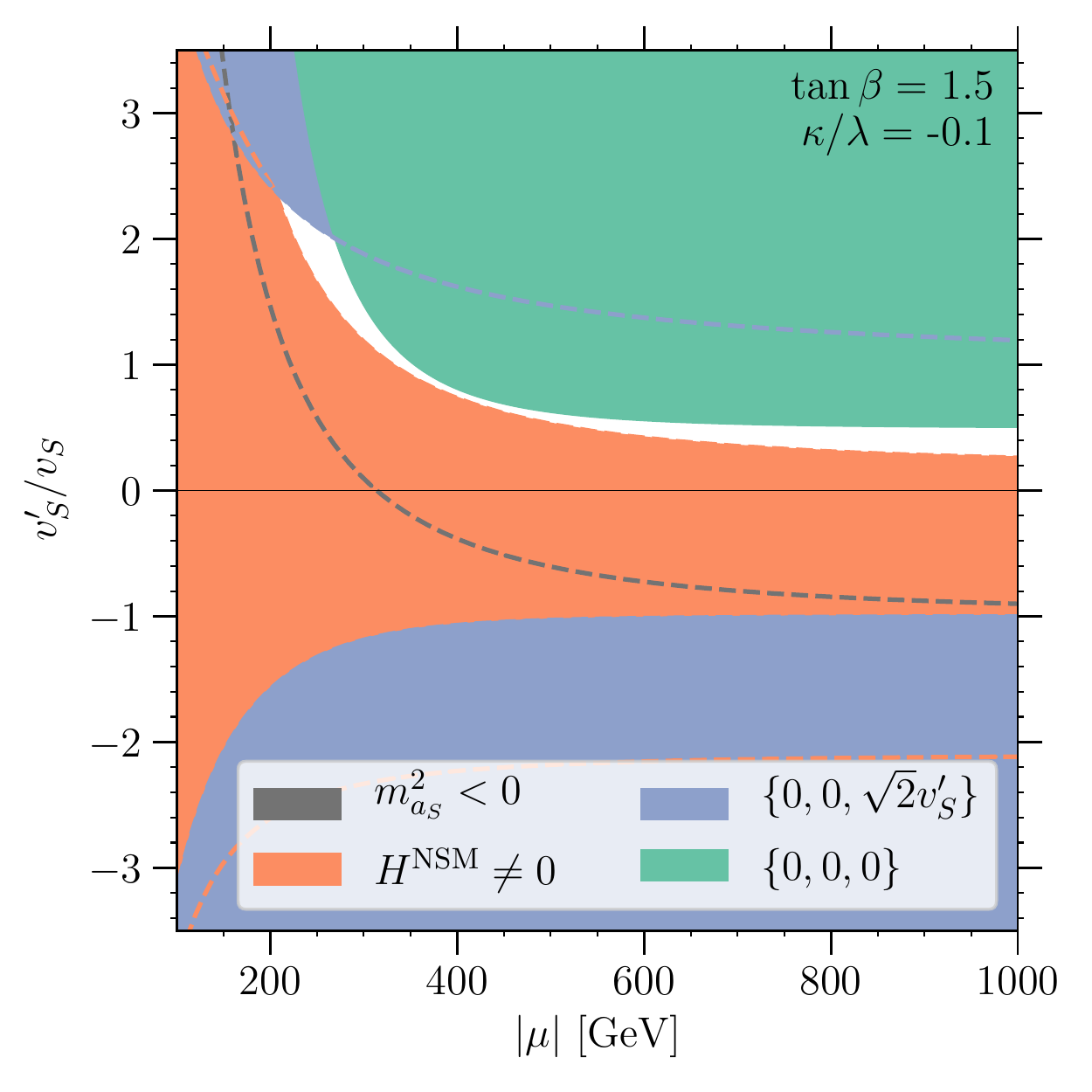}
   \includegraphics[width=0.49\linewidth]{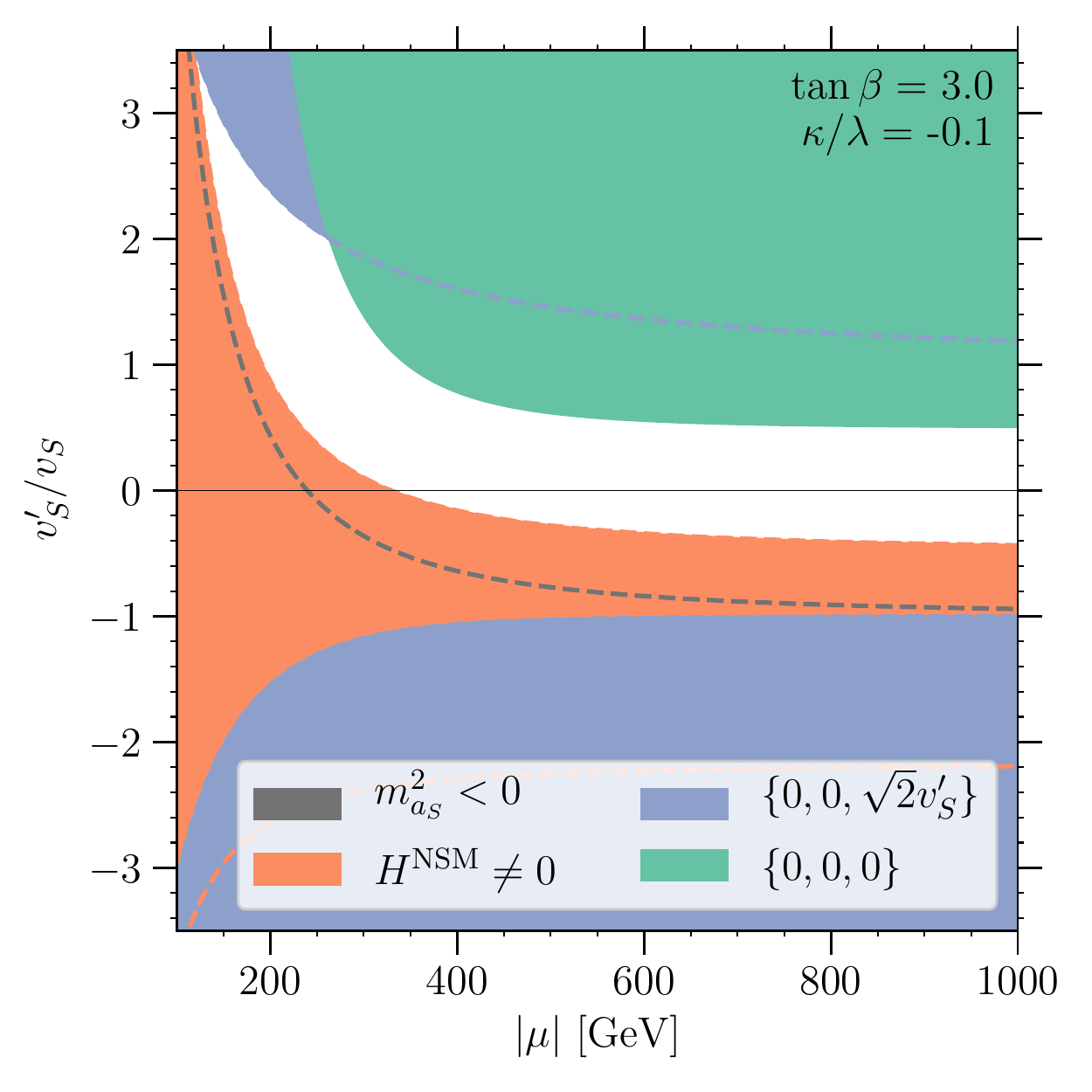}
   \caption{The white region indicates the region of NMSSM parameter space where (in the alignment limit) the physical minimum $\left\{H^{\rm SM}, H^{\rm NSM}, H^{\rm S}\right\} = \sqrt{2} \left\{v, 0, v_S'\right\}$ is the global minimum of the potential. In the gray region labeled as $m_{a_S}^2 < 0$, the singlet-like CP-odd state becomes tachyonic, see eq.~\eqref{eq:a_tachyonic_condition_wpw}. In the orange region labeled as $H^{\rm NSM} \neq 0$, there exist minima with $H^{\rm NSM} \neq 0$ that are deeper than the physical minimum (they are only found numerically). In the blue region labeled $\left\{0,0,\sqrt{2}v_S'\right\}$, there exists a minimum at $\left\{H^{\rm SM}, H^{\rm NSM}, H^{\rm S}\right\} = \left\{0, 0, \sqrt{2}v_S'\right\}$ deeper than the physical minimum, see eq.~\eqref{eq:cond_vS'}. Similarly, in the green region labeled $\left\{0,0,0\right\}$, the trivial minimum is deeper than the physical minimum, see eq.~\eqref{eq:cond_triv}. The regions are shaded on top of each other in the order described in this caption; the dashed lines of the respective colors mark the edges of the respective regions where overlapping. In the figures, we chose $\tan\beta=1.5$ ($\tan\beta = 3$) for the left (right) panel, and $\kappa/\lambda = -0.1$ for both panels.}
   \label{fig:wpw_plane_kaplam-0.1}
\end{figure}

As we noted above, the potential may feature additional stationary points beyond those listed in eq.~\eqref{eq:min_loc}. In particular, minima deeper than the physical minimum can easily appear in the NMSSM for field configurations where $H^{\rm NSM}$ and $H^{\rm SM}$ take non-zero vevs. Such minima break the electroweak symmetry, and, unless $\langle H^{\rm NSM} \rangle = 0$ and $\langle H^{\rm SM} \rangle = \sqrt{2} v$, do not lead to electroweak physics compatible with observations. In general, $V_0^{{\rm eff},3}$ does not have stationary points in the $H^{\rm NSM}$-only direction, $V_0^{{\rm eff},3}(0, H^{\rm NSM}, 0)$, except for the trivial point $H^{\rm SM} = H^{\rm NSM} = H^{\rm S} = 0$. Instead, both $H^{\rm NSM}$ and $H^{\rm SM}$ (and sometimes $H^{\rm S}$) take non-vanishing values at these additional electroweak symmetry breaking minima. Such field configurations are very challenging to identify analytically, thus, we resort to numerical techniques to infer the constraints on the NMSSM parameter space arising from demanding the physical minimum to be deeper than any minima where $H^{\rm NSM} \neq 0$.\footnote{We use the package \texttt{HOM4PS2}~\cite{Lee2008} to solve the system of first derivatives of $V_0^{{\rm eff},3}(H^{\rm SM}, H^{\rm NSM}, H^{\rm S})$ to identify the stationary points, and then check numerically if the global minimum is the physical minimum.}

Finally, the parameter space of the NMSSM is also constrained by avoiding tachyonic masses. As discussed in section~\ref{sec:NMSSM}, the most relevant constraint arises from avoiding the singlet-like neutral CP-odd state, $a_S$, becoming tachyonic. In terms of $v_S'/v_S$, the constraint arising from eq.~\eqref{eq:a_tachyonic_condition} can be rewritten as
\begin{equation} \label{eq:a_tachyonic_condition_wpw}
   \frac{v_S'}{v_S} + 1 \gtrsim - \frac{3}{4} \frac{\lambda^2 v^2}{\mu^2} \sin(2\beta) \left[ \frac{\lambda}{\kappa} - \frac{\sin(2\beta)}{2} \right] \;.
\end{equation}

\begin{figure}
   \includegraphics[width=0.49\linewidth]{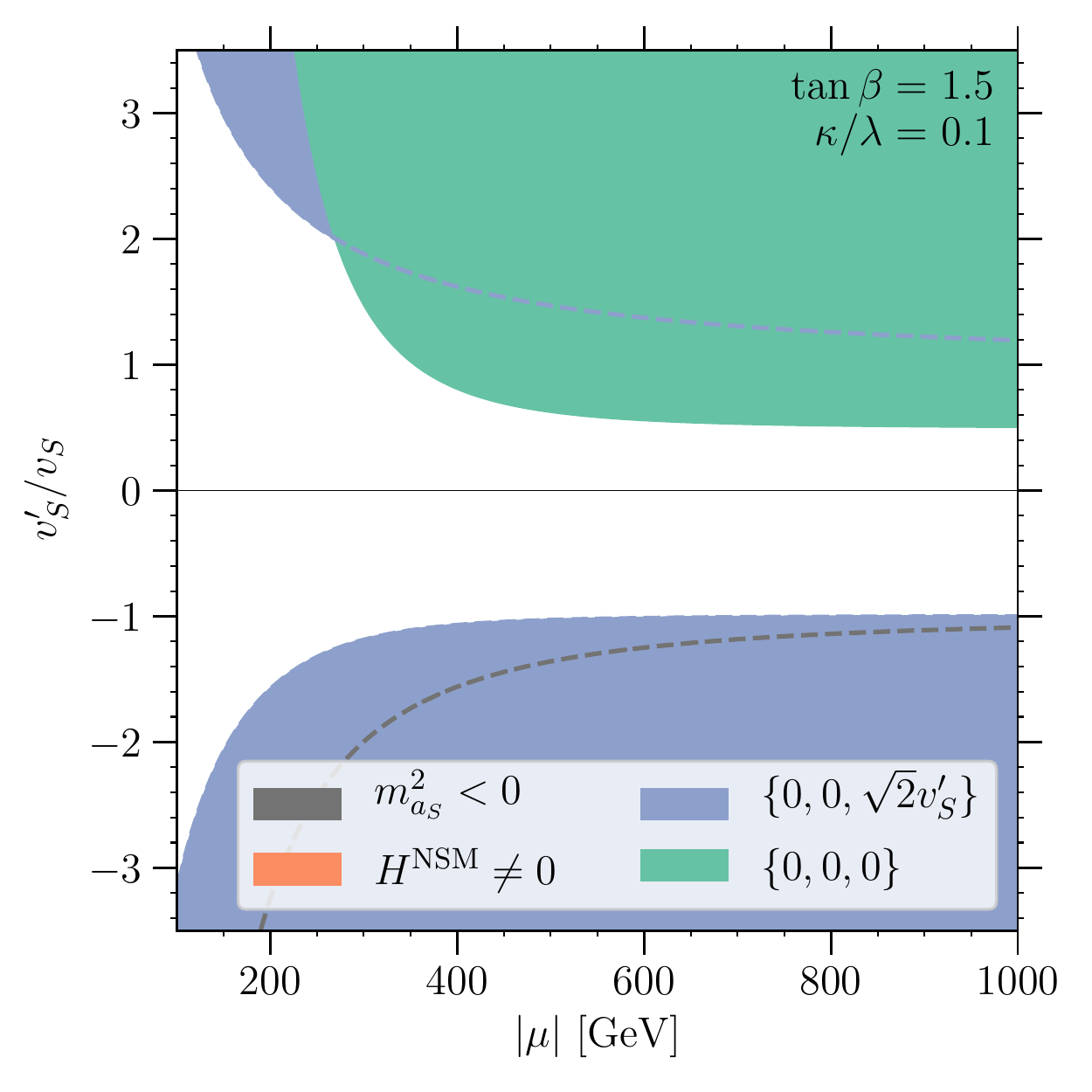}
   \includegraphics[width=0.49\linewidth]{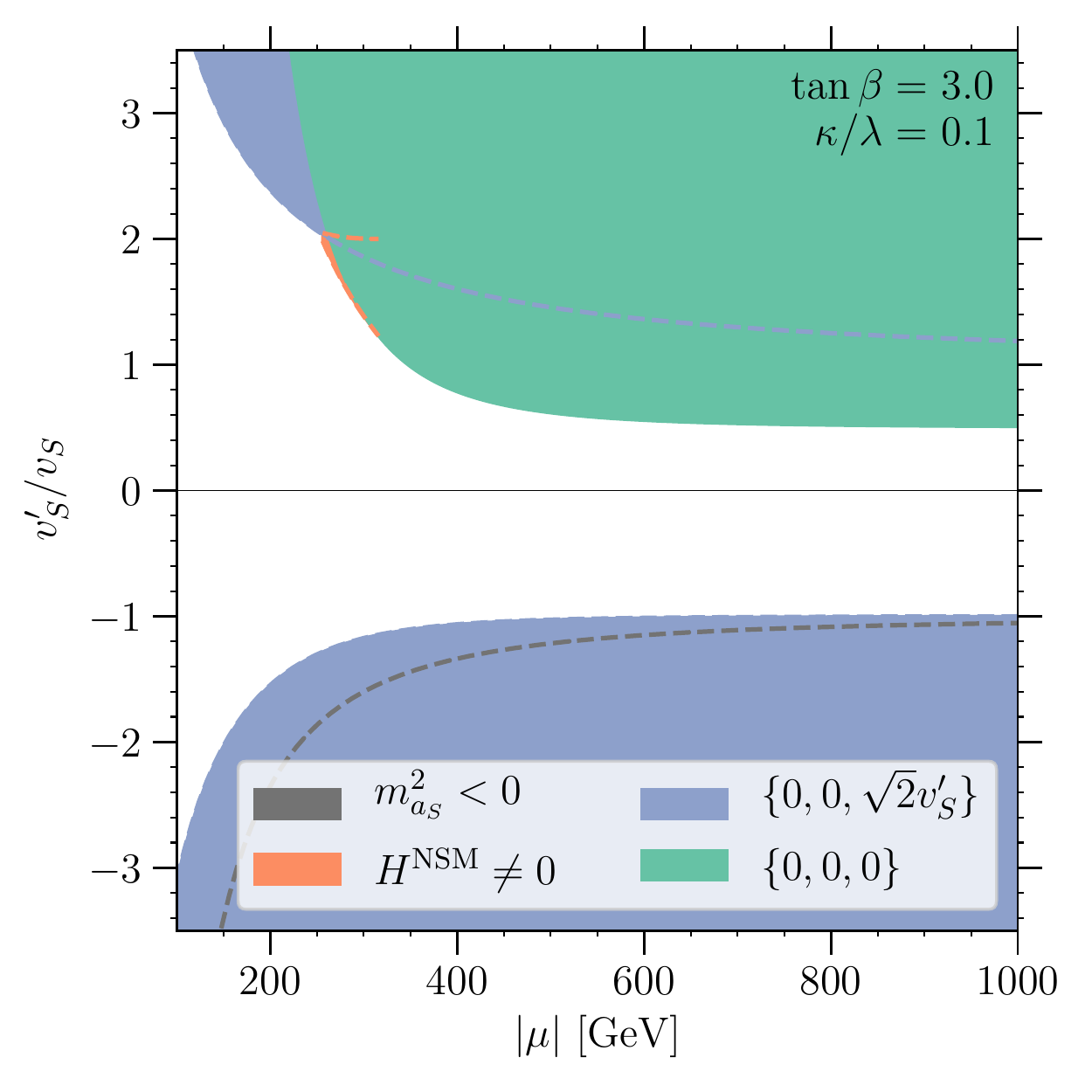}
   \caption{Same as figure~\ref{fig:wpw_plane_kaplam-0.1} but for $\kappa/\lambda = 0.1$.}
   \label{fig:wpw_plane_kaplam0.1}
\end{figure}

\begin{figure}
   \includegraphics[width=0.49\linewidth]{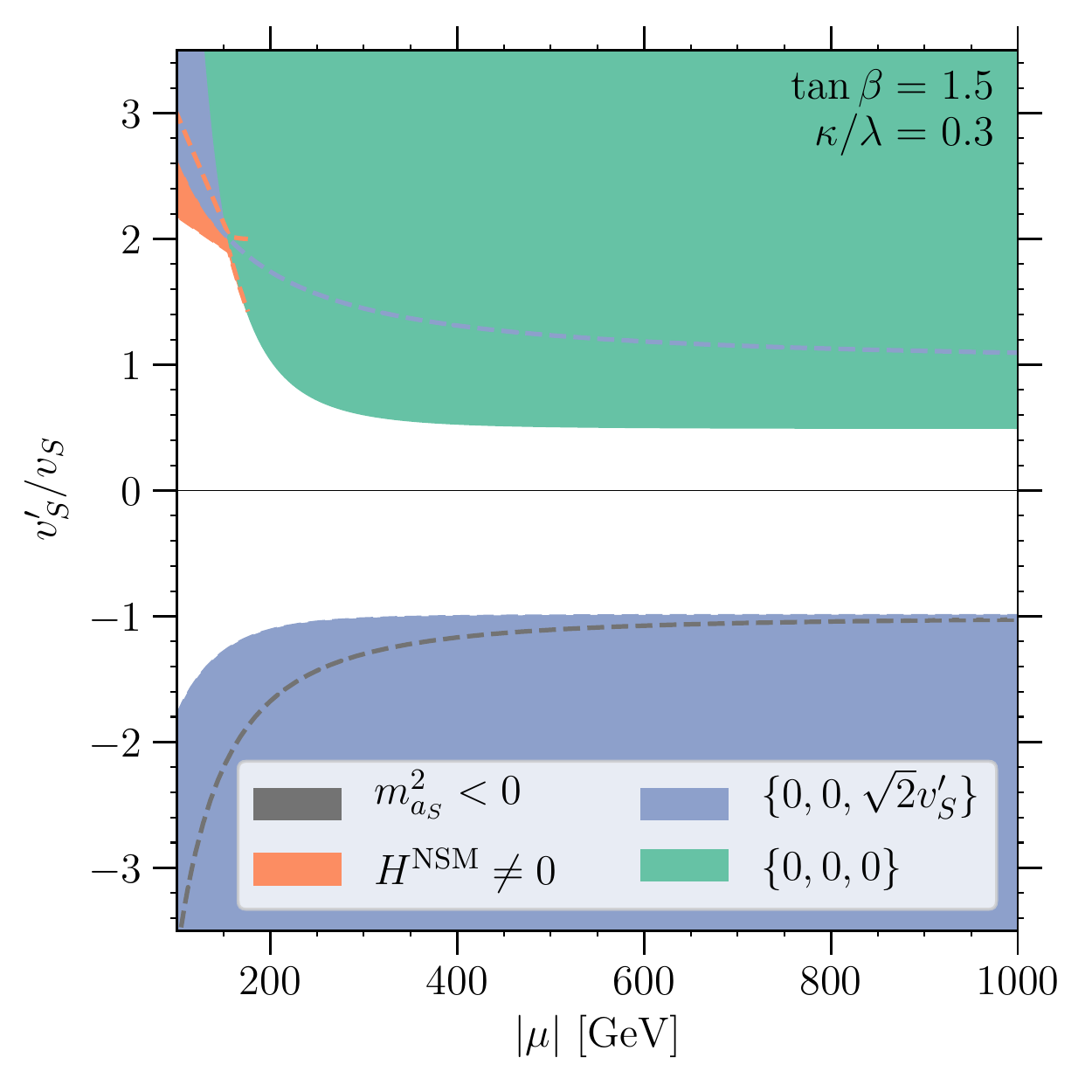}
   \includegraphics[width=0.49\linewidth]{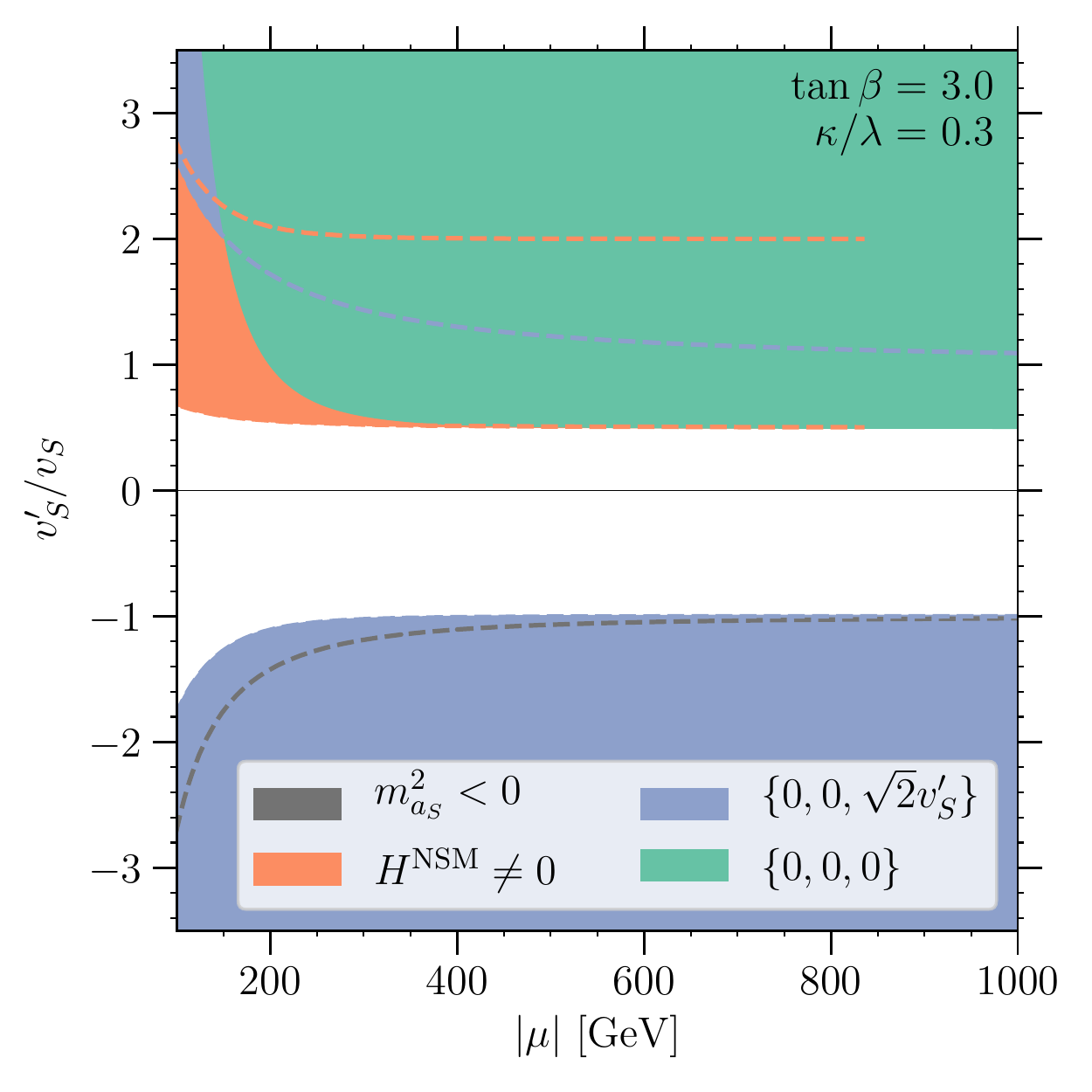}
   \caption{Same as figure~\ref{fig:wpw_plane_kaplam-0.1} but for $\kappa/\lambda = 0.3$.}
   \label{fig:wpw_plane_kaplam0.3}
\end{figure}

Figures~\ref{fig:wpw_plane_kaplam-0.1}--\ref{fig:wpw_plane_kaplam0.3} show the allowed region of parameter space in the $\left|\mu\right|$ vs. $v_S'/v_S$ plane for values of $\tan\beta = \left\{1.5, 3\right\}$ and $\kappa/\lambda = \left\{-0.1, 0.1, 0.3\right\}$. The different shaded regions are excluded by the constraints from eq.~\eqref{eq:cond_triv} (green shade), eq.~\eqref{eq:cond_vS'} (blue shade), and numerical results (orange shade). Correspondingly, these constraints come from avoiding the trivial minimum, the minimum at $\left\{H^{\rm SM}, H^{\rm NSM}, H^{\rm S}\right\} = \left\{0, 0, \sqrt{2}v_S'\right\}$, or minima with $H^{\rm NSM} \neq 0$, becoming deeper than the physical minimum. We also show the region where the singlet-like CP-odd mass eigenstate $a_S$ becomes tachyonic, eq.~\eqref{eq:a_tachyonic_condition_wpw}, with the gray shade. Note that overlapping regions are marked by dashed lines of the corresponding colors. In all figures, we truncate the $x$-axis at $\left|\mu\right| = 100\,$GeV; smaller values of $\left|\mu\right|$ are disfavored by null results of chargino searches at LEP. Since we imposed alignment (without decoupling), the scalar potential is uniquely specified by $v_S'/v_S$ (see eq.~\eqref{eq:vsprime}), $\mu$, $\tan\beta$, and $\kappa/\lambda$, and the potential is insensitive to the sign of $\mu$. As we can see from eqs.~\eqref{eq:cond_triv} and~\eqref{eq:cond_vS'} (the green and blue shaded regions, respectively), the conditions stemming from the trivial minimum and the minimum at $\left\{H^{\rm SM}, H^{\rm NSM}, H^{\rm S}\right\} = \left\{0, 0, \sqrt{2}v_S'\right\}$ becoming deeper than the physical minimum do not depend on the sign of $\kappa$ and are relatively insensitive to the value of $\left|\kappa\right|$. 

For large values of $\left|\kappa \mu^2\right|$, eq.~\eqref{eq:cond_triv} implies that the physical minimum is deeper than the trivial minimum for $v_S'/v_S < 1/2$. Equation~\eqref{eq:cond_vS'} on the other hand implies, for large $\left|\kappa \mu^2\right|$, that $v_S'/v_S > -1$ to avoid the minimum at $\left\{H^{\rm SM}, H^{\rm NSM}, H^{\rm S}\right\} = \left\{0,0,\sqrt{2}v_S'\right\}$ becoming deeper than the physical minimum. These constraints relax for smaller values of $\left|\kappa \mu^2\right|$, i.e. where the term proportional to $m_{h_{125}}^2 v^2$ in eq.~\eqref{eq:V3phys} becomes relevant. As can also be seen from eqs.~\eqref{eq:cond_triv} and \eqref{eq:cond_vS'}, the range of $v_S'/v_S$ opens up for $\left|\kappa \mu^2\right| \lesssim \lambda^2 m_{h_{125}} v$. With $\lambda \sim 0.65$ in the alignment limit, we find $\sqrt{\lambda m_{h_{125}} v} \sim 120\,$GeV. In figures~\ref{fig:wpw_plane_kaplam-0.1}--\ref{fig:wpw_plane_kaplam0.3} we can observe the corresponding change in the blue and green shaded bounds for $\left|\mu\right| \lesssim 120\,{\rm GeV} / \sqrt{\left|\kappa/\lambda\right|}$.

Finally, as discussed above, the region of parameter space where minima with $H^{\rm NSM} \neq 0$ are deeper than the physical minimum can only be inferred by numerically investigating the vacuum structure. From figures~\ref{fig:wpw_plane_kaplam-0.1}--\ref{fig:wpw_plane_kaplam0.3} we see that such constraints become more stringent with larger $\left|\kappa\right|$ and depend on the value of $\tan\beta$. Furthermore, the constraints arising from avoiding such minima are sensitive to the sign of $\kappa$; for $\kappa < 0$, avoiding minima with $H^{\rm NSM} \neq 0$ effectively sets a lower limit on the value of $v_S'/v_S$, while for $\kappa > 0$, avoiding these minima sets an upper bound on the value of $v_S'/v_S$.

\subsection{Thermal History: Analytical Understanding} \label{sec:thermal_ana}

In this section, we explore the possible phase transition patterns in the NMSSM. We first discuss the effective potential at very high temperatures, which gives guidance on the starting point of the thermal evolution. Then, we discuss the requirements a phase transition must satisfy to provide favorable conditions for electroweak baryogenesis via a SFOEWPT. We continue by discussing specific phase transition patterns which appear in the NMSSM, and fix a shorthand notation we will use to identify them. We close this section by discussing the regions in parameter space where we expect to observe different transition patterns, in particular, the regions in which we expect the nucleation probabilities of first order phase transitions to be sufficiently large for such transitions to complete.

Let us start with the vacuum structure at very high temperatures. In the limit $T^2 \gg \widehat{m}_i^2$, and neglecting the Daisy coefficients, the finite temperature potential, eq.~\eqref{eq:V1Tonly}, can be written as
\begin{equation} \label{eq:Vthermal_hT}
   V_{\rm 1-loop}^{T\neq0} \xrightarrow[T^2 \gg \widehat{m}_i^2]{} T^4 \left[\ldots\right] + \frac{T^2}{48} \left( 2 \sum_{i=B} n_i \widehat{m}_i^2 + \sum_{i=F} n_i \widehat{m}_i^2 \right) + T^4 \times \mathcal{O}\left( \left| \frac{\widehat{m}_i^2}{T^2} \right|^{3/2} \right) \;.
\end{equation}
The ellipsis $\left[\ldots\right]$ in eq.~\eqref{eq:Vthermal_hT} indicates terms which are independent of the field values. It is straightforward to see that in this limit, the field-dependent terms of the thermal potential are parameterized by the Daisy coefficients, (see eq.~\eqref{eq:Daisycoeff})
\begin{equation} \begin{split}
   V_{\rm 1-loop}^{T\neq0} &\to \left[ \frac{c_{H^{\rm SM} H^{\rm SM}}}{2} \left(H^{\rm SM}\right)^2 + c_{H^{\rm SM} H^{\rm NSM}} H^{\rm SM} H^{\rm NSM} + \frac{c_{H^{\rm NSM} H^{\rm NSM}}}{2} \left(H^{\rm NSM}\right)^2 \right. \\
   & \qquad \left. + \frac{c_{H^{\rm S} H^{\rm S}}}{2} \left(H^{\rm S}\right)^2 \right] T^2 + \ldots
\end{split} \end{equation}
where the ellipsis now includes both the field-independent and higher-order terms. Explicit expressions for the $c_{ij}$ can be found in appendix~\ref{app:Daisy}. Note that the symmetries of the NMSSM enforce this particular form of the high-temperature potential. In particular, the $\mathbb{Z}_3$ symmetry (and gauge symmetry) ensures that terms linear in the fields (such as $\mu_i H^i T^2$, where $\mu_i$ is a coefficient of dimension mass) cancel, while gauge symmetry forbids terms mixing one doublet with one singlet state, i.e. $H^{\rm SM} H^{\rm S} T^2$ and $H^{\rm NSM} H^{\rm S} T^2$.

Since all coefficients $c_{ij}$ are positive, and $c_{H^{\rm SM} H^{\rm SM}} c_{H^{\rm NSM} H^{\rm NSM}} > c_{H^{\rm SM} H^{\rm NSM}}^2$ throughout the parameter space, the trivial minimum $\left\{H^{\rm SM}, H^{\rm NSM}, H^{\rm S}\right\} = \left\{0, 0, 0\right\}$ is guaranteed to be the global minimum of the effective potential at very high temperatures. Thus, any phase transition patterns in the NMSSM will begin in the trivial phase. In order to give rise to acceptable phenomenology, the (chain of) phase transition(s) must end in the physical minimum, $\left\{H^{\rm SM}, H^{\rm NSM}, H^{\rm S}\right\} = \sqrt{2} \left\{v, 0, v_S \right\}$. If the transition pattern involves multiple steps, the most relevant property of the intermediate phase(s) for electroweak baryogenesis is if the electroweak symmetry is broken, i.e. if $H^{\rm SM}$ or $H^{\rm NSM}$ acquires a non-trivial vev, or if, instead, $H^{\rm SM} = H^{\rm NSM} = 0$ and the electroweak symmetry is conserved in the intermediate phase(s). 

A phase transition must satisfy certain requirements in order to give rise to favorable conditions for electroweak baryogenesis: In order for a baryon asymmetry to be produced in the transition, and such asymmetry not to be subsequently washed out in the low temperature phase, electroweak sphalerons must be active in the high-temperature phase and suppressed in the low temperature phase. Estimating the rate of the sphaleron suppression is a notorious problem in the perturbative approach to the phase transition calculation, see, for example, refs.~\cite{Patel:2011th, Garny:2012cg, Morrissey:2012db}, and even more so if the electroweak symmetry is broken in multiple steps, see, for example, ref.~\cite{Blinov:2015sna}. 

We shall demand 
\begin{equation} \label{eq:SFOEWPT}
   \left( \frac{\sqrt{ \left\langle H^{\rm SM}_{lT} \right\rangle^2 + \left\langle H^{\rm NSM}_{lT} \right\rangle^2}}{T} > 1 \right) \quad \land \quad \left( \frac{\sqrt{ \left\langle H^{\rm SM}_{hT} \right\rangle^2 + \left\langle H^{\rm NSM}_{hT} \right\rangle^2}}{T} < 0.5 \right) \;,
\end{equation} 
as conditions for a SFOEWPT. Here, $\left\langle \Phi_{hT} \right\rangle$ ($\left\langle \Phi_{lT} \right\rangle $) is the value of $\Phi$ in the high (low) temperature phase at the temperature $T$ where the phase transition occurs. The first condition ensures that electroweak sphalerons are inactive in the low-temperature phase, while the second condition requires the sphalerons to not be unduly suppressed in the high temperature phase. We stress that while the numerical thresholds for the order parameters we chose in eq.~\eqref{eq:SFOEWPT} are indicative for the possibility of generating the baryon asymmetry through a SFOEWPT~\cite{Patel:2011th}, obtaining the exact conditions would require a gauge-invariant evaluation of the sphaleron profile through the bubble wall which is beyond the scope of this work.

In the remainder of this paper, we use a shorthand notation to classify the phase transition patterns we observe in the NMSSM: 
\begin{itemize}
   \item We use an integer $(1, 2, \ldots)$ to denote the number of steps in the transition patterns.
   
   \item For 2-step transitions (we don't observe transition patterns with more than 2 steps in our data) we use a roman number to classify the intermediate phase:
   \begin{itemize}
      \item ``(I)'' denotes an intermediate phase in the singlet-only direction, i.e. where $\left\langle H^{\rm SM} \right\rangle = \left\langle H^{\rm NSM} \right\rangle = 0$ and electroweak symmetry is conserved,
      \item ``(II)'' denotes an intermediate phase in which electroweak symmetry is broken, i.e. where at least one of the fields $H^{\rm NSM}$ or $H^{\rm SM}$ acquires non-trivial vev.
      \end{itemize}
   
   \item We use a lower case letter to denote the strength of any transitions in which electroweak symmetry is broken in the low-temperature phase, 
      \begin{itemize}
         \item ``a'' denotes a SFOEWPT,
         \item ``b'' denotes a first order phase transition that is not a SFOEWPT, i.e does not satisfy one (or both) of the conditions in eq.~\eqref{eq:SFOEWPT},
         \item ``c'' denotes a second order phase transition.
      \end{itemize} 
\end{itemize}
Thus, for example, ``1-a'' denotes a direct one-step SFOEWPT from the trivial phase to the electroweak phase. ``2(I)-b'' denotes a two-step transition pattern, where the first step is from the trivial phase to a singlet-only phase (since electroweak symmetry is not broken in this intermediate phase, we do not differentiate the pattern with respect to the strength of this first transition), and the second step is a first order (but not SFOEWPT) transition from the singlet-only to the electroweak phase. ``2(II)-ca'' on the other hand denotes a two-step phase transition pattern, where the first transition is a second order phase transition into a phase in which electroweak symmetry is broken (but which is distinct from the electroweak phase), and the second transition is a SFOEWPT from this intermediate phase to the electroweak phase. 

We can get some intuition about the different regions of parameter space suitable for the respective phase transition patterns from the shape of the effective potential. While thermal effects alter the shape of the potential at finite temperatures, the zero-temperature vacuum structure still indicates the relative importance of the different possible local minima for the thermal history. Thus, we expect the results from section~\ref{sec:VacStr} to be indicative for the transition patterns suggested by the critical temperature calculation. For example, we can expect direct one-step transition patterns to most prominently be realized in the parameter region close to where the trivial minimum becomes the global minimum at zero temperature (green shade in figures~\ref{fig:wpw_plane_kaplam-0.1}--\ref{fig:wpw_plane_kaplam0.3}). Similarly, we can expect ``2(I)'' transition patterns to appear in the parameter regions adjacent to where $\left\{H^{\rm SM}, H^{\rm NSM}, H^{\rm S}\right\} = \left\{0, 0, \sqrt{2} v_S'\right\}$ becomes the global minimum at zero temperature (blue shade), and ``2(II)'' transitions are expected to appear in regions close to those where the global minimum has non-trivial vev of $H^{\rm NSM} \neq 0$ (orange shade).

The vacuum structure gives however little information about the tunneling probability from one local minimum to another, i.e. if a first order phase transition suggested by the critical temperature calculation can actually nucleate. The tunneling rate is controlled by the height of the barrier and the distance (in field space) between the respective local minima. The higher the barrier, and the larger the distance between the minima, the lower the nucleation probability. Although the shape of the potential is modified by thermal effects, we can learn some lessons from the zero-temperature potential. As discussed above, the trivial minimum is the global minimum of the effective potential at very high temperatures. Thus, any phase transition pattern starts at $H^{\rm SM} = H^{\rm NSM} = H^{\rm S} = 0$. The distance between the trivial and the physical minimum (at zero temperature) is given by $\sqrt{2 v^2 + 2 \mu^2/\lambda^2}$. Since the values of $v = 174\,$GeV and $\lambda \sim 0.65$ are fixed by electroweak precision data and the alignment conditions, respectively, the distance between the trivial and the physical minimum is controlled by $\left|\mu\right|$. The distance increases with the value of $\left|\mu\right|$, hence, nucleation proceeds more easily for small $\left|\mu\right|$.

The height of the barrier around the trivial minimum can be inferred from the squared mass parameters of the fields $H^{\rm SM}$, $H^{\rm NSM}$, and $H^{\rm S}$ around the trivial point, i.e. the field-dependent masses given in appendix~\ref{app:field_masses} at $H^{\rm SM} = H^{\rm NSM} = H^{\rm S} = 0$. In order for a phase transition to occur, the smallest of the eigenvalues of the squared mass matrix should be approximately zero, implying a flat direction around the trivial point at zero temperature. If the smallest eigenvalue is too large, the barrier around the trivial minimum is large, and hence the tunneling rate will be too small to allow for successful nucleation. If the smallest squared mass eigenvalue is negative, the trivial minimum is a saddle point of the potential (at zero temperature). Finite temperature effects can still give rise to a barrier between the trivial and the physical minimum required for a SFOEWPT in this situation, but only if the absolute value of the smallest squared mass parameter is not too large, such that thermal effects can overcome the zero-temperature shape of the potential.

At the trivial point $H^{\rm SM} = H^{\rm NSM} = H^{\rm S} = 0$, the matrix of the squared mass parameters is diagonal in the basis $\left\{H_d, H_u, S\right\}$, see eq.~\eqref{eq:V0}. Thus, we can directly infer the presence and height of the barrier around the trivial point from the parameters $m_{H_d}^2$, $m_{H_u}^2$, and $m_S^2$. In the alignment limit, $m_{H_u}^2 - m_{H_d}^2 = M_A^2 \cos (2\beta)$. Note that $\cos (2\beta) < 0$ for $\tan\beta > 1$ and hence, $m_{H_u}^2$ is the smaller of the doublet-like eigenvalues. In the alignment limit,
\begin{equation} \label{eq:mHu_condition}
   m_{H_u}^2 = M_A^2 \cos^2\beta - \mu^2 - \frac{m_{h_{125}}^2}{2} \approx \frac{\mu^2}{\tan^2\beta} \left( 1 - \frac{\kappa}{\lambda} \tan\beta \right) - \frac{m_{h_{125}}^2}{2} \;.
\end{equation}
This equation yields a critical value of $\left|\mu\right|$, for which $m_{H_u}^2 \approx 0$. This critical value of $\left|\mu\right|$ is increasing with larger values of $\tan\beta$ and of $\kappa/\lambda$. For example, for $\tan\beta = 1.5$ and $\kappa/\lambda = -0.1$, the critical value is $\left|\mu\right| \approx 125\,$GeV, while for the larger value $\kappa/\lambda = 0.3$ eq.~\eqref{eq:mHu_condition} implies $m_{H_u}^2 \approx 0$ for $\left|\mu\right| \approx 180\,$GeV. Instead, for a larger value of $\tan\beta = 3$ and $\kappa/\lambda = -0.1$, the critical value is $\left|\mu\right| \approx 235\,$GeV. For values of $\left|\mu\right|$ larger than the critical value, we expect large barriers around the trivial minimum in the $H_u$ direction, while for smaller values of $\left|\mu\right|$, $m_{H_u}^2$ becomes negative and the trivial point becomes a saddle point at zero temperature.

A flat direction can also arise in the $H^{\rm S}$ direction. The squared mass parameter of $H^{\rm S}$ at $H^{\rm SM} = H^{\rm NSM} = H^{\rm S} = 0$, see eq.~\eqref{eq:V3eff_align}, is
\begin{equation} \label{eq:mS_condition}
   m_S^2 = 2 \frac{\kappa^2}{\lambda^2} \mu^2 \frac{v_S'}{v_S} \;.
\end{equation}
The alignment conditions enforce sizable values of $\lambda \sim 0.65$, thus, the value of $m_S^2$ is controlled by $\kappa^2 \mu^2 (v_S'/v_S)$. Since the temperature corrections to $m_S^2$, eq.~(\ref{eq:Daisy}), are of order $0.2\,T^2$, one would expect that at the characteristic temperature of the EWPT of order 100\,GeV, the tunneling rate could only be large enough for successful nucleation if the squared mass parameter controlling the barrier $m_S^2 \ll (100\,{\rm GeV})^2$. This condition can be achieved in two ways: either, $\left|v_S'/v_S\right| \ll 1$, or $\left|\kappa \mu\right| \ll 100$\,GeV. 

Note that the conditions $m_{H_u}^2 \approx 0$ or $m_S^2 \approx 0$ are indicative for the possibility of a first order phase transition to successfully nucleate at finite temperature since they imply the presence of an approximately flat direction around the trivial minimum at zero temperature. However, this analysis does not predict the transition pattern, which is determined by the shape of the potential away from the trivial minimum (at the transition temperature). The bounce solution of the fields (the trajectory in field space connecting the local minima) is, in general, not a straight line in field space; in particular, $m_S^2 \approx 0$ does not necessarily lead to ``2(I)'' transition patters, and $m_{H_u}^2 \approx 0$ does not directly imply ``2(II)'' patterns.

\section{Numerical Results} \label{sec:Numerical}

In order to explore the EWPT in the NMSSM, and, in particular, find which regions of parameter space give rise to phase transition patterns suitable for electroweak baryogenesis, we perform an extensive numerical study using \texttt{CosmoTransitions\_v2.0.5}~\cite{Wainwright:2011kj}. In this section, we first describe our implementation of the NMSSM in \texttt{CosmoTransitions}\footnote{Our code is available at \url{https://github.com/sbaum90/NMSSM_CosmoTrans.git}.} and sketch the steps of the calculations \texttt{CosmoTransitions} performs. As discussed in section~\ref{sec:NMSSM}, in the alignment limit, the Higgs sector of the NMSSM can be described by the four parameters $\left\{\tan\beta, \kappa/\lambda, \mu, v_S'/v_S\right\}$, and we perform random scans in this parameter space. We show the results of our numerical scans in figures~\ref{fig:NumRes_kaplam-0.1_tb_1.5}--\ref{fig:joined_myvspvs_mHmhs}. In section~\ref{sec:Num_BC} we discuss the regions of the parameter space where points satisfy the boundary conditions we implement in our \texttt{CosmoTransitions} calculation. In section~\ref{sec:Num_Tcrit} we discuss the phase transition patterns suggested by the critical temperature calculation and we compare these results with the thermal histories obtained by calculating the nucleation rate. As we shall see, the phase transition patterns obtained from the nucleation calculation differ substantially from those indicated by the critical temperature calculation, and thus, computing only the critical temperatures provides a misleading picture of the regions of parameter space favorable for electroweak baryogenesis. In section~\ref{sec:Num_pheno} we comment on the collider and dark matter phenomenology in the region of parameter space promising for baryogenesis via a SFOEWPT.

The \texttt{CosmoTransitions} package provides a framework for calculating phase transitions in single- and multi-field models (in the perturbative approach). The implementation of a model into \texttt{CosmoTransitions} proceeds via the specification of the effective potential. We have described the effective (temperature-dependent) potential of the NMSSM in section~\ref{sec:NMSSM}, in particular, it consists of the terms
\begin{equation} \label{eq:V1T_sec3}
   V_1(T) = V_0^{\rm eff} + V_{\rm 1-loop}^{\rm CW}(\widetilde{m}_i^2) + V_{\rm 1-loop}^{T\neq0}(\widetilde{m}_i^2) \;.
\end{equation}
$V_0^{\rm eff}$ is the tree-level potential of the effective theory obtained after integrating out the sfermions and gluinos, $V_{\rm 1-loop}^{\rm CW}$ is the Coleman-Weinberg potential (including counterterms as shown in eq.~\eqref{eq:ct_lag}), and $V_{\rm 1-loop}^{T\neq0}$ contains the thermal corrections to one-loop order, see section~\ref{sec:TCorr}. Explicit formulae for the field-dependent masses, the counterterm coefficients, and the Daisy coefficients are collected in appendices~\ref{app:field_masses},~\ref{app:ct_coeff}, and~\ref{app:Daisy}, respectively. 

The calculation of the phase transition pattern with \texttt{CosmoTransitions} proceeds in multiple steps:
\begin{itemize}
   \item First, we compute the locations of the local minima at zero temperature\footnote{Note that it is crucial to find all relevant local minima at $T=0$. To this end, we use a large number of initial guesses (a three-dimensional grid spanned by each of the three fields $H^{\rm SM}, H^{\rm NSM}, H^{\rm S}$ taking values $\Phi_i = \left\{-1000, -100, -10, 0, 10, 100, 1000\right\}\,$GeV) as input for \texttt{CosmoTransitions} default routines for minimizing the effective potential.}. 

   \item Second, the {\it phases}, i.e. the temperature-dependent locations in field space and values of the effective potential at the local minima, are computed from the list of zero-temperature minima. Note that if a phase {\it ends} at some temperature, i.e. ceases to be a local minimum, \texttt{CosmoTransitions} tries to find other local minima nearby in field space and then traces the corresponding phases as well. Thus, \texttt{CosmoTransitions} attempts to include phases which cannot be obtained from the list of zero-temperature minima because they exist only at finite temperatures. 

   \item Third, using the phases as input, \texttt{CosmoTransitions} analyzes the temperature-dependent vacuum structure of the potential. The most relevant output from this step is a list of {\it critical temperatures}, the temperatures at which two distinct local minima of the potential have the same potential value. At the critical temperatures, the role of the global minimum of the effective potential passes from one phase to another, suggesting the phase transition pattern.

   \item Finally, for possible first order phase transitions indicated by the analysis of the vacuum structure, \texttt{CosmoTransitions} allows to compute the probability of the transition taking place. First order phase transitions proceed via bubble nucleation, and the nucleation rate is commonly parameterized via the {\it bounce action} $S_{\rm E}$, the Euclidean space-time integral over the (effective) Lagrangian density. In practice, it typically suffices to compute the three-dimensional effective Euclidean action, $S_{\rm E} \simeq S_3/T$. The technically most challenging part of this computation is finding the {\it bounce solution} for the scalar fields, i.e. the trajectory in field space connecting the two local minima which minimizes the Euclidean effective action.\footnote{\texttt{CosmoTransitions} uses a path deformation method to find the bounce solution, see ref.~\cite{Wainwright:2011kj}. Other publicly available codes for finding the bounce solution in multi-field potentials include \texttt{AnyBubble}~\cite{Masoumi:2016wot}, \texttt{BubbleProfiler}~\cite{Athron:2019nbd}, and \texttt{FindBounce}~\cite{Guada:2020xnz}.} The bubble nucleation rate per unit volume at finite temperature $T$ is given by $\Gamma/V \propto T^4 e^{-S_3/T}$; requiring the nucleation probability (for the EWPT) to be approximately one per Hubble volume and Hubble time leads to the {\it nucleation condition}~\cite{Linde:1981zj} (see, e.g., ref.~\cite{Mazumdar:2018dfl} for a review),
   \begin{equation} \label{eq:NucC}
   \frac{S_3(T)}{T} \simeq 140 \;.
   \end{equation} 
   The {\it nucleation temperature} $T_n$ is the (highest) temperature for which $S_3/T \lesssim 140$. If $S_3/T > 140$ for all $T > 0$, the corresponding transition does not occur because the tunneling probability through the barrier separating the respective local minima is too small. Typically, this is caused by a too high barrier and/or a too large distance (in field space) between the local minima.

\end{itemize}

Since the calculation of the nucleation temperature (involving the computation of the bounce action) is numerically expensive, to date such calculations have only been presented for a few benchmark points in the NMSSM, see refs.~\cite{Huber:2000mg, Carena:2011jy, Kozaczuk:2014kva, Bian:2017wfv, Athron:2019teq}. Here, we present results based on the full nucleation calculation for a broad scan of the parameter space.

We focus our study on the region of parameter space where alignment without decoupling is realized, i.e. the region of parameter space for which the NMSSM features a Higgs mass eigenstate which (at tree-level) couples to SM particles like the SM Higgs boson. As discussed in section~\ref{sec:NMSSM}, the alignment conditions fix the values of $\lambda$ and $M_A^2$ (or, equivalently, $A_\lambda$), leaving $\left\{\tan\beta, \mu, \kappa, A_\kappa \right\}$ as the four free parameters which control the effective potential. We fix the mass and mixing parameters of the stop sector (parameterized by the threshold correction $\Delta\lambda_2$ in $V_0^{\rm eff}$, see section~\ref{sec:RadCorr}) to obtain $m_{h_{125}} \simeq 125\,$GeV for the mass of the SM-like Higgs boson. As discussed in section~\ref{sec:VacStr}, we use $v_S'/v_S$ to re-parameterize $A_\kappa$. Here, $v_S = \mu/\lambda$ is the vev of the CP-even singlet interaction state at the physical minimum, $\left\langle H^{\rm S} \right\rangle = \sqrt{2} \mu/\lambda$, and $v_S' = - \left( \mu/\lambda + A_\kappa / 2\kappa \right)$ is the location of an extremum of $V_0^{\rm eff}$ in the singlet-only direction, $\left\{ H^{\rm SM}, H^{\rm NSM}, H^{\rm S}\right\} = \left\{0, 0,\sqrt{2} v_S'\right\}$. In summary, we use 
\begin{equation}
   \tan\beta \;,\quad \mu \;,\quad \frac{\kappa}{\lambda} \;,\quad \frac{v_S'}{v_S} \;,
\end{equation}
as input parameters for our numerical evaluation. Note that throughout our calculations, we fix the bino and wino mass parameters, which enter the radiative corrections from the charginos and neutralinos (see eqs.~\eqref{eq:mneuhat} and~\eqref{eq:mchar_field}), to $M_1 = M_2 = 1\,$TeV.

The $\left|\mu\right|$ vs. $v_S'/v_S$ plane lends itself particularly well to characterizing the vacuum structure of the NMSSM as discussed in section~\ref{sec:VacStr}. We perform two-dimensional scans over slices of the parameter spaces for fixed values of $\tan\beta$ and $\kappa/\lambda$, varying the values of $\mu$ and $v_S'/v_S$ by means of (linear-)flat distributions. While we have included counterterms to maintain the location of the physical minimum after including the Coleman-Weinberg potential (including $\left\langle H^{\rm S} \right\rangle = \sqrt{2} v_S = \sqrt{2} \mu/\lambda$), we have not included a counterterm which would similarly keep the location of the tree-level extremum at $\left\{H^{\rm SM}, H^{\rm NSM}, H^{\rm S}\right\} = \left\{ 0, 0, \sqrt{2} v_S' \right\}$ fixed. As a result, the location of the corresponding minimum of the effective potential after including $V^{\rm CW}_{\rm 1-loop}$ is no longer $\left\{H^{\rm SM}, H^{\rm NSM}, H^{\rm S}\right\} = \left\{ 0, 0, \sqrt{2} v_S'\right\}$, but changes to a new location we denote by $\left\{H^{\rm SM}, H^{\rm NSM}, H^{\rm S}\right\} = \left\{ 0, 0, \sqrt{2} \vspCW\right\}$. We find the value of $\vspCW$ by numerically solving 
\begin{equation} \label{eq:vspCW}
   \left. \frac{\partial V_1(T=0)}{\partial H^{\rm S}} \right|_{\substack{H^{\rm SM} = 0 \\ H^{\rm NSM} = 0}} = 0 \;.
\end{equation}
This equation yields three solutions: $H^{\rm S} = 0$ and two non-trivial solutions. Of these two non-trivial solutions, we identify the one further away (in $H^{\rm S}$ space) from $v_S= \mu/\lambda$ as $\vspCW$. We plot our numerical results in the $\left|\mu\right|$ vs. $\vspCW/v_S$ plane.

For each randomly drawn parameter point, we first demand a number of {\it boundary conditions}:
\begin{itemize}
   \item We check compatibility with the phenomenology of the observed SM-like 125\,GeV Higgs boson by checking that (after including the radiative corrections and the counterterms discussed in section~\ref{sec:RadCorr}) the parameter point features a CP-even Higgs mass eigenstate with mass $122 < m_{h_{125}}/{\rm GeV} < 128$, and admixtures of the non-SM-like interactions states less than $\left|C_{h_{125}}^{\rm NSM}\right| \tan\beta < 0.05$ and $\left|C_{h_{125}}^{\rm S}\right| < 0.1$,\footnote{Admixtures of $H^{\rm NSM}$ and $H^{\rm S}$ of this size modify the production cross sections and branching ratios of $h_{125}$ by $\lesssim 10\,\%$ compared to the SM prediction. The currently best-measured production cross section of the observed Higgs boson is via the gluon-fusion mode with a $1\sigma$ uncertainty of $\sim 15\,\%$~\cite{Sirunyan:2018koj,Aad:2019mbh}. Similarly, the largest branching ratios of the observed Higgs bosons are measured with $\sim 15\,\%$ uncertainty~\cite{Sirunyan:2018koj,Aad:2019mbh}.} where the $C_i^j$ denote the mixing angles in the extended Higgs basis, 
   \begin{equation} \label{eq:h125mix}
      h_{125} = C_{h_{125}}^{\rm SM} H^{\rm SM} + C_{h_{125}}^{\rm NSM} H^{\rm NSM} + C_{h_{125}}^{\rm S} H^{\rm S} \;.
   \end{equation}
   Note that since we fix $\lambda$ and $M_A^2$ via the alignment conditions, eqs.~\eqref{eq:Align1} and~\eqref{eq:Align2}, and include a counterterm to preserve the $H^{\rm SM}$--$H^{\rm S}$ alignment after including the Coleman-Weinberg corrections, see eq.~\eqref{eq:ct_lag}, most of our parameter points have admixtures of $H^{\rm NSM}$ and $H^{\rm S}$ to $h_{125}$ much smaller than these thresholds. The exception are points where the mass parameters of the interaction eigenstates $H^{\rm SM}$ and $H^{\rm S}$ are approximately degenerate; in this case, relatively small off-diagonal entries in the CP-even squared mass matrix can still lead to sizable mixing of $H^{\rm SM}$ and $H^{\rm S}$.

   \item In order to ensure compatibility with the null-results from chargino searches at the Large Electron Positron collider (LEP) (see, for example, refs.~\cite{LEPchargino1, LEPchargino2}) we exclude the parameter region $\left|\mu\right| < 100\,$GeV. Recall that the alignment conditions lead to a mass scale of the doublet-like Higgs bosons of $\left|M_A\right| \sim 2 \left|\mu\right| / \sin(2\beta)$. Thus, such values of $\left|\mu\right|$ allow for doublet-like Higgs bosons as light as $\left|M_A\right| \sim 200\,$GeV if $\tan\beta \simeq 1$, which potentially are in conflict with null results from direct searches for non-SM-like Higgs bosons at the LHC. We will return to this issue in section~\ref{sec:Num_pheno}. Note that searches for neutralinos and charginos at the LHC do not constrain the parameter space for $\left|\mu\right| \gtrsim 100\,$GeV in a relevant way, see, for example, ref.~\cite{Liu:2020muv}.

   \item We check that, at zero temperature, the physical minimum is the global minimum of the effective potential.\footnote{Thus, in this study we exclude the region of parameter space where the physical minimum is a metastable vacuum (with sufficiently long lifetimes to allow for feasible cosmology). While interesting in its own right, considering this scenario is beyond the scope of this work.} 
\end{itemize}
For each point satisfying all boundary conditions, we compute the phase transition pattern with \texttt{CosmoTransitions} as discussed above.

Figures~\ref{fig:NumRes_kaplam-0.1_tb_1.5}--\ref{fig:NumRes_kaplam0.3_tb_3.0}, to be discussed in detail in sections~\ref{sec:Num_BC} and \ref{sec:Num_Tcrit}, show the results from our parameter scans for $\tan\beta = \left\{1.5, 3\right\}$ and $\kappa/\lambda = \left\{-0.1, 0.1, 0.3\right\}$ in the $\left|\mu\right|$ vs. $\vspCW/v_S$ plane; these are the same slices of parameter space for which we have shown constraints from the zero-temperature vacuum structure of the effective tree-level potential, $V_0^{\rm eff}$, in figures~\ref{fig:wpw_plane_kaplam-0.1}--\ref{fig:wpw_plane_kaplam0.3}. In order to compare the results of the respective calculations, we color-code the points according to the transition patterns indicated by the critical temperature calculations in the left panels of figures~\ref{fig:NumRes_kaplam-0.1_tb_1.5}--\ref{fig:NumRes_kaplam0.3_tb_3.0}, while in the right panels, points are color-coded according to the thermal history obtained from the full nucleation calculation; see section~\ref{sec:thermal_ana} for our shorthand notation of the phase transition patterns. Points violating the boundary conditions described above are labeled ``failed BC'' in figures~\ref{fig:NumRes_kaplam-0.1_tb_1.5}--\ref{fig:NumRes_kaplam0.3_tb_3.0}. Points which satisfy all boundary conditions, but for which \texttt{CosmoTransitions} fails to return a phase transition pattern starting from the trivial minima at high temperature and ending in the physical minimum at zero temperature are labeled ``no transitions''. Note that the left and right panels show the same set of points in parameter space, the only difference is the color-coding of the points.

\subsection{Boundary Conditions} \label{sec:Num_BC}

Let us begin the discussion of the results of our parameter scans with the regions of parameter space where points fail to satisfy the boundary conditions. The boundary conditions are independent of the thermal calculation, hence, the same points are labeled ``failed BC'' in the left and right panels of figures~\ref{fig:NumRes_kaplam-0.1_tb_1.5}--\ref{fig:NumRes_kaplam0.3_tb_3.0}.

We observe that, for large values of $\left|\mu\right|$, the range of $\vspCW/v_S$ where points satisfy the boundary conditions is $-1 \lesssim \vspCW/v_S \lesssim 0.5$. This range is only weakly dependent on the values of $\tan\beta$ and $\kappa/\lambda$; only in the case of $\kappa/\lambda = -0.1$, shown in figures~\ref{fig:NumRes_kaplam-0.1_tb_1.5} and \ref{fig:NumRes_kaplam-0.1_tb_3.0}, we observe a different lower bound on $\vspCW/v_S$ at large $\left|\mu\right|$, being $\vspCW/v_S \gtrsim -0.5$ for $\tan\beta =1.5$ and $\vspCW/v_S \gtrsim -0.8$ for $\tan\beta = 3$. The range of $\vspCW/v_S$ where points satisfy the boundary conditions widens at small values of $\left|\mu\right|$, and here, the behavior depends more strongly on the values of $\kappa/\lambda$ and $\tan\beta$, as we can see by comparing the different slices of parameter space shown in figures~\ref{fig:NumRes_kaplam-0.1_tb_1.5}--\ref{fig:NumRes_kaplam0.3_tb_3.0}. We note that the boundary conditions widen for values of $\left|\mu\right| \lesssim 120\,{\rm GeV} / \sqrt{\left|\kappa/\lambda\right|}$. Furthermore, we observe that for $\kappa/\lambda = -0.1$ and $\tan\beta = 1.5$ (figure~\ref{fig:NumRes_kaplam-0.1_tb_1.5}), points fail the boundary conditions for $\left|\mu\right| \lesssim 150\,$GeV regardless of the value of $\vspCW/v_S$, while we do not observe such a lower bound on the value of $\left|\mu\right|$ for the other slices of parameter space.

This behavior can largely be understood from the discussion of the zero-temperature vacuum structure in section~\ref{sec:VacStr}, see also figures~\ref{fig:wpw_plane_kaplam-0.1}--\ref{fig:wpw_plane_kaplam0.3}. The analysis of the vacuum structure in section~\ref{sec:VacStr} was based on $V_0^{\rm eff}$, the potential of our effective model after integrating out all sfermions and the gluinos, but prior to including the Coleman-Weinberg corrections. We indicate the region of parameter space for which, per the analysis in section~\ref{sec:VacStr}, the physical minimum is the global minimum of $V_0^{\rm eff}$ at zero temperature with the thin black contours in figures~\ref{fig:NumRes_kaplam-0.1_tb_1.5}--\ref{fig:NumRes_kaplam0.3_tb_3.0}. Since these contours are derived from $V_0^{\rm eff}$, the $y$-axis for these contours is $v_S'/v_S$, where $v_S' = -\left( \mu/\lambda + A_\kappa/2\kappa \right)$ is the tree-level value. We see that, although these contours are derived from $V_0^{\rm eff}$, they describe well many of the features of the boundary conditions seen in our parameter scan, which incorporates radiative corrections. The largest deviations appear for $\kappa/\lambda = -0.1$, see figures~\ref{fig:NumRes_kaplam-0.1_tb_1.5} and~\ref{fig:NumRes_kaplam-0.1_tb_3.0}. While the contours here allow only a narrow range of $v_S'/v_S$ values, we see that the points from our parameter scan satisfy the boundary conditions for a much wider range of values of $\vspCW/v_S$ than what the contours suggest. Comparing with figure~\ref{fig:wpw_plane_kaplam-0.1}, we see that this discrepancy occurs in regions of parameter space where the analysis of $V_0^{\rm eff}$ suggested that a minimum with $\langle H^{\rm NSM} \rangle \neq 0$ was the global minimum of the potential (indicated by the orange shade in figure~\ref{fig:wpw_plane_kaplam-0.1}). This constraint was derived numerically in section~\ref{sec:VacStr}, and hence is challenging to understand quantitatively. However, it is not surprising that the region of parameter space disfavored by vacua with $\langle H^{\rm NSM} \rangle \neq 0$ becoming the global minimum of the potential changes considerably after including the Coleman-Weinberg corrections: the potential is subject to larger radiative corrections in the doublet-like directions of the effective potential than in the singlet-like direction, and furthermore, the $H^{\rm NSM}$ direction is affected by the counterterms we have included to maintain the location of the physical minimum, $\left\{H^{\rm SM}, H^{\rm NSM}, H^{\rm S}\right\} = \sqrt{2} \left\{v, 0, \mu/\lambda \right\}$. 

Before moving to the discussion of the phase transition patterns we observe for points satisfying the boundary conditions in section~\ref{sec:Num_Tcrit}, let us briefly mention a few features visible in figures~\ref{fig:NumRes_kaplam-0.1_tb_1.5}--\ref{fig:NumRes_kaplam0.3_tb_3.0}. First, we can see a gap in the points around $\vspCW/v_S \approx 1$, which widens for small values of $\left|\mu\right|$. This gap is due to numerical difficulties in our algorithm to find $\vspCW$ if $\vspCW \approx v_S$. Identifying the value of $\vspCW$ is particularly challenging for small $\left|\mu\right|$, because $\left|\mu\right|$ controls the size of $v_S = \mu/\lambda$. 

Second, an arc of points failing the boundary conditions crosses the region of parameter space consistent with the physical vacuum being the global minimum at zero temperature, starting at small values of $\left|\mu\right|$ and negative $\vspCW/v_S$ and ending at larger values of $\left|\mu\right|$ and positive $\vspCW/v_S$. This feature is particularly pronounced for $\tan\beta = 1.5$, and is due to the mass parameters of the interaction states $H^{\rm SM}$ and $H^{\rm S}$ becoming approximately degenerate for those points. As discussed below eq.~\eqref{eq:h125mix}, in this situation, even small deviations from the alignment conditions lead to a sizable $H^{\rm S}$ component of $h_{125}$, and thus, these points are forbidden by our requirement $\left| C_{h_{125}}^{\rm S} \right| < 0.1$. 

Neither of these issues is related to the thermal history of a given parameter point, and these issues do not occur in regions of parameter space which are of special interest for the phase transition calculation. Hence, we ignore them in the following.

We also note that in the left panels of figures~\ref{fig:NumRes_kaplam-0.1_tb_1.5}--\ref{fig:NumRes_kaplam0.3_tb_3.0}, where we show the results of the critical temperature calculation, points labeled ``no transition'' appear. As discussed in section~\ref{sec:thermal_ana}, the trivial minimum is guaranteed to be the global minimum of the potential at high temperatures, and for any point passing the boundary conditions, the physical minimum is the global minimum at zero temperature. For points labeled ``no transition'', \texttt{CosmoTransitions} failed to return a transition pattern starting in the trivial minimum at high temperatures and ending in the physical minimum at zero temperatures. This is due to numerical errors arising in the second step of the numerical calculation described above, i.e. the step in which \texttt{CosmoTransitions} attempts to trace the local minima of the effective potential with changing temperatures. We have investigated these numerical issues, and have not found any indication that they bias our results towards particular regions of parameter space. Thus, we expect that our scanning over a large number of points throughout the parameter space gives an accurate picture of the regions of parameter space suitable for electroweak baryogenesis.

\subsection{Comparison of Critical Temperature and Nucleation Results} \label{sec:Num_Tcrit}

\begin{figure}
   \includegraphics[width=.49\linewidth]{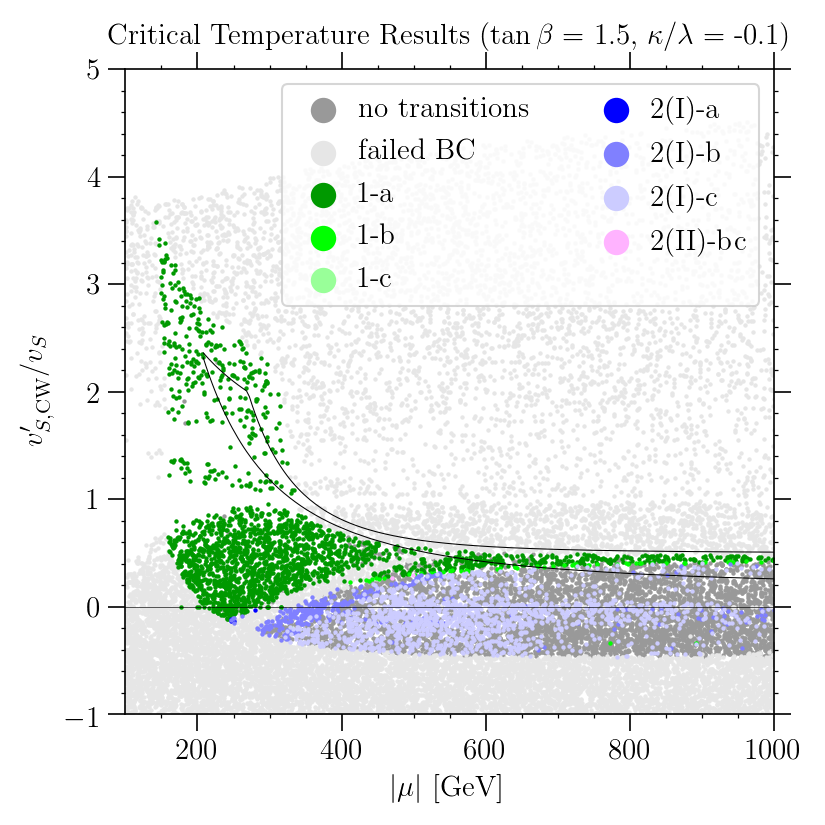}
   \includegraphics[width=.49\linewidth]{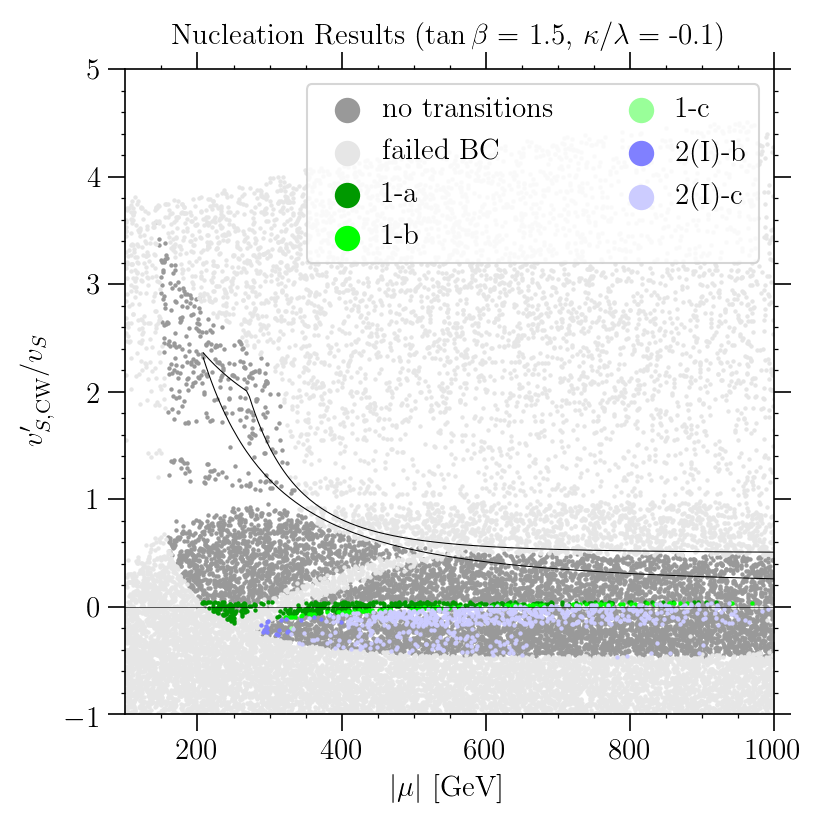}
   \caption{Results from our parameter scans in the $\left|\mu\right|$ vs. $\vspCW/v_S$ plane for the same slice of parameter space as shown in the left panel of figure~\ref{fig:wpw_plane_kaplam-0.1}: $\tan\beta = 1.5$ and $\kappa/\lambda = -0.1$. The left panel shows points categorized according to the phase transition patterns suggested by the critical temperature calculation. In the right panel, points are instead categorized by the thermal histories obtained from the nucleation calculation.  For points labeled ``no transition'', \texttt{CosmoTransitions} did not return a transition chain starting in the trivial minimum at high temperatures and ending in the physical minimum at zero temperature, and points labeled ``failed BC'' do not satisfy our boundary conditions defined in the text. The solid lines enclose the region of parameter space for which we find feasible zero-temperature vacuum structure in section~\ref{sec:VacStr}. These bounds are obtained from tree-level relations, hence, for these bounds, the $y$-axis is $v_S'/v_S$, where $v_S' = - \left( \mu/\lambda + A_\kappa/2\kappa \right)$ is the location of an extremum of $V_0^{\rm eff}$ in the singlet-only direction.}
    \label{fig:NumRes_kaplam-0.1_tb_1.5}
 \end{figure}

\begin{figure}
   \includegraphics[width=.49\linewidth]{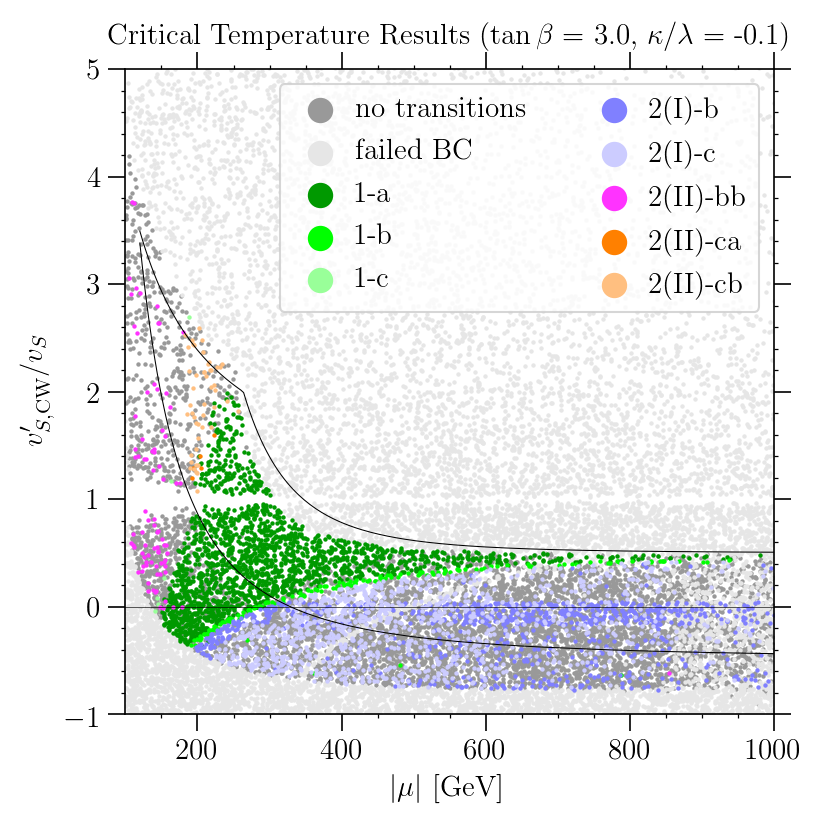}
   \includegraphics[width=.49\linewidth]{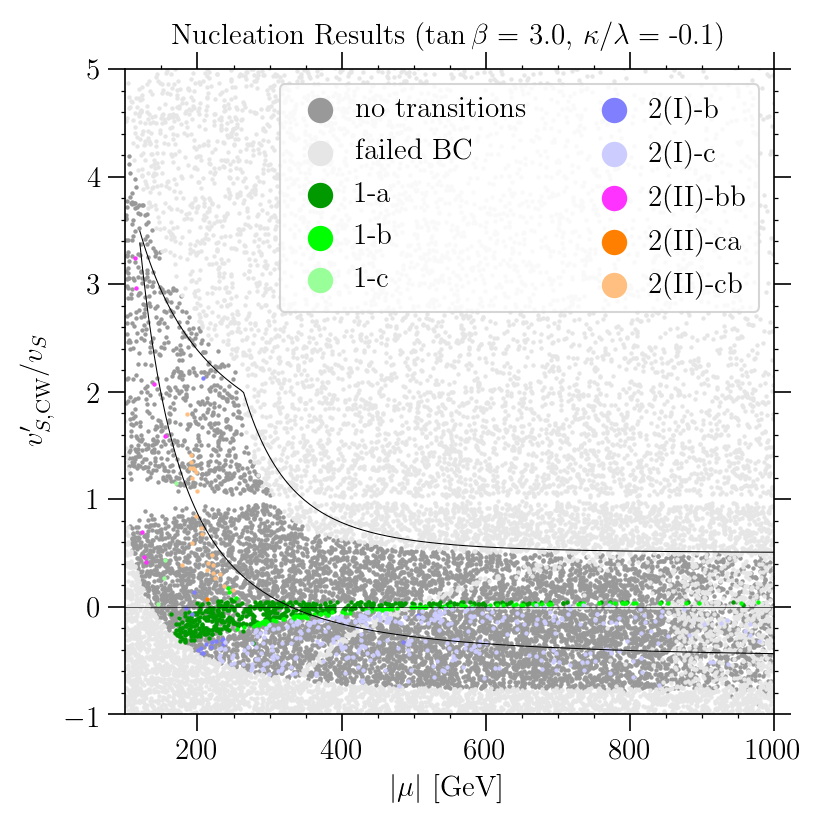}
   \caption{Same as figure~\ref{fig:NumRes_kaplam-0.1_tb_1.5}, but for $\tan\beta = 3$ and $\kappa/\lambda = -0.1$}
   \label{fig:NumRes_kaplam-0.1_tb_3.0}
\end{figure}

\begin{figure}
   \includegraphics[width=.49\linewidth]{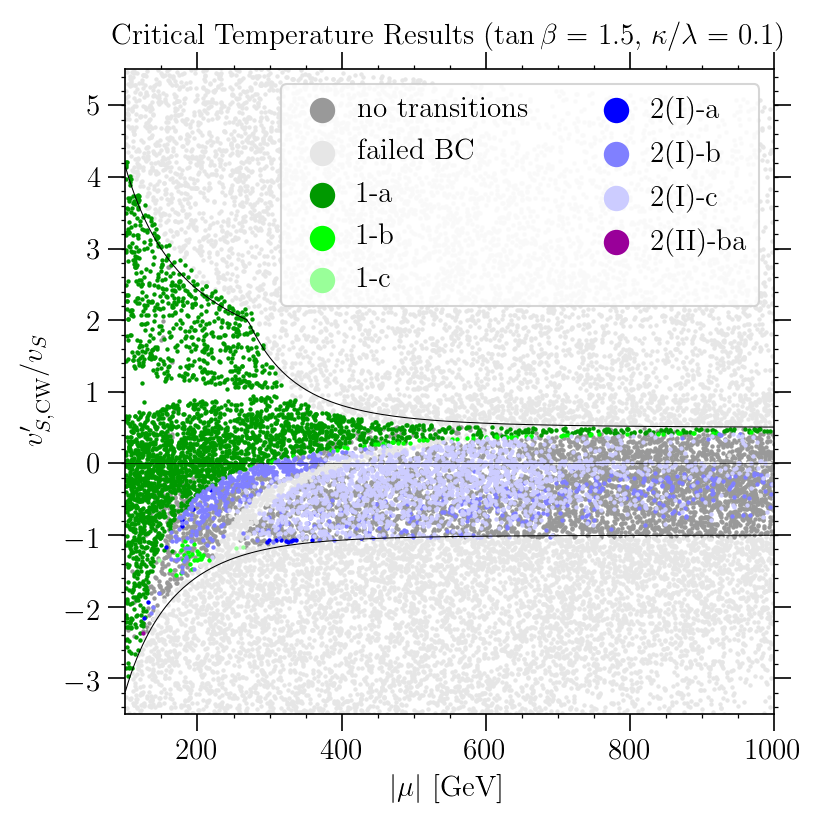}
   \includegraphics[width=.49\linewidth]{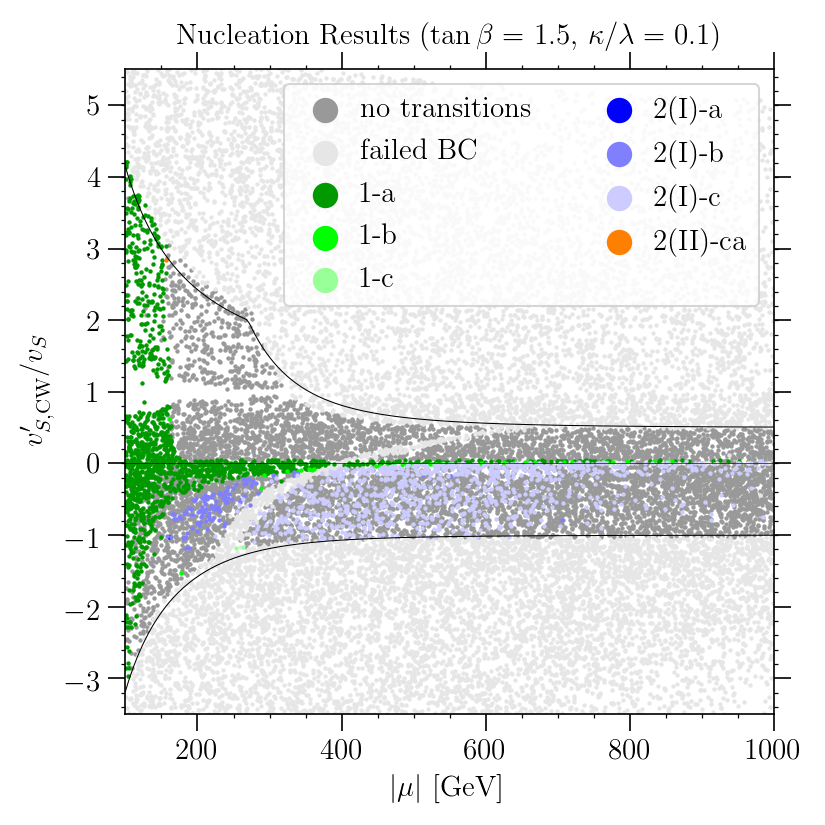}
   \caption{Same as figure~\ref{fig:NumRes_kaplam-0.1_tb_1.5}, but for $\tan\beta = 1.5$ and $\kappa/\lambda = 0.1$.}
   \label{fig:NumRes_kaplam0.1_tb_1.5}

   \includegraphics[width=.49\linewidth]{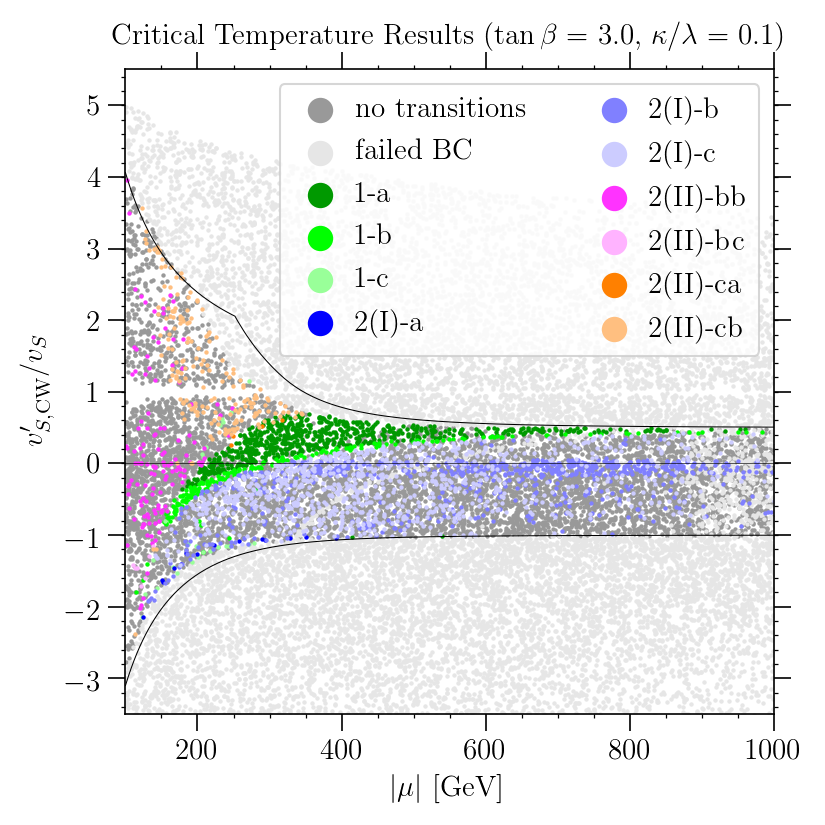}
   \includegraphics[width=.49\linewidth]{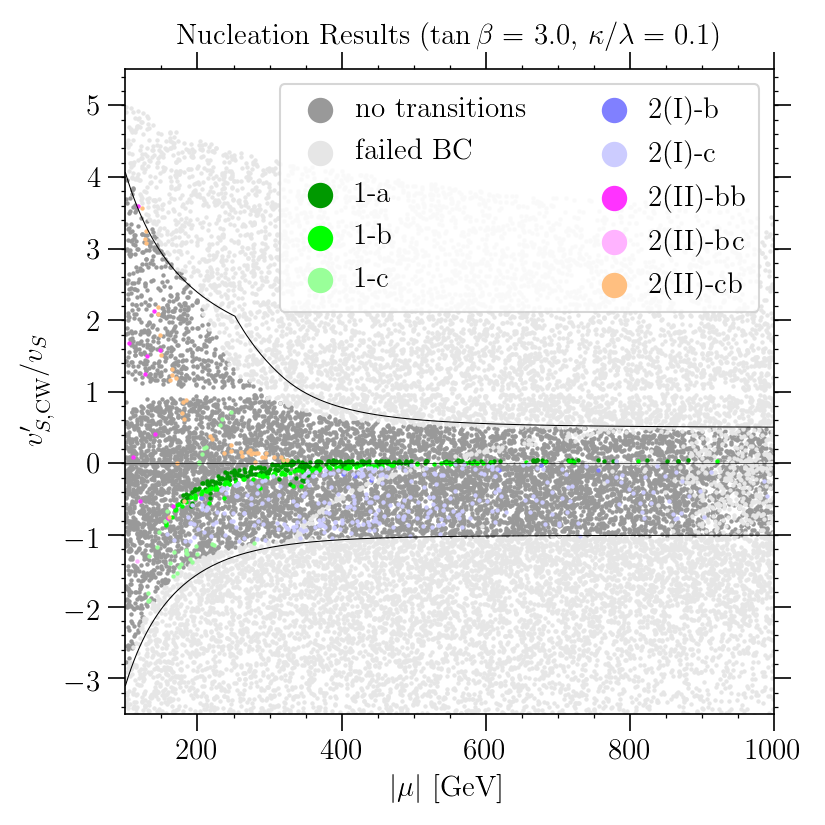}
   \caption{Same as figure~\ref{fig:NumRes_kaplam-0.1_tb_1.5}, but for $\tan\beta = 3.0$ and $\kappa/\lambda = 0.1$.} 
   \label{fig:NumRes_kaplam0.1_tb_3.0}
\end{figure}

\begin{figure}
   \includegraphics[width=.49\linewidth]{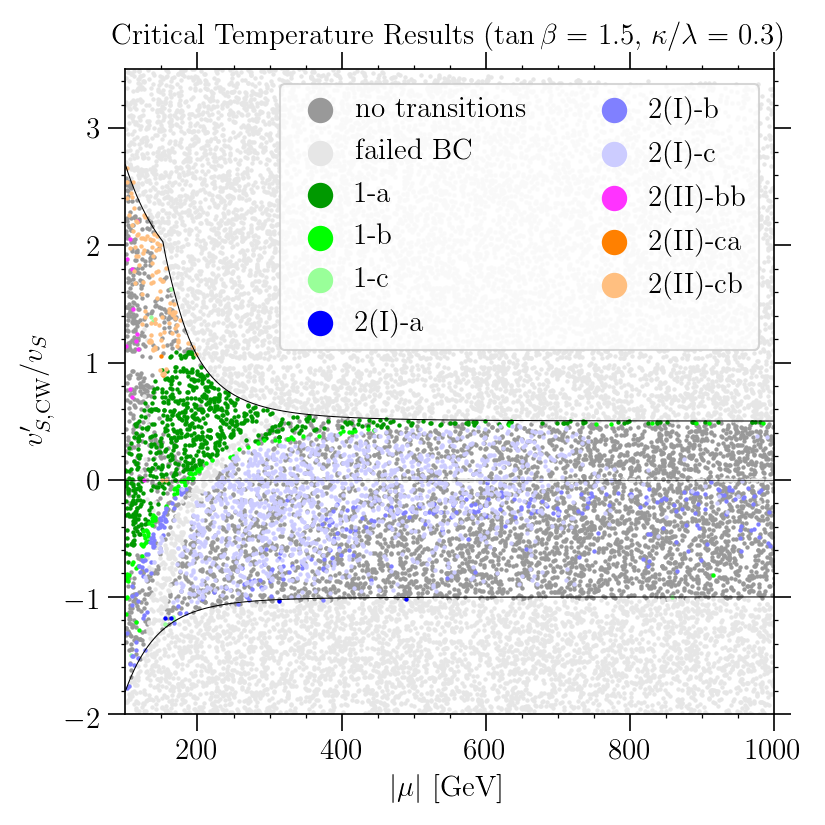}
   \includegraphics[width=.49\linewidth]{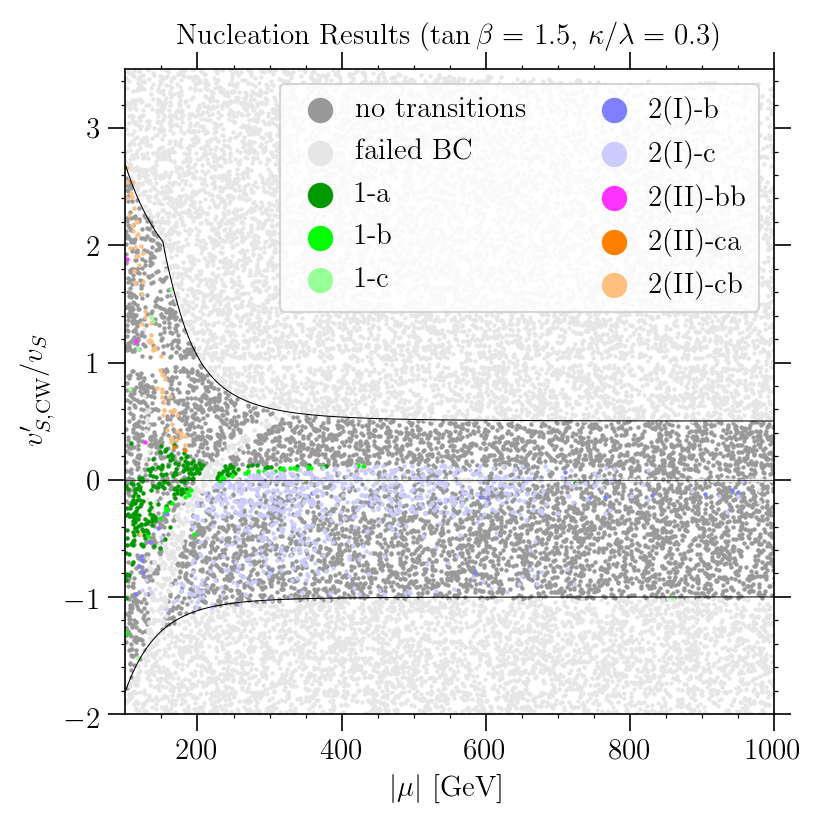}
   \caption{Same as figure~\ref{fig:NumRes_kaplam-0.1_tb_1.5}, but for $\tan\beta = 1.5$ and $\kappa/\lambda = 0.3$.} 
   \label{fig:NumRes_kaplam0.3_tb_1.5}

   \includegraphics[width=.49\linewidth]{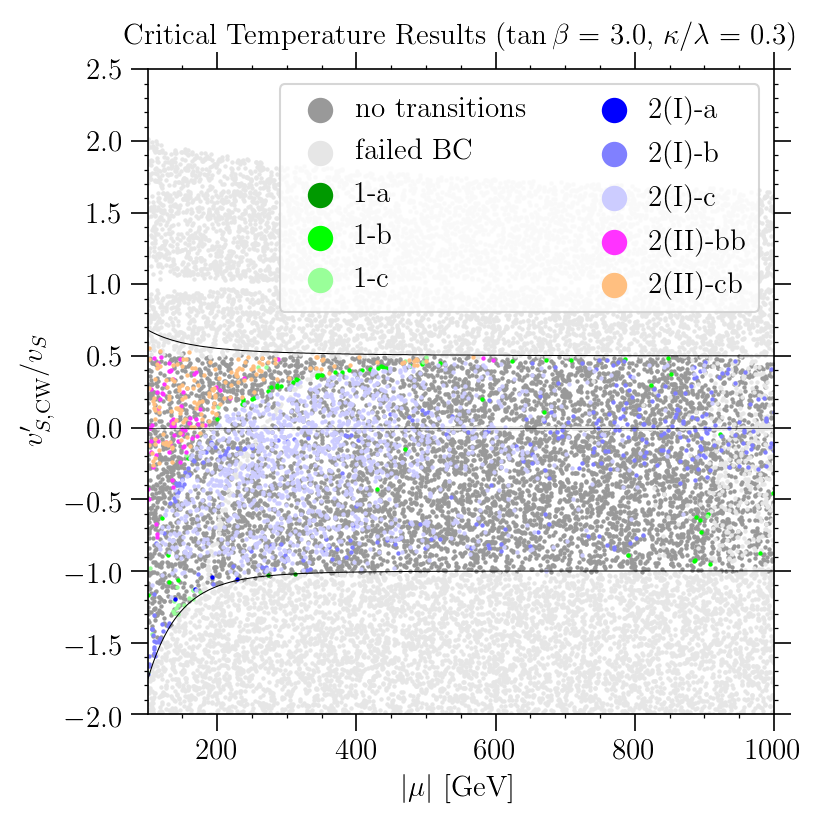}
   \includegraphics[width=.49\linewidth]{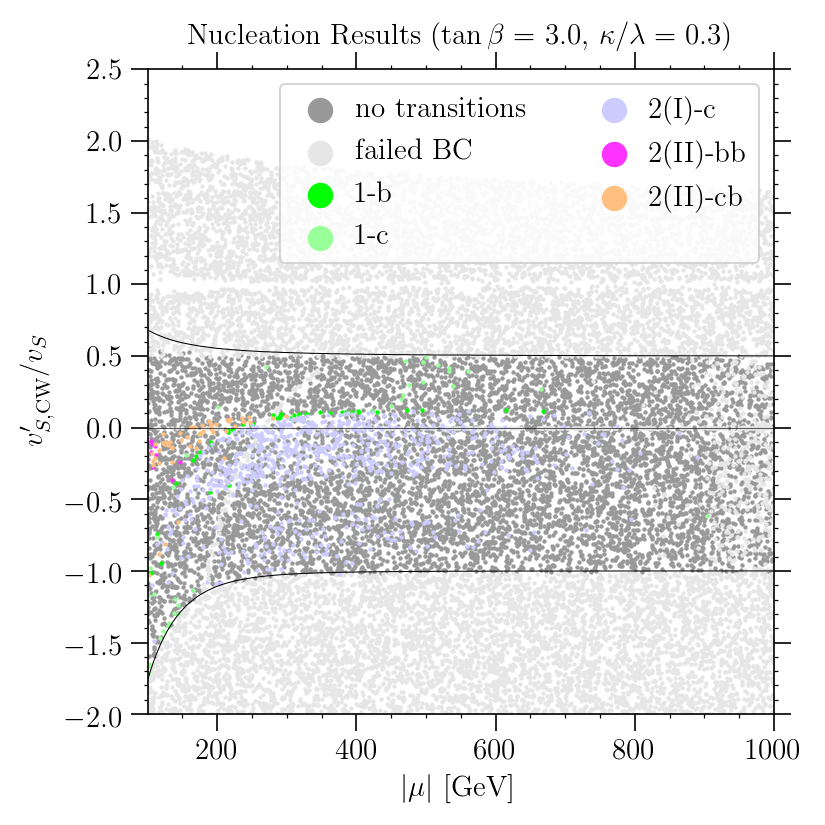}
   \caption{Same as figure~\ref{fig:NumRes_kaplam-0.1_tb_1.5}, but for $\tan\beta = 3.0$ and $\kappa/\lambda = 0.3$.} 
   \label{fig:NumRes_kaplam0.3_tb_3.0}
\end{figure}

In this section, we compare the phase transition patterns obtained from the nucleation calculation with the ones suggested by the analysis of the temperature-dependent vacuum structure at the critical temperatures. In figures~\ref{fig:NumRes_kaplam-0.1_tb_1.5}--\ref{fig:NumRes_kaplam0.3_tb_3.0}, the color-coding of the points in the left panels shows the phase transition patterns suggested by the critical temperature calculation. In the right panels of figures~\ref{fig:NumRes_kaplam-0.1_tb_1.5}--\ref{fig:NumRes_kaplam0.3_tb_3.0}, we color-code the points according to the thermal histories obtained from the nucleation calculation. Comparing the left and right panels, we see that the thermal histories obtained from the nucleation calculation differ significantly from those the critical temperature analysis suggests, leading to a marked shift in the regions of parameter space which allows for a SFOEWPT.

Let us begin by discussing the results for $\tan\beta = 1.5$ and $\kappa/\lambda = -0.1$, shown in figure~\ref{fig:NumRes_kaplam-0.1_tb_1.5}. For the critical temperature results, shown in the left panel, we observe that one-step SFOEWPT patterns (``1-a'', dark green points) occur at the upper range of the values of $\vspCW/v_S$ allowed by the boundary conditions, and that the range of $\vspCW/v_S$ for which we find such ``1-a'' transition patterns becomes wider for smaller values of $\left|\mu\right|$. For smaller values of $\vspCW/v_S$ and larger values of $\left|\mu\right|$, we find two-step transition patterns where the intermediate phase is in the singlet-only direction (``2(I)'', blue points). However, except for a few ``2(I)-a'' points at values of $\mu \simeq 250$--$300\,$GeV and small values of $\left|\vspCW/v_S\right|$, the EWPT for these points is weakly first order (``2(I)-b'') or a second order transition (``2(I)-c'') as indicated by the lighter blue shades of the points.

Qualitatively, the patterns suggested by the critical temperature calculation can mostly be understood from the discussion of the zero-tem\-pe\-ra\-ture vacuum structure in section~\ref{sec:VacStr}. The left panel of figure~\ref{fig:wpw_plane_kaplam-0.1} shows the different constraints on the zero-temperature vacuum structure (at tree level) for the same slice of parameter space as figure~\ref{fig:NumRes_kaplam-0.1_tb_1.5}. At large values of $v_S'/v_S$, the trivial minimum is deeper than the physical minimum, indicated by the green shade in figure~\ref{fig:wpw_plane_kaplam-0.1}. Thus, towards large $\vspCW/v_S$, we expect the trivial minimum to play a large role in the thermal history, and accordingly, we find one-step transitions from the trivial to the physical minimum in this region of parameter space in the left panel of figure~\ref{fig:NumRes_kaplam-0.1_tb_1.5}. Similarly, for small values of $v_S'/v_S$, the minimum in the singlet-only direction is deeper than the physical minimum (blue shaded region in figure~\ref{fig:wpw_plane_kaplam-0.1}), hence, the singlet-only phase plays a larger role in the thermal history, explaining the appearance of ``2(I)'' transition patterns for smaller values of $\vspCW/v_S$. 

Focusing now on the results of the nucleation calculation, we should recall that electroweak baryogenesis requires a SFOEWPT, i.e. one of the phase transition patterns labeled with an ``a'' in our shorthand notation. The only such patterns we observe for $\tan\beta = 1.5$ and $\kappa/\lambda = -0.1$ in the right panel of figure~\ref{fig:NumRes_kaplam-0.1_tb_1.5} are direct one-step transitions (``1-a'', dark green points), that occur for a narrow range of values $\vspCW/v_S \sim 0$. At small values of $\left|\mu\right|$, the range of values of $\vspCW/v_S$ for which we find SFOEWPTs widens slightly, before being truncated by the boundary conditions. For values of $\vspCW/v_S$ just below the ``1-a'' patterns, we find one-step transitions from the trivial to the physical minimum which are not strong first order (``1-b'' and ``1-c'', lighter green colors). For even smaller values of $\vspCW/v_S$, we find two-step transitions where the intermediate phase is in the singlet-only direction and where the second transition step, in which electroweak symmetry is broken, is weakly first order or second order (``2(I)-b'' or ``2(I)-c'', light blue points). Note that outside of these bands in $\vspCW/v_S$, we do not find points for which the nucleation calculation indicates thermal histories ending in the physical minimum. This should be contrasted with the phase transition patterns suggested by the critical temperature calculation, where we observe ``1-a'' patterns at much larger values of $\vspCW/v_S$. The nucleation calculation points to a very different region of parameter space for SFOEWPTs than the critical temperature calculation, except for a small overlap of the ``1-a''-regions at $\left|\vspCW/v_S\right| \ll 1$ and the smallest values of $\left|\mu\right|$ allowed by the boundary conditions. 

The reason for the mismatch between the critical temperature and nucleation results was discussed in section~\ref{sec:thermal_ana}: While the behavior of the critical temperatures can be understood from the zero-temperature vacuum structure, the nucleation probability is controlled by the height of the barrier separating the local minima, and the distance in field space between the local minima. For all parameter points, the thermal evolution starts in the trivial minimum at high temperatures. For large values of $v_S'/v_S$, the barriers around the trivial minimum are large, making the tunneling probability prohibitively small. Hence, for larger values of $v_S'/v_S$, the fields are ``stuck'' at $H^{\rm SM} = H^{\rm NSM} = H^{\rm S} = 0$, even if at zero temperature the trivial minimum is no longer the global minimum of the potential as required by the boundary conditions. For $v_S'/v_S \to 0$, the zero-temperature effective potential becomes flat in the singlet direction around the trivial point, and for $v_S'/v_S < 0$ the trivial point turns into a saddle point of the potential, see eq.~\eqref{eq:mS_condition}. For small values of $\left|v_S'/v_S\right|$, thermal effects can still give rise to a barrier around the trivial minimum at finite temperatures, while for large negative values of $v_S'/v_S$, thermal effects can no longer overcome the zero-temperature shape of the potential to give rise to the barrier required for a SFOEWPT. This behavior of the barrier explains why the nucleation calculation singles out the region around $\vspCW/v_S = 0$ for a SFOEWPT in the right panel of figure~\ref{fig:NumRes_kaplam-0.1_tb_1.5}.

For $\tan\beta = 3$, shown in figure~\ref{fig:NumRes_kaplam-0.1_tb_3.0}, we find similar results as for $\tan\beta = 1.5$. Beginning with the critical temperature results (left panel), the main difference is that for the larger values of $\tan\beta$, we observe that two-step transition patterns (``2(II)'', orange and magenta points) appear at small values of $\left|\mu\right|$. This is somewhat difficult to understand from the analysis in section~\ref{sec:VacStr}. The constraints coming from local minima in the doublet-like directions (orange shade in figure~\ref{fig:wpw_plane_kaplam-0.1}) are the only vacuum structure constraints depending on the value of $\tan\beta$. However, as mentioned in section~\ref{sec:Num_BC}, the doublet-like directions are subject to large radiative corrections, explaining the mismatch between the region where ``2(II)'' patterns appear in our numerical results and the orange shaded region of the tree-level vacuum structure analysis in figure~\ref{fig:wpw_plane_kaplam-0.1}. The appearance of the ``2(II)'' patterns can however be understood from the region of parameter space for which $m_{H_u}^2 < 0$, eq.~\eqref{eq:mHu_condition}. In section~\ref{sec:thermal_ana}, this condition was discussed in the context of the zero-temperature barrier in the $H_u$-direction disappearing for $m_{H_u}^2 \lesssim 0$, leading to large tunneling rates at finite temperature. To understand the critical temperature results, it is more relevant to note that for $m_{H_u}^2 < 0$, the trivial point $H^{\rm SM} = H^{\rm NSM} = H^{\rm S} = 0$ becomes a saddle point in the $H_u$-direction, suggesting that a local minimum should appear in the doublet-like direction. For $\tan\beta = 3$ and $\kappa/\lambda = -0.1$, at tree-level, $m_{H_u}^2 \lesssim 0$ for $\left|\mu\right| \lesssim 230\,$GeV, explaining the appearance of ``2(II)'' patterns in the small-$|\mu|$ region of the left panel of figure~\ref{fig:NumRes_kaplam-0.1_tb_3.0}. For $\tan\beta = 1.5$ and $\kappa/\lambda = -0.1$, shown in figure~\ref{fig:NumRes_kaplam-0.1_tb_1.5}, instead, $m_{H_u}^2 \lesssim 0$ for $\left|\mu\right| \lesssim 125\,$GeV. Such small values of $\left|\mu\right|$ are forbidden by the boundary conditions, and thus, we do not see ``2(II)'' patterns appear in figure~\ref{fig:NumRes_kaplam-0.1_tb_1.5}.

Comparing the nucleation calculation results for $\kappa/\lambda = -0.1$ and $\tan\beta = 1.5$ with those for $\tan\beta = 3$, shown in the right panel of figures~\ref{fig:NumRes_kaplam-0.1_tb_1.5} and \ref{fig:NumRes_kaplam-0.1_tb_3.0}, respectively, we see that the preferred region of parameter space for a SFOEWPT is almost independent of the value of $\tan\beta$. The main difference is that for $\tan\beta = 3$, points with smaller values of $\left|\mu\right|$ satisfy the boundary conditions, leading to the band of points around $\vspCW/v_S = 0$ for which we find SFOEWPTs (``1-a'', dark green points) extending to lower values of $\left|\mu\right|$ than for $\tan\beta = 1.5$. For $\tan\beta = 3$, we also see the emergence of two-step transition patterns, where electroweak symmetry is broken in the intermediate phase, (``2(II)'', orange and magenta points) for positive values of $\vspCW/v_S$ and small values of $\left|\mu\right|$. As discussed around eq.~\eqref{eq:mHu_condition}, for small values of $\left|\mu\right|$, the barrier around the trivial point in the $H_u$ direction disappears. Note however that these points (except for one parameter point at $\vspCW/v_S \sim 0$) do not feature a SFOEWPT step, but both steps are weakly first order or second order.

Let us now discuss the results for $\kappa/\lambda = 0.1$, shown in figures~\ref{fig:NumRes_kaplam0.1_tb_1.5} and~\ref{fig:NumRes_kaplam0.1_tb_3.0} for $\tan\beta =1.5$ and $\tan\beta=3$, respectively. Comparing the $\kappa/\lambda = -0.1$ critical temperature results (left panels) with those for $\kappa/\lambda = 0.1$, we find that many of the features remain the same. The two main differences are that the boundary conditions relax for small values of $\left|\mu\right|$, allowing a larger range of values for $\vspCW/v_S$, and that for $\tan\beta = 3$, ``2(II)'' patterns appear even more prominently in the low $\left|\mu\right|$ region. The behavior of the boundary conditions is discussed in section~\ref{sec:Num_BC}, hence, we focus on the latter difference here. As for the $\kappa/\lambda = -0.1$ case, the appearance of ``2(II)'' patterns can be understood from the region of parameter space where $m_{H_u}^2 < 0$. From eq.~\eqref{eq:mHu_condition}, we find that, for $\tan\beta = 3$ and $\kappa/\lambda = -0.1$, the mass parameter for $H_u$ becomes tachyonic for $\left|\mu\right| \lesssim 230\,$GeV, while for $\kappa/\lambda = 0.1$, this critical value increases to $\left|\mu\right| \lesssim 320\,$GeV. Accordingly, we see that ``2(II)'' patterns appear for larger values of $\left|\mu\right|$ for $\tan\beta = 3$ and $\kappa/\lambda = 0.1$ (left panel of figure~\ref{fig:NumRes_kaplam0.1_tb_3.0}) than for $\kappa/\lambda = -0.1$ (left panel of figure~\ref{fig:NumRes_kaplam-0.1_tb_3.0}). 

Let us now concentrate on the nucleation results for $\kappa/\lambda = 0.1$. For $\tan\beta = 1.5$, see the right panel of figure~\ref{fig:NumRes_kaplam0.1_tb_1.5}, we find SFOEWPTs in the same regions of parameter space as for $\kappa/\lambda=-0.1$ (figure~\ref{fig:NumRes_kaplam-0.1_tb_1.5}), with the exception of the $\left|\mu\right| \lesssim 150\,$GeV region, in which points failed the boundary conditions for $\kappa/\lambda = -0.1$. For $\kappa/\lambda = 0.1$, the boundary conditions are satisfied in this region of parameter space, and we see that for these small values of $\left|\mu\right|$, one-step SFOEWPT patterns (``1-a'', dark green points) appear for virtually the entire range of $\vspCW/v_S$ allowed by the boundary conditions. As discussed above, for small values of $\left|\mu\right|$, the barrier in the $H_u$ direction can become small. More important for the small $\left|\mu\right|$ region in this slice of the parameter space, the barrier in the singlet direction also becomes small for $\left|\kappa \mu\right| \ll 100\,$GeV, since $m_S^2 \propto \kappa^2 \mu^2 (v_S'/v_S)$, see eq.~\eqref{eq:mS_condition}, allowing for a SFOEWPT even if $\vspCW/v_S$ takes values far from zero. 

For $\tan\beta = 3$, we likewise find similar behavior for $\kappa/\lambda = 0.1$ (right panel of figure~\ref{fig:NumRes_kaplam0.1_tb_3.0}) and for $\kappa/\lambda = -0.1$ (right panel of figure~\ref{fig:NumRes_kaplam-0.1_tb_3.0}). Here, the main difference is that for $\kappa/\lambda = 0.1$, two-step transition patterns where electroweak symmetry is broken in the intermediate phase (``2-II'') play a larger role than for $\kappa/\lambda = -0.1$, restricting the values for which we find SFOEWPTs to a narrower band of values of $\vspCW/v_S$. This can again be understood from the range of values for which $H_u$ becomes tachyonic around the trivial point. Note that the presence of this tachyonic direction in the effective potential (at zero temperature) makes it more difficult to achieve transition patterns favorable for baryogenesis, which we see reflected in the absence of ``1-a'' transition patterns for $\left|\mu\right| \lesssim 200\,$GeV in the right panel of figure~\ref{fig:NumRes_kaplam0.1_tb_3.0}. 
 
Considering finally the critical temperature results for $\kappa/\lambda = 0.3$ (left panels of figures~\ref{fig:NumRes_kaplam0.3_tb_1.5} and~\ref{fig:NumRes_kaplam0.3_tb_3.0}), we find that compared to the results for smaller values of $\kappa/\lambda$, two-step transition patterns play a much larger role. Comparing eq.~\eqref{eq:V3phys} with eq.~\eqref{eq:V3vsp}, we see that the depth of the singlet-like minimum is much more sensitive to the value of $\kappa/\lambda$ than the depth of the physical minimum, and thus, the minimum in the singlet-only direction plays a larger role in the thermal history for larger values of $\kappa/\lambda$, leading to ``2(I)'' patterns (blue points) appearing more prominently for $\kappa/\lambda = 0.3$ than for $\kappa/\lambda = -0.1$ and~0.1. Likewise, we see ``2(II)'' patterns (orange and magenta points) appearing more prominently in the region of parameter space not ruled out by the boundary conditions. For $\tan\beta = 1.5$ and $\kappa/\lambda = 0.3$, we find from eq.~\eqref{eq:mHu_condition} that $m_{H_u}^2 < 0$ (at zero temperature) for $\left|\mu\right| \lesssim 180\,$GeV, while for $\kappa/\lambda = 0.3$, the critical value is $\left|\mu\right| \lesssim 840\,$GeV. 

Regarding the nucleation results, for $\tan\beta = 1.5$ and $\kappa/\lambda = 0.3$, shown in the right panel of figure~\ref{fig:NumRes_kaplam0.3_tb_1.5}, we find SFOEWPTs for small values of $\left|\mu\right|$ and $\left|\vspCW/v_S\right|$. The scaling of the depths of the respective local minima with $v_S'/v_S$ becomes faster the larger the value of $\left|\kappa/\lambda\right|$, making the change in phase transition behavior with the value of $\vspCW/v_S$ more rapid for this larger value of $\kappa/\lambda$ than what we have observed for lower values of $\kappa/\lambda$. Thus, the range of $\vspCW/v_S$ leading to (one-step) SFOEWPTs is smaller for all values of $\left|\mu\right|$ than what we found for $\kappa/\lambda = \pm 0.1$. Furthermore, we observe that ``2(II)'' transition patterns appear for small values of $\left|\mu\right|$ due to the disappearance of the barrier in the $H_u$ direction. This behavior is even more pronounced for $\tan\beta = 3$ and $\kappa/\lambda = 0.3$, see the right panel of figure~\ref{fig:NumRes_kaplam0.3_tb_3.0}. In this slice of parameter space, $m_{H_u}^2 < 0$ (at zero temperature) for $\left|\mu\right| \lesssim 840\,$GeV, and we do not find any parameter points with a SFOEWPT.

We stress that for all slices of parameter space shown in figures~\ref{fig:NumRes_kaplam-0.1_tb_1.5}--~\ref{fig:NumRes_kaplam0.3_tb_3.0}, the region providing favorable conditions for electroweak baryogenesis via a SFOEWPT differs markedly when the thermal history is inferred from the nucleation calculation instead of the simpler calculation of studying only the vacuum structure at the critical temperatures. While the critical temperature results can be explained from the zero-temperature vacuum structure, the regions of parameter space where SFOEWPTs actually nucleate can only be understood when considering the barriers of the effective potential. We find that SFOEWPTs can only nucleate if $\left|\vspCW/v_S\right| \ll 1$ and $\left|\kappa/\lambda\right|$ is not too large, leading to a small barrier in the singlet direction. If $\left|\kappa \mu\right|$ is significantly smaller than the weak scale, larger values of $\vspCW/v_S$ can still lead to a small barrier in the singlet direction and a successful SFOEWPT. For larger values of $\kappa/\lambda$ and $\tan\beta$, the barrier in the $H_u$ direction disappears in the small $\left|\mu\right|$ region, leading to multi-step phase transition patterns where electroweak symmetry is broken in the intermediate phase, and typically, no SFOEWPT is realized. 

A collection of five benchmark points which exemplify the different types of phase transition behavior we observe in the different regions of the parameter space is provided in appendix~\ref{app:BP_points}.

In figure~\ref{fig:joined_myvspvs_mHmhs}, we collect the results of our scans over the different slices of parameter space shown separately in figures~\ref{fig:NumRes_kaplam-0.1_tb_1.5}--~\ref{fig:NumRes_kaplam0.3_tb_3.0}. As before, we classify points based on the thermal histories suggested by the critical temperature calculation in the left panels, while in the right panels, parameter points are color-coded according to the results of the nucleation calculation. In order to highlight the region of parameter space for which the respective calculations indicate a SFOEWPT, we show only the points falling in one of the transition patterns ``1-a'', ``2(I)-a'', ``2(II)-aa'', ``2(II)-ab'', ``2(II)-ac'', ``2(II)-ba'', or ``2(II)-ca'' in figure~\ref{fig:joined_myvspvs_mHmhs}. In the upper panels, we show results in the $\left|\mu\right|$ vs. $\vspCW/v_S$ plane. Comparing the left and the right panels, it is evident that the critical temperature calculation gives a misleading picture of the parameter space favorable for electroweak baryogenesis. We note also that a one-step SFOEWPT (``1-a'', green points) is by far the most generic possibility to realize a SFOEWPT in the NMSSM. While multi-step transitions including a SFOEWPT step can occur in the NMSSM, our results suggest that such transition patterns require very particular combinations of parameters, making them rare in a (random) parameter scan.

\subsection{Collider and Dark Matter Phenomenology} \label{sec:Num_pheno}

\begin{figure}
   \includegraphics[width=0.49\linewidth]{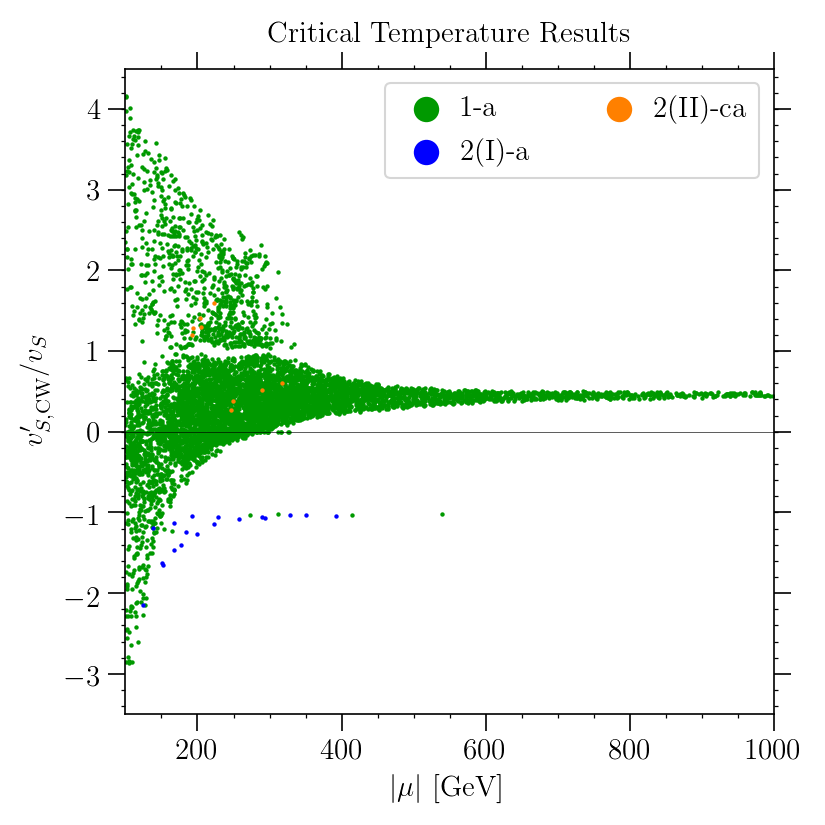}
   \includegraphics[width=0.49\linewidth]{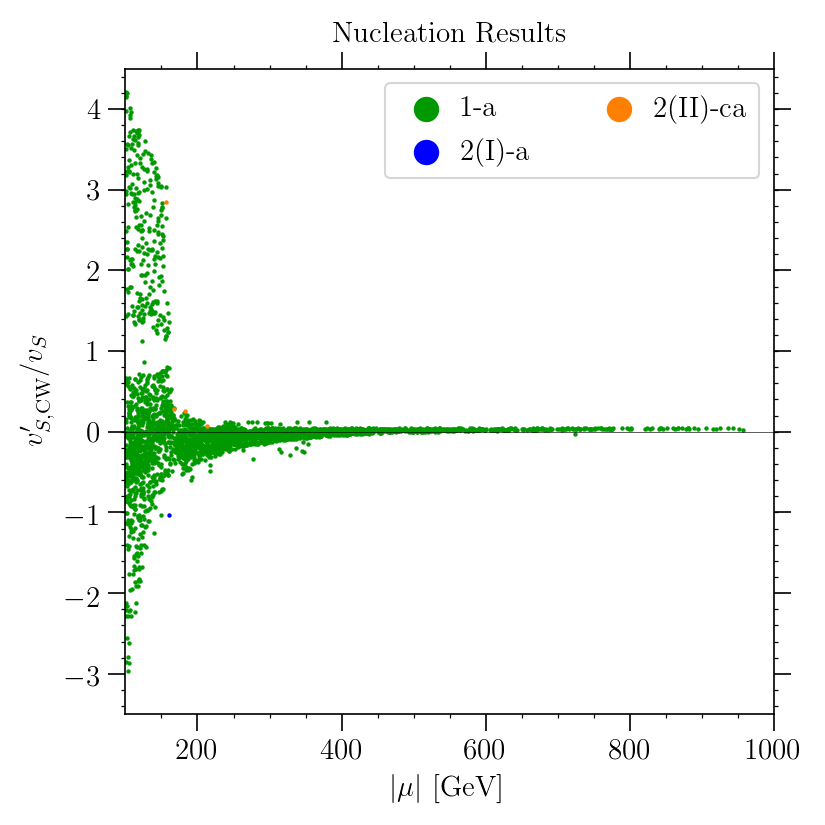}
   
   \includegraphics[width=0.49\linewidth]{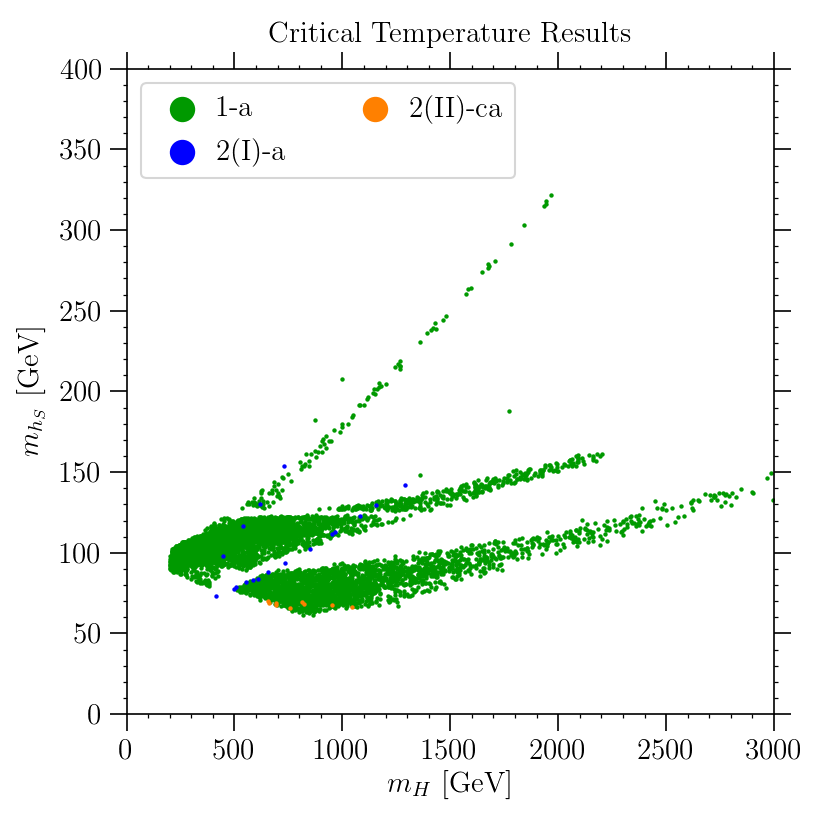}
   \includegraphics[width=0.49\linewidth]{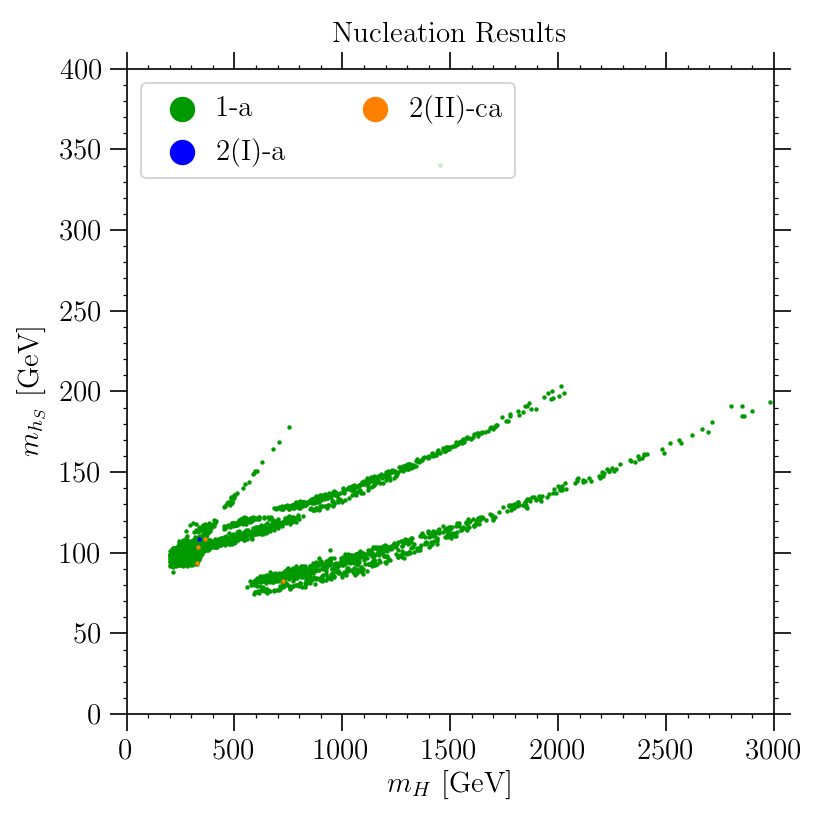}
   \caption{Points collected from our combined parameter scans ($\tan\beta = 1.5,3$ and $\kappa/\lambda = -0.1, 0.1, 0.3$) for which the critical temperature calculation (left panels) or the nucleation calculation (right panels) indicates a SFOEWPT. In the upper panels, we plot the points in the same plane as in figures~\ref{fig:NumRes_kaplam-0.1_tb_1.5}--\ref{fig:NumRes_kaplam0.3_tb_3.0}, while in the lower panels we show parameter points in the plane of the masses of the non-SM-like CP-even Higgs mass eigenstates.}
   \label{fig:joined_myvspvs_mHmhs}
\end{figure}

In this section we discuss the prospects for collider searches to cover the region of parameter space where we find SFOEWPTs and comment on the possibility of realizing a dark matter candidate in this parameter space.

In the lower panels of figure~\ref{fig:joined_myvspvs_mHmhs}, we show the points from our parameter scans for which we find a SFOEWPT in the plane of the masses of the two non-SM-like neutral CP-even Higgs bosons. Recall that we denote the state with the largest $H^{\rm S}$ component by $h_S$, and the state with the largest $H^{\rm NSM}$ component by $H$. Comparing the left and the right panels, we observe that, similar to what we saw in the $\left|\mu\right|$ vs.~$\vspCW/v_S$ plane, the results based on the full nucleation calculation lead to a considerably tighter relation between $m_H$ and $m_{h_S}$ for points with SFOEWPTs than the results of the critical temperature calculation, as well as a significant shift of the preferred region of parameter space. As we have seen above, SFOEWPTs occur in the region of parameter space where $\left|v_S'/v_S\right| \ll 1$, or $\left|\kappa \mu\right| \ll 100\,$GeV. In this limit, the mass of the singlet-like mass eigenstate (at tree-level and in the alignment limit), is approximately given by
\begin{equation} \label{eq:mhS}
   m_{h_S}^2 \approx \sin^2(2\beta) \left\{ \frac{\kappa^2}{\lambda^2} \frac{M_A^2}{2} + \lambda^2 v^2 \left[ 1 - \frac{\kappa}{\lambda} \left( \frac{1 + 2 \cos^2(2\beta)}{\sin(2\beta)} \right) \right] \right\} \;,
\end{equation} 
while the mass of the doublet-like mass eigenstate is approximately 
\begin{equation} \label{eq:mH}
   m_H^2 \sim M_A^2 \sim 4\mu^2/\sin^2(2\beta)\;.
\end{equation}
Due to the overall dependence $m_{h_S} \propto M_A \sin(2\beta)| \kappa/\lambda|$, the mass of $h_S$ decreases with growing values of $\tan\beta$. Furthermore, $m_{h_S}$ grows faster with $m_H$ for larger $\left|\kappa/\lambda\right|$ values, and the dependence of $m_{h_S}$ on the sign of $\kappa/\lambda$ is small unless $\left|\kappa/\lambda\right|$ takes large values. These properties, together with the distribution of points in the $\left|\mu\right|$--$\vspCW/v_S$ plane for the respective values of $\kappa/\lambda$ and $\tan\beta$ shown in figures~\ref{fig:NumRes_kaplam-0.1_tb_1.5}--\ref{fig:NumRes_kaplam0.3_tb_3.0}, allow us to understand the relation between $m_H$ and $m_{h_S}$ visible in the lower right panel of figure~\ref{fig:joined_myvspvs_mHmhs}. For instance, points on the right, for which one obtains the largest values of $M_H$ and the smallest values of $m_{h_S}$ for a given $M_H$, correspond to $\tan\beta = 3$ and $\kappa/\lambda = \pm 0.1$. The points on the left, which correspond to $\tan\beta = 1.5$, separate in two branches. The branch with the lowest values of $m_{h_S}$ corresponds to $\kappa/\lambda = \pm 0.1$, while the branch with the largest values of $m_{h_S}$ correspond to $\kappa/\lambda = 0.3$.

While we leave a study of the collider phenomenology of the region of parameter space where we find a SFOEWPT for future work, we can make some broad statements. As we have seen in section~\ref{sec:Num_Tcrit}, see also figures~\ref{fig:NumRes_kaplam-0.1_tb_1.5}--\ref{fig:joined_myvspvs_mHmhs}, SFOEWPTs can be realized in the NMSSM for small values of $\left|\kappa/\lambda\right|$ and $\tan\beta$, and not too large values of $\left|\mu\right|$, leading to relatively light non-SM-like Higgs bosons. From eq.~\eqref{eq:mH} we find that the doublet-like state can be as light as $m_H \sim 200\,$GeV for $\tan\beta = 1.5$ and $\left|\mu\right| \sim 100\,$GeV, as shown in the lower right panel of figure~\ref{fig:joined_myvspvs_mHmhs}. Similarly, the singlet-like state can be as light as $m_{h_S} \sim 70\,$GeV for $\tan\beta = 3$, $\kappa/\lambda = 0.1$, and $\left|\mu\right| \sim 100\,$GeV. Despite the relatively small masses, this region of parameter space is challenging to probe at colliders: The direct production cross section of the singlet-like state is suppressed by its small doublet component, we find $\left|C_{h_S}^{\rm NSM}\right| \lesssim 10\,\%$ for the points featuring a SFOEWPT. The doublet-like state $H$ has sizable production cross sections. However, its decay patterns make it challenging to probe for the small values of $\tan\beta$ preferred by a SFOEWPT. Considering the decays into pairs of SM fermions, due to the small value of $\tan\beta$, the decay mode into top-quark pairs will dominate if kinematically accessible. Thus, for $m_H \gtrsim 350\,$GeV, the branching ratio into pairs of top quarks will be large and this final state is very challenging to probe at hadron colliders such as the LHC~\cite{Dicus:1994bm,Barcelo:2010bm,Barger:2011pu,Bai:2014fkl,Jung:2015gta,Craig:2015jba,Gori:2016zto,Carena:2016npr}. For $m_H \lesssim 350\,$GeV on the other hand, the $H \to \tau^+\tau^-$ channel could provide some sensitivity. However, due to the preference for small values of $\left|\mu\right|$ and $\left|\kappa/\lambda\right|$, the Higgsinos and singlinos will be relatively light; their mass parameters are $\mu$ and $2\kappa\mu/\lambda$, respectively. Thus, decays of $H$ into pairs of neutralinos will be kinematically allowed in the parameter region preferred by a SFOEWPT, and the associated branching ratios will be sizable, suppressing $H \to \tau^+\tau^-$ decays. The final states arising from decays of $H$ into neutralinos are challenging to probe at the LHC, see, for example, refs.~\cite{Moortgat:529361,Denegri:2001pn,Han:2013gba,Craig:2015jba,Gori:2018pmk,Liu:2020muv}. Out of the di-boson final states, decays of $H$ into two SM(-like) states, e.g. $h_{125} h_{125}$, $ZZ$, and $W^+ W^-$ will be strongly suppressed due to alignment~\cite{Carena:2015moc,Baum:2018zhf}. However, the branching ratios into final states containing at least one singlet-like boson, such as $h_{125} h_S$ or $a_S Z$, will be sizable if kinematically allowed~\cite{Kang:2013rj,King:2014xwa,Carena:2015moc,Ellwanger:2015uaz,Costa:2015llh,vonBuddenbrock:2016rmr,Baum:2017gbj,Ellwanger:2017skc,Heng:2018kyd,Baum:2018zhf,Baum:2019uzg}, making these channels a promising means to explore the region of parameter space preferred for a SFOEWPT.

Considering the neutralino sector, we find that the region of parameter space where a SFOEWPT is realized features light singlinos. However, a singlino-like neutralino is only a good dark matter candidate if its spin-independent cross section is suppressed by the so-called blind-spot cancellations, see, for example, refs.~\cite{Cheung:2014lqa,Badziak:2015exr,Baum:2017enm}. For a singlino-like dark matter candidate, the blind-spot condition in the NMSSM is $2 \kappa / \lambda \approx \sin 2\beta$, requiring larger values of $\kappa/\lambda$ or $\tan\beta$ than those for which we find SFOEWPTs. On the other hand, the value of the bino mass parameter $M_1$ has practically no influence on the SFOEWPT\footnote{In our calculation, $M_1$ enters only via the radiative corrections, see eq.~\eqref{eq:mneuhat}. Any effect on the phase transition pattern of a given parameter point from changing the value of $M_1$ can be counteracted by, e.g., modifying the value of $M_2$.}. Thus, the most promising dark matter scenario in the region of parameter space where we find SFOEWPTs is a bino-like lightest neutralino. The interaction cross sections of such a bino-like neutralino can be sufficiently small to be compatible with the null results from direct detection type experiments without requiring additional (blind-spot) cancellations~\cite{Huang:2014xua,Baum:2017enm}. However, its couplings are too small to provide the correct dark matter relic density via standard thermal production. For $\left|M_1\right| \gtrsim m_t$, the correct relic density for a bino-like lightest neutralino can be achieved via co-annihilation with the singlino-like neutralino in the so-called new well-tempered scenario, where $\left|M_1\right| \sim |2\kappa \mu/\lambda|$~\cite{Baum:2017enm}. The bulk of the region of parameter space where we find SFOEWPTs features smaller values of $\left|\mu\right|$. There, the correct relic density for a bino-like lightest neutralino could be achieved via resonant annihilation through the singlet-like CP-even or CP-odd states, $h_S$ or $a_S$, requiring the mass of the lightest neutralino $\chi_1$ to satisfy $m_{\chi_1} \simeq m_{h_S}/2$ or $m_{\chi_1} \simeq m_{a_S}/2$, respectively. Alternatively, the NMSSM neutralinos may be unstable (on cosmological scales) and the dark matter may be provided by particles not included in the NMSSM, like axions and axinos (see, for example, ref.~\cite{Baer:2010wm}).

\section{Summary and Conclusions} \label{sec:conc}

Electroweak baryogenesis is a compelling scenario for the generation of the baryon asymmetry of the Universe. It relies on the presence of a Strong First Order Electroweak Phase Transition (SFOEWPT). The Standard Model (SM) of particle physics does not provide appropriate conditions for electroweak baryogenesis, hence, new physics is needed for realizing a SFOEWPT. Calculating the phase transitions in models of new physics is numerically expensive, and hence, most studies in the literature content themselves with studying the vacuum structure at the critical temperatures. At the critical temperature, the role of the global minimum of the potential passes from one local minimum to another, hence, this calculation ensures that a necessary condition for a first order phase transition is met. However, the critical temperature calculation does not ensure that the (quantum-mechanical) tunneling rate through the barrier separating the false from the true vacuum is large enough for such a first order phase transition to occur. In this work, we have investigated if a more complete calculation including the computation of the nucleation probability is necessary to understand the phase transition patterns in models of new physics. As an example model, we chose the Next-to-Minimal Supersymmetric extension of the Standard Model (NMSSM). 

We focused our case study of the NMSSM on the region of parameter space where alignment-without-decoupling is realized. For the purposes of the phase transition, the remaining four-dimensional parameter space is well described by the set of parameters $\kappa/\lambda$, $\tan\beta$, $\left|\mu\right|$, and $v_S'/v_S$, where $v_S'$ is the vev of the singlet $H^{\rm S}$ at an extremum of the effective potential in the singlet-only direction, and $v_S$ is the vev of $H^{\rm S}$ at the physical minimum.\footnote{We will suppress the subscript ``CW'' which we use to differentiate between the vev of the tree-level potential ($v_S'$) and of the effective potential after including radiative corrections ($\vspCW$) in the main text here. We refer the reader to section~\ref{sec:Numerical} for a more detailed discussion of our results.} 

Using extensive parameter scans, we have demonstrated that successful nucleation of a SFOEWPT occurs mostly in a narrow region of parameter space where $\left|v_S'/v_S\right|$ takes small values, and that the range of $v_S'/v_S$ leading to a SFOEWPT becomes increasingly narrow for larger values of $\kappa/\lambda$, $\tan\beta$, and $\left|\mu\right|$. This region of parameter space differs markedly from what one would have inferred from the critical temperature calculation alone, that, in general, suggests a SFOEWPT for much larger values of $v_S'/v_S$. The difference between the two results can be understood from the shape of the effective potential. In the region of the parameter space suggested by the critical temperature calculation, the barriers around the trivial minimum, where the thermal evolution of the model begins at very high temperatures, are large, leading to prohibitively small tunneling rates. However, the barrier in the singlet direction diminishes for small values of $\left|v_S'/v_S\right|$, enabling tunneling from the trivial minimum. As we have shown, the requirement on the values for $v_S'/v_S$ loosens for values of $\left|\kappa \mu\right|$ far below the weak scale. The dependence of the parameter region where we find a SFOEWPT on the value of $\tan\beta$ arises mainly from the disappearance of the barrier in the $H_u$-direction, triggering a phase transition which tends to lead to thermal histories incompatible with electroweak baryogenesis. The barrier in the $H_u$-direction disappears for small values of $\left|\mu\right|$, and the range of values of $\left|\mu\right|$ for which this occurs is broader for larger values of $\tan\beta$ and $\kappa/\lambda$.

Note that our findings are obtained in a perturbative expansion of the effective potential (to one loop, improved by relevant resummations), and, hence, may be affected by the well-known shortcomings associated with this expansion~\cite{Linde:1980ts, Ginsparg:1980ef, Appelquist:1981vg, Kajantie:1995dw, Laine:1998qk, Kajantie:1995kf, Laine:2017hdk, Kainulainen:2019kyp, Niemi:2020hto, Ekstedt:2020abj, Croon:2020cgk}. Nonetheless, our results stress the relevance of computing the nucleation probability to obtain the regions of parameter space promising for electroweak baryogenesis via a SFOEWPT. Our computations strongly rely on the accuracy of \texttt{CosmoTransitions}, thus, they would profit from corroboration with an independent calculation of the tunneling rate. 

While we have focused on the phase transitions, the region of parameter space where a SFOEWPT occurs also leads to interesting collider and dark matter phenomenology. We find masses of the singlet-like state $70\,{\rm GeV} \lesssim m_{h_S} \lesssim 200\,$GeV. The mass of the new doublet-like Higgs $H$, on the other hand, depends more strongly on $\tan\beta$. At moderate values of $\tan\beta$, we find $m_H \gtrsim 350\,$GeV, and hence, $H$ decays prominently into pairs of top quarks. For smaller values of $\tan\beta \sim 1.5$, $H$ can be lighter than the top pair production threshold. Although in principle this enhances the branching ratio into tau leptons, collider searches in conventional SM decay modes of these non-SM-like Higgs bosons are quite challenging due to the presence of decays into light non-standard Higgs, neutralino, and chargino states. The most promising search channels seem to arise via the so-called Higgs cascade decays, e.g., $H \to h_{125} + h_S$. We reserve a more detailed study of the collider phenomenology for future investigation.

The preference for small values of $\kappa/\lambda$ for a SFOEWPT implies the presence of a light singlino in the spectrum. While the spin-independent cross section of such a singlino is too large to be compatible with the null results from direct detection experiments in the region of parameter space where we find a SFOEWPT, a viable dark matter candidate could be realized via a bino-like lightest neutralino, if its annihilation cross section is enhanced through co-annihilation or resonant annihilation. 

In closing, we would like to stress that arguably the most important result of this work is that the nucleation calculation yields qualitatively different results for the phase transition patterns in the NMSSM than what the simpler analysis based only on the vacuum structure at the critical temperatures suggests. While our numerical results are obtained in the NMSSM, we expect similar behavior to appear in other models where multiple scalar fields participate in the EWPT. Our results emphasize that, in order to infer the regions of parameter space where electroweak baryogenesis can be realized, it is critical to compute the thermal histories based on the nucleation probabilities. 

\acknowledgments

SB thanks the University of Chicago for hospitality during various stages of this work.
SB is supported in part by NSF Grant PHY-1720397, DOE HEP QuantISED award \#100495, the Gordon and Betty Moore Foundation Grant GBMF7946, and the Vetenskapsr\r{a}det (Swedish Research Council) through contract No. 638-2013-8993 and the Oskar Klein Centre for Cosmoparticle Physics. 
NRS is supported in part by U.S. Department of Energy under Contract No. DE-SC0007983 and Wayne State University. 
This manuscript has been authored by Fermi Research Alliance, LLC under Contract No. DE-AC02-07CH11359 with the U.S. Department of Energy, Office of Science, Office of High Energy Physics. 
Work at University of Chicago is supported in part by U.S. Department of Energy grant number DE-FG02-13ER41958. 
Work at ANL is supported in part by the U.S. Department of Energy under Contract No. DE-AC02-06CH11357. 
The work of MC, NRS and CW was partially performed at the Aspen Center for Physics, which is supported by National Science Foundation grant PHY-1607611.

\appendix
\section{Benchmark Points} \label{app:BP_points}
\FloatBarrier


\begin{table}[tbh]
   \centering
   \setlength{\extrarowheight}{4pt}
   \begin{tabular}{c|cccc|ccc|c}
      \hline\hline
      & $\tan\beta$ & $\kappa/\lambda$ & $\mu$ [GeV] & $v_S'/v_S$ & $\lambda$ & $A_\kappa$ [GeV] & $A_\lambda$ [GeV] & $\vspCW/v_S$ \\
      \hline
      BP1 & $3.00$ & $-0.10$ & $250$ & $1.82$ & $0.639$ & $141$ & $885$ & $-9.21 \times 10^{-3}$ \\
      BP2 & $3.00$ & $-0.10$ & $243$ & $3.22$ & $0.639$ & $205$ & $858$ & $0.513$\\
      BP3 & $3.00$ & $-0.10$ & $500$ & $1.53$ & $0.639$ & $253$ & $1770$ & $2.93 \times 10^{-2}$ \\
      BP4 & $1.50$ & $0.10$ & $123$ & $4.74$ & $0.670$ & $-141$ & $242$ & $2.50$ \\
      BP5 & $3.00$ & $0.10$ & $-141$ & $2.38$ & $0.639$ & $95.2$ & $-441$ & $-0.049$ \\
      \hline\hline
   \end{tabular}
   \caption{Parameters controlling the scalar potential for our benchmark points. The left-most block of parameters contains the input parameters for our \texttt{CosmoTransitions} calculation. The middle block contains derived parameters. In particular, we note that $\lambda$ and $A_\lambda$ are fixed by the alignment conditions, see eqs.~\eqref{eq:Align1}--\eqref{eq:MAdef}, and $A_\kappa$ is fixed by $v_S'$, see eq.~\eqref{eq:vsprime}. In the right-most column we show the ratio $\vspCW/v_S$ which is obtained from the effective potential after including radiative corrections, see eq.~\eqref{eq:vspCW} and associated discussion.}
   \label{tab:BP_params}
\end{table}

\begin{table}[tbh]
   \centering
   \setlength{\extrarowheight}{4pt}
   \begin{tabular}{c|ccc|ccc}
      \hline\hline
      & $m_{h_{125}}$ [GeV] & $m_H$ [GeV] & $m_{h_S}$ [GeV] & $C_{h_{125}}^{\rm SM}$ & $C_{h_{125}}^{\rm NSM}$ & $C_{h_{125}}^{\rm S}$ \\ [3pt]
      \hline
      BP1 & $125$ & $848$ & $87.4$ & $1.0$ & $-7.2 \times 10^{-5}$ & $-6.4 \times 10^{-3}$\\
      BP2 & $125$ & $822$ & $76.7$ & $1.0$ & $-1.8 \times 10^{-4}$ & $-5.5 \times 10^{-3}$ \\
      BP3 & $125$ & $1690$ & $124$ & $1.0$ & $4.9 \times 10^{-3}$ & $-0.097$ \\
      BP4 & $125$ & $257$ & $94.1$ & $1.0$ & $-8.8 \times 10^{-3}$ & $-0.024$ \\
      BP5 & $125$ & $469$ & $63.7$ & $1.0$ & $-3.5 \times 10^{-4}$ & $5.1 \times 10^{-3}$ \\
      \hline\hline
      & $C_H^{\rm SM}$ & $C_H^{\rm NSM}$ & $C_H^{\rm S}$ & $C_{h_S}^{\rm SM}$ & $C_{h_S}^{\rm NSM}$ & $C_{h_S}^{\rm S}$ \\ [3pt]
      \hline
      BP1 & $7.4 \times 10^{-4}$ & $0.99$ & $0.10$ & $6.3 \times 10^{-3}$ & $-0.10$ & $0.99$ \\
      BP2 & $7.7 \times 10^{-4}$ & $0.99$ & $0.11$ & $5.4 \times 10^{-3}$ & $-0.11$ & $0.99$ \\
      BP3 & $2.2 \times 10^{-4}$ & $1.0$ & $0.052$ & $0.097$ & $-0.052$ & $0.99$ \\
      BP4 & $0.014$ & $0.98$ & $0.21$ & $0.022$ & $-0.21$ & $0.98$ \\
      BP5 & $1.3 \times 10^{-3}$ & $0.98$ & $-0.20$ & $-4.9 \times 10^{-3}$ & $0.20$ & $0.98$ \\
      \hline\hline
   \end{tabular}
   \caption{Physical masses of the CP-even Higgs bosons and composition in terms of the interaction states of the extended Higgs basis for our benchmark points.}
   \label{tab:BP_mass}
\end{table}

\begin{table}[tbh]
   \centering
   \setlength{\extrarowheight}{4pt}
   \begin{tabular}{c|cccrcl}
      \hline\hline
      & $i$ & pattern & $T_i$ [GeV] & $\{H^{\rm SM}_{hT}, H^{\rm NSM}_{hT}, H^{\rm S}_{hT}\}$ & $\xrightarrow[\phantom{{\rm SFOEW}}]{{\rm order}}$ & $\{H^{\rm SM}_{lT}, H^{\rm NSM}_{lT}, H^{\rm S}_{lT}\}$ [GeV] \\
      \hline
      \multirow{2}{*}{BP1} & $T_c$ & 1-a & $130$ & $\{ 0.0 ,~~ 0.0 ,~~ 0.0 \}$ & $\xrightarrow[\phantom{{\rm SFOEW}}]{{\rm SFOEW}}$ & $\{ 157 ,~~ 5.1 ,~~ 479 \}$\\
         \cline{2-7}
         & $T_n$ & 1-a & $75$ & $\{ 0.0 ,~~ 0.0 ,~~ 0.0 \}$ & $\xrightarrow[\phantom{{\rm SFOEW}}]{{\rm SFOEW}}$ & $\{ 235 ,~~ 0.8 ,~~ 546 \}$\\
      \hline
      \multirow{2}{*}{BP2} & $T_c$ & 1-a & $102$ & $\{ 0.0 ,~~ 0.0 ,~~ 0.0 \}$ & $\xrightarrow[\phantom{{\rm SFOEW}}]{{\rm SFOEW}}$ & $\{211 ,~~ 3.1 , ~~ 502\}$\\
         \cline{2-7} 
         & $T_n$ & --- & --- & & ------ & \\
      \hline
      \multirow{3}{*}{BP3} & \multirow{2}{*}{$T_c$} & \multirow{2}{*}{2(I)-c} & $218$ & $\{ 0.0 ,~~ 0.0 ,~~ 0.0 \}$ & $\xrightarrow[\phantom{{\rm SFOEW}}]{{\rm FO}}$ & $\{ 0.0 ,~~ 0.0 ,~~ 850 \}$\\
            & & & $154$ & $\{ -0.1 ,~~ 0.0 ,~~  1040 \}$ & $\xrightarrow[\phantom{{\rm SFOEW}}]{{\rm 2nd}}$ & $\{ 25 ,~~ 0.4 ,~~ 1040 \}$\\
         \cline{2-7} 
         & $T_n$ & 1-a & $105$ & $\{ 0.0 ,~~ 0.0 ,~~ 0.0 \}$ & $\xrightarrow[\phantom{{\rm SFOEW}}]{{\rm SFOEW}}$ & $\{ 210 ,~~ 0.7 ,~~  1090 \}$ \\
      \hline
      \multirow{2}{*}{BP4} & $T_c$ & 1-a & $115$ & $\{ 0.0 ,~~ 0.0 ,~~ 0.0 $ & $\xrightarrow[\phantom{{\rm SFOEW}}]{{\rm SFOEW}}$ & $\{ 182 ,~~ 2.7 ,~~ 221 \}$\\
         \cline{2-7} 
         & $T_n$ & 1-a & $100$ & $\{ 0.0 ,~~ 0.0 ,~~ 0.0 \}$ & $\xrightarrow[\phantom{{\rm SFOEW}}]{{\rm SFOEW}}$ & $\{ 210 ,~~ 1.0 ,~~ 241 \}$ \\
      \hline
      \multirow{3}{*}{BP5} & \multirow{2}{*}{$T_c$} & \multirow{2}{*}{2(II)-bb} & $137$ & $\{ 0.0 ,~~ 0.0 ,~~ 0.0 \}$ & $\xrightarrow[\phantom{{\rm SFOEW}}]{{\rm FO}}$ & $\{ 18 ,~~ 5.9 ,~~ 0.0 \}$\\
            & & & $117$ & $\{ 135 ,~~ 45 ,~~ 0.0 \}$ & $\xrightarrow[\phantom{{\rm SFOEW}}]{{\rm FO}}$ & $\{ 170 ,~~ 17 ,~~ -196 \}$\\
         \cline{2-7}
         & $T_n$ & --- & --- & & ------ & \\
      \hline\hline
   \end{tabular}
   \caption{Information detailing the phase transition behavior of our benchmark points. Note that for each benchmark point, the values listed in the rows marked ``$T_c$'' contain the phase transition patterns suggested by the critical temperature calculation, while the rows labeled ``$T_n$'' show the phase transition behavior we obtain from the nucleation calculation. See section~\ref{sec:thermal_ana} for the short-hand notation we use for the phase transition patterns. In the right-most column, we also denote the order of the transition(s) - ``SFOEW'' denotes a strong first order electroweak phase transition as defined in section~\ref{sec:thermal_ana}, ``FO'' denotes a transition that is first order but not a SFOEWPT, and ``2nd'' denotes a second order transition.}
   \label{tab:BP_phase}
\end{table}

In this appendix, we present five benchmark points, BP1--BP5, from our parameter scan that exemplify different types of phase transition behavior we observe in our calculations. For these benchmark points, we present the parameters entering the scalar potential in table~\ref{tab:BP_params}, the physical masses and mixing angles of the CP-even Higgs bosons in table~\ref{tab:BP_mass}, and information detailing the phase transition behavior in table~\ref{tab:BP_phase}.

BP1 is an example drawn from the points in our parameter scan for which the critical temperature calculation suggests a direct (one-step) SFOEWPT (``1-a'') pattern and where we find the same transition pattern at the level of the nucleation calculation. In our parameter scan, we observe such points most readily at relatively small values of $\left|\mu\right| \lesssim 250\,$GeV, small values of $\left|\kappa/\lambda\right|$, and $\vspCW/v_S \approx 0$, see e.g. figure~\ref{fig:NumRes_kaplam-0.1_tb_3.0}. Note that such values correspond to rather small values $\left|A_\kappa\right|$; in the case of BP1 we find $A_\kappa = 141\,$GeV (see table~\ref{tab:BP_params}). Considering the associated masses of the CP-even Higgs bosons shown in table~\ref{tab:BP_mass}, we find that the singlet-like CP-even state $h_S$ is relatively light with $m_{h_S} = 87.4\,$GeV, while the doublet like mass eigenstate has mass $m_H = 848\,$GeV. This large mass gap suppresses the $H^{\rm NSM} - H^{\rm S}$ mixing; the singlet-like state has a $H^{\rm NSM}$ component of $\left(C_{h_S}^{\rm NSM}\right)^2 = 1\,\%$ only, and similarly $H$ is almost purely composed of $H^{\rm NSM}$. Note also that as enforced by our alignment conditions, $h_{125}$ is very well aligned with $H^{\rm SM}$ and hence has practically identical couplings to pairs of SM particles as the SM Higgs boson. From the phase transition details presented in table~\ref{tab:BP_phase} we see sizable supercooling: while the critical temperature for the phase transition is $T_c = 130\,$GeV, the tunneling rate becomes sufficiently large to allow for nucleation only at $T_n = 75\,$GeV. 

BP2 is an example from the parameter region where the critical temperature calculation suggests a ``1-a'' transition pattern, but where the nucleation calculation shows that the tunneling rate is too suppressed to allow for nucleation, leaving the model trapped at the (meta-stable) trivial minimum, see e.g. figure~\ref{fig:NumRes_kaplam-0.1_tb_3.0}. BP2 is a parameter point with very similar parameters as BP1, except that it features a larger value of $v_S'/v_S$, leading to $\vspCW/v_S = 0.5$ instead of $\vspCW/v_S \approx 0$ for BP1. In terms of the usual NMSSM parameters, this corresponds to a larger value of $\left|A_\kappa\right|$; BP2 features $A_\kappa = 205\,$GeV. As discussed in the main text, in this parameter region the barrier around the trivial minimum becomes too high, and correspondingly the phase transition never nucleates.

BP3 is instead an example of a point where we find successful nucleation of a (one-step) SFOEWPT in the nucleation calculation for larger values of $\left|\mu\right|$ than BP1. As we can see from figures~\ref{fig:NumRes_kaplam-0.1_tb_1.5}--\ref{fig:NumRes_kaplam0.3_tb_3.0}, we observe successful SFOEWPTs only for a very narrow range of $\vspCW/v_S \approx 0$ for such larger values of $\left|\mu\right|$, and the critical temperature calculation typically indicates ``2(I)'' two-step phase transition patterns where the intermediate phase is in the singlet-only direction for these points. From table~\ref{tab:BP_phase} we see that while the critical temperature calculation suggests a ``2(I)-c'' transition pattern for BP3 with a critical temperature of $T_c = 218\,$GeV for the first step, we find from the nucleation calculation that the tunneling rate corresponding to this transition is so small that instead at $T_n = 105\,$GeV a direct transition from the trivial to the physical phase occurs. Comparing the spectrum of the CP-even Higgs bosons of BP3 to those of BP1 and BP2, see table~\ref{tab:BP_mass}, we observe that, due to the larger value of $\left|\mu\right|$, BP3 features a much heavier $H^{\rm NSM}$-like state, $m_H = 1.7\,$TeV. We also note that due to the choice of $\tan\beta = 3$, $\kappa/\lambda = -0.1$, $\mu = 500\,$GeV, and a value of $v_S'/v_S$ which yields $\vspCW/v_S \approx 0$, the singlet-like CP even mass eigenstate and $h_{125}$ are almost mass degenerate. This leads to somewhat larger $H^{\rm SM} - H^{\rm S}$ mixing than for BP1 and BP2 since the small numerical deviations from perfect alignment occurring in our parameter scan are no longer suppressed by the large difference of the diagonal entries of the CP-even mass matrix. However, the $H^{\rm S}$ component of $h_{125}$ we find for BP3, $\left(C_{h_{125}}^{\rm S}\right)^2 = 0.9\,\%$, is still so small that the phenomenology of $h_{125}$ is practically indistinguishable from that of the SM Higgs boson at the LHC. Note also that because of the small doublet components of $h_S$, $\left(C_{h_S}^{\rm SM}\right)^2 = 0.9\,\%$ and $\left(C_{h_S}^{\rm NSM}\right)^2 = 0.3\,\%$, the direct production cross sections of $h_S$ are strongly suppressed compared to those of the SM-like Higgs boson. Hence, $h_S$ does not lead to significant effects on measurements relying on direct production of a 125\,GeV Higgs boson at the LHC; such measurements are instead dominated by the contribution from $h_{125}$ which has SM-like phenomenology.

BP4 is an example from the region of parameter space at smaller values of $|\mu| \lesssim 150\,$GeV where we found ``1-a'' SFOEWPTs in the full nucleation calculation even for sizable values of $\vspCW/v_S$ in the $\tan\beta = 1.5$, $\kappa/\lambda = 0.1$ slice of the parameter space, see figure~\ref{fig:NumRes_kaplam0.1_tb_1.5}. BP4 features a value of $\mu = 123\,$GeV and a sizable value of $\vspCW/v_S = 2.5$. Note also that for BP4's input parameters, the alignment conditions enforce a much smaller value of $\left|A_\lambda\right|$ than what we see for the other benchmark points, suggestive of a relatively small barrier in directions involving the doublet-like fields. Considering the physical Higgs bosons of BP4, see table~\ref{tab:BP_mass}, the most notable feature compared to the other benchmark points is a relatively small mass of the $H^{\rm NSM}$-like CP-even state, $m_H = 257\,$GeV. Such a light doublet-like state has sizable production cross sections at the LHC. However, because of the small value of $|\mu| = 123\,$GeV, we expect $H$ to have large branching ratios into neutralinos and charginos, suppressing decays into pairs of SM particles and potentially allowing such a state to efficiently escape detection at the LHC. A more detailed consideration of the collider phenomenology of BP4 is beyond the scope of this work. Considering the phase transitions of BP4, see table~\ref{tab:BP_phase}, we see that as typical in this parameter region, both the critical temperature calculation and the more complete nucleation calculation suggest a direct (``1-a'') SFOEWPT pattern. Compared to BP1, BP4 features much smaller supercooling - we find a critical temperature of $T_c = 115\,$GeV and a nucleation temperature of $T_n = 100\,$GeV. 

BP5 is an example from the parameter region where the critical temperature calculation suggests ``2(II)'' phase transition patterns, i.e. two-step patterns in which electroweak symmetry is broken in the intermediate phase and where the nucleation calculation suggests that no transition has a sufficiently large nucleation rate and hence the model would remain trapped at the (meta-stable) trivial minimum. From figure~\ref{fig:NumRes_kaplam0.1_tb_3.0} we see that such behavior occurs copiously in the small $|\mu|$ region of the $\tan\beta = 3.0$, $\kappa/\lambda = 0.1$ slice of parameter space. We chose BP5 to otherwise have similar parameters to BP1 and BP3, in particular, BP5 has $\vspCW/v_S \approx 0$, but a value of $|\mu|$ ($\mu = -141\,$GeV) much smaller than those of BP1 and BP3. Note that we chose $\kappa/\lambda = 0.1$ for BP5 instead of the value $\kappa/\lambda = -0.1$ BP1 and BP3 feature not because of the phase transition patterns we observe in this region of parameter space (which show very little difference between the two signs of $\kappa/\lambda$), but rather because the boundary conditions remove most of the parameter space around $\vspCW/v_S \approx 0$ for small values of $|\mu|$ for $\kappa/\lambda = -0.1$, see figure~\ref{fig:NumRes_kaplam-0.1_tb_3.0} and the associated discussion. We can also note that $|A_\lambda|$ takes much smaller value for BP5 than for BP1 and BP3. From table~\ref{tab:BP_mass}, we see that compared to BP1 and BP3, we find a much lighter $H^{\rm NSM}$-like mass eigenstate for BP5, $m_H = 469\,$GeV. 

\FloatBarrier

\section{Field-Dependent Masses} \label{app:field_masses}

In this appendix, we present explicit expressions for the field-dependent masses after inclusion of the leading stop corrections, but without corrections from the Coleman-Weinberg potential. As argued in section~\ref{sec:NMSSM}, it suffices to study the potential as a function of the three neutral CP-even degrees of freedom $\left\{H^{\rm SM}, H^{\rm NSM}, H^{\rm S}\right\}$. 

Let us begin by presenting the expression for the field-dependent (squared) masses in the scalar sector. These can be directly obtained from the scalar potential,
\begin{equation}
   \widehat{m}_{\Phi_i,\Phi_j} = \widehat{m}_{i,j}(H^{\rm SM}, H^{\rm NSM}, H^{\rm S}) \equiv \left.\frac{\partial^2 V}{\partial \Phi_i \partial \Phi_j}\right|_{\substack{H^{\rm SM} \neq 0 \\ H^{\rm NSM} \neq 0 \\ H^{\rm S} \neq 0 }} \;.
\end{equation}

The entries involving the CP-even interaction states are
\begin{align} 
   \widehat{m}_{H^{\rm SM}, H^{\rm SM}}^2 &= \left( m_Z^2 c_{2\beta}^2 + \lambda^2 v^2 s_{2\beta}^2 + 2 \Delta\lambda_2 v^2 s_\beta^4 \right) \left\{ 1 + \frac{3 \left[ (H^{\rm SM})^2 - 2 v^2 \right]}{4v^2} \right\} \nonumber\\
      &\quad + \frac{\lambda^2}{2} \left(1 - \frac{\kappa}{\lambda} s_{2\beta}\right) \left[ (H^{\rm S})^2 - \frac{2\mu^2}{\lambda^2} \right] - \frac{\lambda}{\sqrt{2}} \left( \frac{M_A^2}{2\mu} s_{2\beta} - \frac{\kappa\mu}{\lambda} \right) s_{2\beta} \left( H^{\rm S} - \frac{\sqrt{2}\mu}{\lambda} \right) \nonumber\\
      &\quad + \frac{(H^{\rm NSM})^2}{4v^2} \left[ m_Z^2 s_{2\beta}^2 + \lambda^2 v^2 c_{2\beta}^2 - \left( m_Z^2 - \lambda^2 v^2 \right) c_{4\beta} \right] \nonumber\\
      &\quad - \frac{3 H^{\rm SM} H^{\rm NSM}}{4v^2} \left( m_Z^2 - \lambda^2 v^2 \right) s_{4\beta} \nonumber\\
      &\quad + \frac{3 H^{\rm NSM}}{4 v} \Delta\lambda_2 v s_\beta s_{2\beta} \left( 2 H^{\rm SM} s_\beta + H^{\rm NSM} c_\beta \right) \;, 
   \label{eq:hatM2h11}
\end{align}

\begin{align} 
   \widehat{m}_{H^{\rm SM}, H^{\rm NSM}}^2 &= -\left(m_Z^2 - \lambda^2 v^2 - \Delta\lambda_2 v^2 \frac{s_\beta^2}{c_{2\beta}} \right) s_{2\beta} c_{2\beta} \left\{ 1 + \frac{3 \left[ (H^{\rm SM})^2 - 2 v^2 \right]}{4v^2} \right\} \nonumber\\
      &\quad + \frac{H^{\rm SM} H^{\rm NSM}}{4v^2} \left[ m_Z^2 \left( 1 - 3c_{4\beta} \right) + \lambda^2 v^2 \left( 1 + 3 c_{4\beta} \right) + 3 \Delta\lambda_2 v^2 s_{2\beta}^2 \right] \nonumber\\
      &\quad + \frac{3 (H^{\rm NSM})^2}{4v^2} \left( m_Z^2 - \lambda^2 v^2 + \Delta\lambda_2 v^2 \frac{c_\beta^2}{c_{2\beta}} \right) s_{2\beta} c_{2\beta} \nonumber\\
      &\quad - \frac{\lambda}{2} c_{2\beta} \left( \kappa H^{\rm S} + \frac{M_A^2}{\sqrt{2} \mu} s_{2\beta} \right) \left( H^{\rm S} - \frac{\sqrt{2} \mu}{\lambda} \right) \;, \\
\phantom{0}\nonumber\\
   \widehat{m}_{H^{\rm SM}, H^{\rm S}}^2 &= 2\lambda v \mu \left[ \frac{H^{\rm SM}}{\sqrt{2}v} \frac{H^{\rm S}}{\sqrt{2}\mu/\lambda} - \frac{M_A^2}{4\mu^2} \frac{H^{\rm SM}}{\sqrt{2}v} s_{2\beta}^2 \right. \nonumber\\
      &\qquad \left. - \frac{\kappa}{\lambda} \left( \frac{H^{\rm SM}}{\sqrt{2}v} s_{2\beta} + \frac{H^{\rm NSM}}{\sqrt{2}v} c_{2\beta} \right) \left( \frac{1}{2} + \frac{H^{\rm S} - \sqrt{2}\mu/\lambda}{\sqrt{2}\mu/\lambda} \right) \right] \nonumber\\
      &\quad - \frac{\sqrt{2} \lambda M_A^2}{8 \mu} H^{\rm NSM} s_{4\beta} \;, \\
\phantom{0}\nonumber\\
   \widehat{m}_{H^{\rm NSM}, H^{\rm NSM}}^2 &= M_A^2 + \left( m_Z^2 - \lambda^2 v^2 + \frac{\Delta\lambda_2 v^2}{2} \right) s_{2\beta}^2 \nonumber\\
      &\quad + \left[ m_Z^2 \left( 1 - 3 c_{4\beta} \right) + \lambda^2 v^2 \left( 1 + 3 c_{4\beta} \right) + 3 \Delta\lambda_2 v^2 s_{2\beta}^2 \right] \frac{(H^{\rm SM})^2 - 2 v^2}{8 v^2} \nonumber\\
      &\quad + \frac{\lambda^2}{2} \left(1 + \frac{\kappa}{\lambda} s_{2\beta}\right) \left[ (H^{\rm S})^2 - \frac{2\mu^2}{\lambda^2} \right] + \frac{\lambda}{\sqrt{2}} \left( \frac{M_A^2}{2\mu} s_{2\beta} - \frac{\kappa\mu}{\lambda} \right) s_{2\beta} \left( H^{\rm S} - \frac{\sqrt{2}\mu}{\lambda} \right) \nonumber\\
      &\quad + \frac{3 H^{\rm SM} H^{\rm NSM}}{4v^2} \left( m_Z^2 s_{4\beta} - \lambda^2 v^2 s_{4\beta} + 2 \Delta\lambda_2 v^2 s_{2\beta} c_\beta^2 \right) \nonumber\\
      &\quad + \frac{3 (H^{\rm NSM})^2}{4v^2} \left( m_Z^2 c_{2\beta}^2 + \lambda^2 v^2 s_{2\beta}^2 + 2 \Delta\lambda_2 v^2 c_\beta ^4 \right) \;, \\
\phantom{0}\nonumber\\
   \widehat{m}_{H^{\rm NSM}, H^{\rm S}}^2 &= -\frac{H^{\rm SM}}{\sqrt{2}} \lambda \mu c_{2\beta} \left\{ \frac{\kappa}{\lambda} \left[ 1 + \frac{2 \left( H^{\rm S} - \sqrt{2}\mu/\lambda \right)}{\sqrt{2}\mu/\lambda} \right] + \frac{M_A^2}{2\mu^2} s_{2\beta} \right\} \nonumber\\
      &\quad + \sqrt{2} H^{\rm NSM} \lambda \mu \left\{ \frac{M_A^2}{4\mu^2} s_{2\beta}^2 + \frac{\lambda}{\sqrt{2}\mu} H^{\rm S} + \frac{\kappa}{2\lambda} s_{2\beta} \left[ 1 + \frac{2 \left( H^{\rm S} - \sqrt{2}\mu/\lambda \right)}{\sqrt{2}\mu/\lambda} \right] \right\} \;, \\
\phantom{0}\nonumber\\
   \widehat{m}_{H^{\rm S}, H^{\rm S}}^2 &= \frac{\lambda^2 v^2}{2} s_{2\beta} \left\{ \frac{M_A^2}{2\mu^2} s_{2\beta} - \frac{\kappa}{\lambda} \left[ 1 + \frac{(H^{\rm SM})^2 - 2 v^2}{v^2} \right] \right\} \nonumber\\
      &\quad + \frac{\kappa\mu}{\lambda} \left\{ A_\kappa \left[ 1 + \frac{2 \left( H^{\rm S} - \sqrt{2}\mu/\lambda \right)}{\sqrt{2}\mu/\lambda} \right] + 4 \frac{\kappa\mu}{\lambda} \left[ 1 + \frac{(H^{\rm S})^2 - 2\mu^2/\lambda^2}{4\mu^2/3\lambda^2} \right] \right\} \nonumber\\
      &\quad + \lambda^2 v^2 \frac{(H^{\rm SM})^2 - 2 v^2}{2 v^2} - \frac{\lambda \kappa H^{\rm NSM}}{2} \left[ 2 H^{\rm SM} c_{2\beta} - H^{\rm NSM} \left( \frac{\lambda}{\kappa} + s_{2\beta} \right) \right] \;.
\end{align}

The entries involving the CP-odd states are
\begin{align}
   \widehat{m}_{A^{\rm NSM}, A^{\rm NSM}}^2 &= M_A^2 + \left[ \lambda^2 v^2 \left( 3 + c_{4\beta} \right) - 2 m_Z^2 c_{2\beta}^2 + \Delta\lambda_2 v^2 s_{2\beta}^2 \right] \frac{(H^{\rm SM})^2 - 2 v^2}{8v^2} \nonumber\\
      &\quad + \frac{\lambda^2}{2} \left( 1 + \frac{\kappa}{\lambda} s_{2\beta} \right) \left[ (H^{\rm S})^2 - \frac{2\mu^2}{\lambda^2} \right] + \frac{\lambda}{\sqrt{2}} \left( \frac{M_A^2}{2\mu} s_{2\beta} - \frac{\kappa\mu}{\lambda} \right) s_{2\beta} \left( H^{\rm S} - \frac{\sqrt{2}\mu}{\lambda} \right) \nonumber\\
      &\quad + \left( \lambda^2 v^2 s_{2\beta}^2 + m_Z^2 c_{2\beta}^2 + 2 \Delta\lambda_2 v^2 c_\beta^4 \right) \frac{(H^{\rm NSM})^2}{4v^2} \nonumber\\
      &\quad - \left( \lambda^2 v^2 s_{4\beta} - m_Z^2 s_{4\beta} - 2 \Delta\lambda_2 v^2 s_{2\beta} c_\beta^2\right) \frac{H^{\rm SM} H^{\rm NSM}}{4v^2} \;,\\
\phantom{0}\nonumber\\
   \widehat{m}_{A^{\rm NSM}, A^{\rm S}}^2 &= \lambda v \frac{H^{\rm SM}}{\sqrt{2} v} \left[ \frac{M_A^2}{2\mu} s_{2\beta} - 3 \frac{\kappa \mu}{\lambda} \left( 1 + \frac{H^{\rm S} - \sqrt{2} \mu/\lambda}{3\mu/\sqrt{2} \lambda} \right) \right] \;, \\
\phantom{0}\nonumber\\
   \widehat{m}_{A^{\rm NSM}, G^0}^2 &= - \left( m_Z^2 c_{2\beta} - \lambda^2 v^2 c_{2\beta} - \Delta\lambda_2 v^2 s_\beta^2 \right) s_{2\beta} \frac{(H^{\rm SM})^2 - 2v^2}{4v^2} \nonumber\\
      &\quad + \left( m_Z^2 c_{2\beta} - \lambda^2 v^2 c_{2\beta} + \Delta\lambda_2 v^2 c_\beta^2 \right) s_{2\beta} \frac{(H^{\rm NSM})^2}{4v^2} \nonumber\\
      &\quad + \left( m_Z^2 - \lambda^2 v^2 + \frac{\Delta\lambda_2 v^2}{2} \right) s_{2\beta}^2 \frac{H^{\rm SM} H^{\rm NSM}}{2v^2} \nonumber\\
      &\quad - \frac{\lambda\kappa}{2} c_{2\beta} \left[ (H^{\rm S})^2 - \frac{2\mu^2}{\lambda^2} \right] - \frac{\lambda}{\sqrt{2}} \left( \frac{M_A^2}{2\mu} s_{2\beta} - \frac{\kappa\mu}{\lambda} \right) c_{2\beta} \left( H^{\rm S} - \frac{\sqrt{2}\mu}{\lambda} \right) \;, \\
\phantom{0}\nonumber\\
   \widehat{m}_{A^{\rm S}, A^{\rm S}}^2 &= \frac{\lambda^2 v^2}{2} s_{2\beta} \left( \frac{M_A^2}{2\mu^2} s_{2\beta} + \frac{3\kappa}{\lambda} \right) - \frac{3 \kappa \mu A_\kappa}{\lambda} \nonumber\\
      &\quad + \kappa^2 \left[ (H^{\rm S})^2 - \frac{2\mu^2}{\lambda^2} \right] - \sqrt{2} \kappa A_\kappa \left( H^{\rm S} - \frac{\sqrt{2} \mu}{\lambda} \right) \nonumber\\
      &\quad + \lambda^2 v^2 \left( 1 + \frac{\kappa}{\lambda} s_{2\beta} \right) \frac{(H^{\rm SM})^2 - 2 v^2}{2v^2} + \lambda^2 v^2 \left( 1 - \frac{\kappa}{\lambda} s_{2\beta} \right) \frac{(H^{\rm NSM})^2}{2v^2} \nonumber\\
      &\quad + \lambda \kappa c_{2\beta} H^{\rm SM} H^{\rm NSM} \;,\\
\phantom{0}\nonumber\\
   \widehat{m}_{A^{\rm S}, G^0}^2 &= \lambda H^{\rm NSM} \left( \kappa H^{\rm S} + \frac{\kappa \mu}{\sqrt{2} \lambda} - \frac{M_A^2}{2\sqrt{2} \mu} s_{2\beta} \right) \;,\\
\phantom{0}\nonumber\\
   \widehat{m}_{G^0, G^0}^2 &= \left( m_Z^2 c_{2\beta}^2 + \lambda^2 v^2 s_{2\beta}^2 + 2 \Delta\lambda_2 v^2 s_{\beta}^4 \right) \frac{(H^{\rm SM})^2 - 2v^2}{4v^2} \nonumber\\
   &\quad - \left[ 2 m_Z^2 c_{2\beta}^2 - \lambda^2 v^2 \left( 3 + c_{4\beta} \right) - \Delta\lambda_2 v^2 s_{2\beta}^2 \right] \frac{(H^{\rm NSM})^2}{8v^2} \nonumber\\
   &\quad - \left( m_Z^2 c_{2\beta} - \lambda^2 v^2 c_{2\beta} - \Delta\lambda_2 v^2 s_\beta^2 \right) s_{2\beta} \frac{H^{\rm SM} H^{\rm NSM}}{2v^2} \nonumber\\
   &\quad + \frac{\lambda^2}{2} \left( 1 - \frac{\kappa}{\lambda} s_{2\beta} \right) \left[ (H^{\rm S})^2 - \frac{2\mu^2}{\lambda^2} \right] - \frac{\lambda}{\sqrt{2}} \left( \frac{M_A^2}{2\mu} s_{2\beta} - \frac{\kappa\mu}{\lambda} \right) s_{2\beta} \left( H^{\rm S} - \frac{\sqrt{2}\mu}{\lambda} \right) \;.
\end{align}
Note that as required, at the physical minimum, i.e. where $\widehat{m}_{\Phi_i \Phi_j} \left(H^{\rm SM}, H^{\rm NSM}, H^{\rm S}\right) \to \widehat{m}_{\Phi_i \Phi_j} \left(\sqrt{2}v, 0, \sqrt{2}\mu/\lambda\right) \equiv m_{\Phi_i \Phi_j}$,
\begin{equation}
   m_{G^0, G^0} = m_{A^{\rm NSM}, G^0} = m_{A^{\rm S}, G^0} = 0 \;,
\end{equation}
or in words, the neutral Goldstone mode $G^0$ is massless and decouples from the other CP-odd neutral states $A^{\rm NSM}$ and $A^{\rm S}$.

The elements involving the charged states are
\begin{align}
   \widehat{m}_{H^+, H^-}^2 &= M_A^2 - \lambda^2 v^2 + m_W^2 \nonumber\\
      &\quad - \left( m_Z^2 c_{2\beta}^2 + \lambda^2 v^2 s_{2\beta}^2 - 2 m_W^2 - \frac{\Delta\lambda_2 v^2}{2} s_{2\beta}^2 \right) \frac{(H^{\rm SM})^2 - 2v^2}{4v^2} \nonumber\\
      &\quad + \left( m_Z^2 c_{2\beta}^2 + \lambda^2 v^2 s_{2\beta}^2 + 2\Delta\lambda_2 v^2 c_\beta^4 \right) \frac{(H^{\rm NSM})^2}{4v^2} \nonumber\\
      &\quad + \left( m_Z^2 c_{2\beta} - \lambda^2 v^2 c_{2\beta} + \Delta\lambda_2 v^2 c_\beta^2 \right) s_{2\beta} \frac{H^{\rm SM} H^{\rm NSM}}{2v^2} \nonumber\\ 
      &\quad + \frac{\lambda^2}{2} \left( 1 + \frac{\kappa}{\lambda} s_{2\beta} \right) \left[ (H^{\rm S})^2 - \frac{2\mu^2}{\lambda^2} \right] + \frac{\lambda}{\sqrt{2}} \left( \frac{M_A^2}{2\mu} s_{2\beta} - \frac{\kappa\mu}{\lambda} \right) s_{2\beta} \left( H^{\rm S} - \frac{\sqrt{2}\mu}{\lambda} \right)\;, 
\end{align}
\begin{align}
   \widehat{m}_{H^+, G^-}^2 &= \widehat{m}_{H^-, G^+}^2 = - \left( m_Z^2 c_{2\beta} - \lambda^2 v^2 c_{2\beta} - \Delta\lambda_2 v^2 s_\beta^2 \right) s_{2\beta} \frac{(H^{\rm SM})^2 - 2v^2}{4v^2} \nonumber\\
      &\quad + \left( m_Z^2 c_{2\beta} - \lambda^2 v^2 c_{2\beta} + \Delta\lambda_2 v^2 c_\beta^2 \right) s_{2\beta} \frac{(H^{\rm NSM})^2}{4v^2} \nonumber\\
      &\quad + \left( m_Z^2 s_{2\beta}^2 + \lambda^2 v^2 c_{2\beta}^2 - m_W^2 + \frac{\Delta\lambda_2 v^2}{2} s_{2\beta}^2 \right) \frac{H^{\rm SM} H^{\rm NSM}}{2v^2} \nonumber\\
      &\quad - \frac{\lambda\kappa}{2} c_{2\beta} \left[ (H^{\rm S})^2 - \frac{2\mu^2}{\lambda^2} \right] - \frac{\lambda}{\sqrt{2}} \left( \frac{M_A^2}{2\mu} s_{2\beta} - \frac{\kappa\mu}{\lambda} \right) c_{2\beta} \left( H^{\rm S} - \frac{\sqrt{2}\mu}{\lambda} \right) \;,\\
\phantom{0}\nonumber\\
   \widehat{m}_{G^+, G^-}^2 &= \left( m_Z^2 c_{2\beta}^2 + \lambda^2 v^2 s_{2\beta}^2 + 2 \Delta\lambda_2 v^2 s_\beta^4 \right) \frac{(H^{\rm SM})^2 - 2v^2}{4v^2} \nonumber\\
      &\quad - \left( m_Z^2 c_{2\beta}^2 + \lambda^2 v^2 s_{2\beta}^2 - 2 m_W^2 - \frac{\Delta\lambda_2 v^2}{2} s_{2\beta}^2 \right) \frac{(H^{\rm NSM})^2}{4v^2} \nonumber\\
      &\quad - \left( m_Z^2 c_{2\beta} - \lambda^2 v^2 c_{2\beta} - \Delta\lambda_2 v^2 s_\beta^2 \right) s_{2\beta} \frac{H^{\rm SM} H^{\rm NSM}}{2v^2} \nonumber\\ 
      &\quad + \frac{\lambda^2}{2} \left( 1 - \frac{\kappa}{\lambda} s_{2\beta} \right) \left[ (H^{\rm S})^2 - \frac{2\mu^2}{\lambda^2} \right] - \frac{\lambda}{\sqrt{2}} \left( \frac{M_A^2}{2\mu} s_{2\beta} - \frac{\kappa\mu}{\lambda} \right) s_{2\beta} \left( H^{\rm S} - \frac{\sqrt{2}\mu}{\lambda} \right)\;.
\end{align}
At the physical minimum, we again find
\begin{equation}
   m_{G^+ G^-} = m_{H^+, G^-} = m_{H^-, G^+} = 0 \;,
\end{equation}
or in words, the charged Goldstone mode $G^\pm$ is massless and decouples from the charged Higgs $H^\pm$.

The remaining entries of the (symmetric) ($10 \times 10$) matrix of the $\widehat{m}_{\Phi_i,\Phi_j}$ not listed above vanish due to CP- and charge conservation.

The field-dependent masses for the electroweak gauge bosons are given by
\begin{align}
   &\widehat{m}_{W^\pm}^2 = \frac{g_2^2}{4} \left[ (H^{\rm SM})^2 + (H^{\rm NSM})^2 \right] \;,\\
   &\widehat{m}_Z^2 = \frac{g_1^2 + g_2^2}{4} \left[ (H^{\rm SM})^2 + (H^{\rm NSM})^2 \right] \;,
\end{align}
with the weak mixing angle $\cos\theta_W = g_2/\sqrt{g_1^2 + g_2^2} = m_W/m_Z$. The masses of the vector bosons at the physical minimum are related to the gauge couplings as
\begin{equation}
   g_1 = \sqrt{2} \sin\theta_W \frac{m_Z}{v}\:, \quad g_2 = \sqrt{2} \frac{m_W}{v} \;.
\end{equation}

For the 5 neutralinos, the (symmetric) matrix of field-dependent masses in the basis $\left\{\widetilde{B}, \widetilde{W}^3, \widetilde{H}_d^0, \widetilde{H}_u^0, \widetilde{S} \right\}$ can be written as
\begin{equation} \label{eq:mneuhat}
   \widehat{\mathcal{M}}_{\chi^0} = \begin{pmatrix} M_1 & 0 & -\frac{g_1}{2} \left( c_\beta H^{\rm SM} - s_\beta H^{\rm NSM} \right) & \frac{g_1}{2} \left( s_\beta H^{\rm SM} + c_\beta H^{\rm NSM} \right) & 0 \\
                                                        & M_2 & \frac{g_2}{2} \left( c_\beta H^{\rm SM} - s_\beta H^{\rm NSM} \right) & -\frac{g_2}{2} \left( s_\beta H^{\rm SM} + c_\beta H^{\rm NSM} \right) & 0 \\
                                                        & & 0 & -\frac{\lambda}{\sqrt{2}} H^{\rm S} & - \frac{\lambda}{\sqrt{2}} \left( s_\beta H^{\rm SM} + c_\beta H^{\rm NSM} \right) \\
                                                        & & & 0 & - \frac{\lambda}{\sqrt{2}} \left( c_\beta H^{\rm SM} - s_\beta H^{\rm NSM} \right) \\
                                                        & & & & \sqrt{2}\kappa H^{\rm S}
                                    \end{pmatrix} \;.
\end{equation}

In the basis $\psi_i^\pm = \left\{ \widetilde{W}^+, \widetilde{H}_u^+, \widetilde{W}^-, \widetilde{H}_d^-\right\}$ the field-dependent mass terms for the charginos can be written as
\begin{equation} \label{eq:mchar_field}
   \mathcal{L} \supset -\frac{1}{2} (\psi^\pm)^T \begin{pmatrix} 0 & \widehat{X}^T \\ \widehat{X} & 0 \end{pmatrix} \psi^{\pm} + {\rm h.c.} \;,
\end{equation}
where
\begin{equation}
   \widehat{X} = \begin{pmatrix} M_2 & \frac{g_2}{\sqrt{2}} \left( s_\beta H^{\rm SM} + c_\beta H^{\rm NSM} \right) \\ \frac{g_2}{\sqrt{2}} \left( c_\beta H^{\rm SM} - s_\beta H^{\rm NSM} \right) & \frac{\lambda}{\sqrt{2}} H^{\rm S} \end{pmatrix} \;.
\end{equation}
Finally, the field-dependent mass of the top quark is given by
\begin{equation}
   \widehat{m}_t = \frac{1}{\sqrt{2}} h_t \left( s_\beta H^{\rm SM} + c_\beta H^{\rm NSM} \right) \;,
\end{equation}
where the Yukawa coupling $h_t$ is related to the (running) top quark mass $m_t$ via $h_t = m_t/s_\beta v$.

We compute the contributions to the Coleman-Weinberg potential as well as to the thermal potential in the Landau gauge. This is useful since in the Landau gauge the ghosts decouple and we do not have to include them in our calculations. The quantities entering the Coleman-Weinberg and the thermal potential are the eigenvalues of the respective mass matrices. Recall that the number of degrees of freedom are $n_i = 1$ for the three neutral CP-even and three neutral CP-odd states, $n_i = 2$ for the two charged Higgs states, $n_i = 6$ for the $W
^\pm$ bosons, and $n_i = 3$ for the $Z$-boson. Out of the fermions, the top quark has $n_i = 12$ and the five neutralinos have $n_i = 2$ each. Since we wrote the chargino mass matrix, eq.~\eqref{eq:mchar_field}, in terms of four Majorana states (which combine to two physical Dirac fermions), the four eigenvalues of eq.~\eqref{eq:mchar_field} enter with $n_i = 2$ each.

\section{Counterterm Coefficients} \label{app:ct_coeff}
In order to maintain the location of the physical minimum at $\left\{H^{\rm SM}, H^{\rm NSM}, H^{\rm S}\right\} = \sqrt{2} \left\{v,0,\mu/\lambda\right\}$, preserve $m_{h_{125}} = 125\,$GeV, and $\mathcal{M}_{S,13}^2 \to 0$ (i.e. alignment of $H^{\rm S}$ and $H^{\rm S}$) after including the Coleman-Weinberg corrections, we include the counterterms given in eq.~\eqref{eq:ct_lag}. The solutions for the counterterms to satisfy these conditions are 
\begin{align}
   \delta_{m_{H_d}^2} &= -\frac{1}{2v} \left( \sqrt{2} \frac{\partial V_1}{\partial H^{\rm SM}} - \sqrt{2} \tan\beta \frac{\partial V_1}{\partial H^{\rm NSM}} - \frac{\mu}{\lambda \cos^2\beta} \frac{\partial^2 V_1}{\partial H^{\rm SM}\,\partial H^{\rm S}} \right) \;,\\
   \delta_{m_{H_u}^2} &= \frac{1}{2 v \sin^2\beta} \left[ \frac{\cos(2\beta)-2}{\sqrt{2}} \frac{\partial V_1}{\partial H^{\rm SM}} - \frac{\sin(2\beta)}{\sqrt{2}} \frac{\partial V_1}{\partial H^{\rm NSM}} \right. \nonumber \\
      &\qquad \qquad \qquad \left. + v \left( \frac{\partial^2 V_1}{\partial H^{\rm SM}\,\partial H^{\rm SM}} - m_{h_{125}}^2 \right) + \frac{\mu}{\lambda} \frac{\partial^2 V_1}{\partial H^{\rm SM}\,\partial H^{\rm S}} \right] \;,\\
   \delta_{m_S^2} &= -\frac{\lambda}{2\mu} \left( \sqrt{2} \frac{\partial V_1}{\partial H^{\rm S}} - v \frac{\partial^2 V_1}{\partial H^{\rm SM}\,\partial H^{\rm S}} \right) \;,\\
   \delta_{\lambda A_\lambda} &= \frac{1}{v \sin(2\beta)} \frac{\partial^2 V_1}{\partial H^{\rm SM}\,\partial H^{\rm S}} \;,\\
   \delta_{\lambda_2} &= \frac{1}{2 \sqrt{2} v^3 \sin^4\beta} \left[ \frac{\partial V_1}{\partial H^{\rm SM}} + \sqrt{2}v \left( m_{h_{125}}^2 - \frac{\partial^2 V_1}{\partial H^{\rm SM}\,\partial H^{\rm SM}} \right) \right] \;,
\end{align}
where 
\begin{equation}
   V_1 = V_1(T=0) = V_0^{\rm eff} + V_{\rm 1-loop}^{\rm CW} \;,
\end{equation}
is the effective potential including the Coleman-Weinberg corrections $V_{\rm 1-loop}^{\rm CW}$ at zero temperature, all derivatives are evaluated at the physical minimum, $\left\{H^{\rm SM}, H^{\rm NSM}, H^{\rm S}\right\} = \sqrt{2} \left\{v,0,\mu/\lambda\right\}$, and $m_{h_{125}}$ is an input parameter which sets the mass of the $H^{\rm SM}$ interaction eigenstate of the Higgs basis.

\section{Daisy Coefficients} \label{app:Daisy}
The Daisy coefficients $c_i$ for the thermal masses
\begin{equation}
   \widetilde{m}_i^2 = \widehat{m}_i^2 + c_i T^2 \;,
\end{equation}
can be obtained from the high-temperature limit of the thermal corrections to the effective potential,
\begin{equation} \label{eq:Daisycoeff}
   c_{ij} = \frac{1}{T^2} \left.\frac{\partial^2 V_{\rm 1-loop}^{T\neq0}(\widehat{m}^2)}{\partial\phi_i \; \partial\phi_j} \right|_{T \gg \widehat{m}^2} \;.
\end{equation}
Note that for the derivation of the Daisy coefficient, $V_{\rm 1-loop}^{T\neq0} = V_{\rm 1-loop}^{T\neq0}(\widehat{m}_i^2)$ is computed with the temperature independent field-dependent masses $\widehat{m}_i^2$, while when computing the temperature-dependent effective potential, the Daisy-resummation improved thermal masses $\widetilde{m}_i^2$ are inserted in $V_{\rm 1-loop}^{T\neq0}$ as well as in the Coleman-Weinberg potential.

Note also that while we gave explicit expressions for the $\widehat{m}_i^2$ as a function of the three neutral CP-even Higgs boson interaction states, $H^{\rm SM}$, $H^{\rm NSM}$, and $H^{\rm S}$, in appendix~\ref{app:field_masses}, when computing the Daisy coefficients via eq.~\eqref{eq:Daisycoeff}, the field-dependent masses must be inserted as a function of all bosonic fields, i.e.
\begin{equation}
   \widehat{m}_{i,j}^2 = \widehat{m}_{i,j}^2(H^{\rm SM}, H^{\rm NSM}, H^{\rm S}, A^{\rm NSM}, A^{\rm S}, H^\pm, G^0, G^\pm, Z^0, W^\pm) \;.
\end{equation}

The non-vanishing coefficients involving the neutral Higgs bosons are
\begin{align}
   c_{H^{\rm SM} H^{\rm SM}} = c_{G^0 G^0} &= \frac{\lambda^2}{4} + \frac{m_Z^2 + 2 m_W^2}{4 v^2} + \frac{m_t^2}{4v^2} + \frac{\Delta\lambda_2}{4} s_\beta^2 \;, \\
   c_{H^{\rm SM} H^{\rm NSM}} = c_{A^{\rm NSM} G^0} &= \frac{m_t^2}{4v^2} \frac{1}{t_\beta} + \frac{\Delta\lambda_2}{8} s_{2\beta} \;, \\
   c_{H^{\rm NSM} H^{\rm NSM}} = c_{A^{\rm NSM} A^{\rm NSM}} &= \frac{\lambda^2}{4} + \frac{m_Z^2 + 2 m_W^2}{4 v^2} + \frac{m_t^2}{4v^2} \frac{1}{t_\beta^2} + \frac{\Delta\lambda_2}{4} c_\beta^2 \;, \\
   c_{H^{\rm S} H^{\rm S}} = c_{A^{\rm S} A^{\rm S}} &= \frac{\lambda^2 + \kappa^2}{2} \;,
   \label{eq:Daisy}
\end{align}
and those involving the charged Higgs bosons are
\begin{align}
   c_{H^+ H^-} &= \frac{\lambda^2}{6} + \frac{m_Z^2 + 2 m_W^2}{6 v^2} + \frac{m_t^2}{4v^2} \frac{1}{t_\beta^2} + \frac{\Delta\lambda_2}{4} c_\beta^2 \;, \\
   c_{H^+ G^-} &= c_{H^{\rm SM} H^{\rm NSM}} \; \\
   c_{G^+ G^-} &= \frac{\lambda^2}{6} + \frac{m_Z^2 + 2 m_W^2}{6 v^2} + \frac{m_t^2}{4v^2} + \frac{\Delta\lambda_2}{4} s_\beta^2 \;.
\end{align}

The Daisy coefficients for the longitudinal modes of the gauge bosons are~\cite{Carrington:1991hz,Comelli:1996vm}
\begin{equation} 
   c_{W^+_L W^-_L} = c_{W^3_L W^3_L} =\frac{5}{2} g_2^2 = 5 \frac{m_W^2}{v^2} \;, \quad c_{B_L B_L} = \frac{13}{6} g_1^2 \;.
\end{equation}
Note that the photon gets a temperature-dependent mass. In order to properly account for this appearance of the longitudinal degree of freedom of the photon, the Daisy resummation improved thermal masses of the neutral electroweak gauge bosons must thus be included as the eigenvalues of mass matrix,
\begin{equation}
   \widetilde{m}^2_{Z_L,A_L}(H^{\rm SM}, H^{\rm NSM}, H^{\rm S}; T) = \frac{(H^{\rm SM})^2 + (H^{\rm NSM})^2}{4} \begin{pmatrix} g_2^2 & - g_1 g_2 \\ - g_1 g_2 & g_1^2 \end{pmatrix} + T^2 \begin{pmatrix} 5 g_2^2 /2 & 0 \\ 0 & 13 g_1^2 /6 \end{pmatrix} \;.
\end{equation}

After removing the contribution from the neutralinos and charginos to $V_{\rm 1-loop}^{T\neq0}(\widehat{m}^2)$, these results agree with the results in ref.~\cite{Athron:2019teq} (where the neutralino and chargino contribution were neglected).

\bibliographystyle{JHEP.bst}
\bibliography{nmssmbib}

\end{document}